Fmi
\documentclass[preprint]{aastex62}
\usepackage{float}
\usepackage{epsfig}
\usepackage{array}
\usepackage{graphicx}
\usepackage{graphics}
\usepackage{subfigure}
\usepackage{latexsym}
\usepackage{amsmath}
\newcommand{\nd}{\multicolumn{1}{c}{$\dots$}}

\shorttitle{SN~Ia 2018oh} \shortauthors{Li et al.}

\def\gsim{\;\lower4pt\hbox{${\buildrel\displaystyle >\over\sim}$}\;}
\def\lsim{\;\lower4pt\hbox{${\buildrel\displaystyle <\over\sim}$}\;}
\def\grls{\;\lower4pt\hbox{${\buildrel\displaystyle >\over <}$}\;}
\begin{document}
\title{Photometric and Spectroscopic Properties of Type Ia Supernova 2018oh with Early Excess Emission from the $Kepler$ 2 Observations}


\author{W.~Li}
\affil{Physics Department and Tsinghua Center for Astrophysics (THCA), Tsinghua University, Beijing, 100084, China}
\affil{Las Cumbres Observatory, 6740 Cortona Dr Ste 102, Goleta, CA 93117-5575, USA}
\author{X.~Wang}
\affil{Physics Department and Tsinghua Center for Astrophysics (THCA), Tsinghua University, Beijing, 100084, China}
\author{J.~Vink\'o}
\affil{Konkoly Observatory, MTA CSFK, Konkoly Thege M. ut 15-17, Budapest, 1121, Hungary}
\affil{Department of Optics \& Quantum Electronics, University of Szeged, Dom ter 9, Szeged, 6720 Hungary}
\affil{Department of Astronomy, University of Texas at Austin, Austin, TX, 78712, USA}
\author{J.~Mo}
\affil{Physics Department and Tsinghua Center for Astrophysics (THCA), Tsinghua University, Beijing, 100084, China}
\author{G. Hosseinzadeh}
\affil{Las Cumbres Observatory, 6740 Cortona Dr Ste 102, Goleta, CA 93117-5575, USA}
\affiliation{Department of Physics, University of California, Santa Barbara, CA 93106-9530, USA}
\affil{Harvard-Smithsonian Center for Astrophysics, 60 Garden Street, Cambridge, MA 02138, USA}
\author{D. J. Sand}
\affil{Steward Observatory, University of Arizona,  933 North Cherry Avenue, Rm. N204, Tucson, AZ 85721-0065, USA}

\author{J. Zhang}
\affil{Yunnan Observatories (YNAO), Chinese Academy of Sciences, Kunming 650216, China}
\affil{Key Laboratory for the Structure and Evolution of Celestial Objects, Chinese Academy of Sciences, Kunming 650216, China}
\affil{Center for Astronomical Mega-Science, Chinese Academy of Sciences, 20A Datun Road, Chaoyang District, Beijing, 100012, China}
\author{H. Lin}
\affil{Physics Department and Tsinghua Center for Astrophysics (THCA), Tsinghua University, Beijing, 100084, China}

\nocollaboration

\collaboration{PTSS / TNTS}
\author{T. Zhang}
\affil{National Astronomical Observatory of China, Chinese Academy of Sciences, Beijing, 100012, China}
\author{L. Wang}
\affil{Purple Mountain Observatory, Chinese Academy of Sciences, Nanjing 210034, China}
\affil{George P. and Cynthia Woods Mitchell Institute for Fundamental Physics $\&$ Astronomy, Texas A. $\&$ M. University, Department of Physics and Astronomy, 4242 TAMU, College Station, TX 77843, USA}
\author{J. Zhang}
\affil{Physics Department and Tsinghua Center for Astrophysics (THCA), Tsinghua University, Beijing, 100084, China}
\author{Z. Chen}
\affil{Physics Department and Tsinghua Center for Astrophysics (THCA), Tsinghua University, Beijing, 100084, China}
\author{D. Xiang}
\affil{Physics Department and Tsinghua Center for Astrophysics (THCA), Tsinghua University, Beijing, 100084, China}
\author{L. Rui}
\affil{Physics Department and Tsinghua Center for Astrophysics (THCA), Tsinghua University, Beijing, 100084, China}
\author{F. Huang}
\affil{Physics Department and Tsinghua Center for Astrophysics (THCA), Tsinghua University, Beijing, 100084, China}
\affil{Department of Astronomy, School of Physics and Astronomy, Shanghai Jiao Tong University, Shanghai 200240, China}
\author{X. Li}
\affil{Physics Department and Tsinghua Center for Astrophysics (THCA), Tsinghua University, Beijing, 100084, China}
\author{X. Zhang}
\affil{Physics Department and Tsinghua Center for Astrophysics (THCA), Tsinghua University, Beijing, 100084, China}
\author{L. Li}
\affil{Physics Department and Tsinghua Center for Astrophysics (THCA), Tsinghua University, Beijing, 100084, China}
\author{E. Baron}
\affil{Homer L. Dodge Department of Physics and Astronomy, University of Oklahoma, Norman, OK}
\author{J. M. Derkacy}
\affil{Homer L. Dodge Department of Physics and Astronomy, University of Oklahoma, Norman, OK}
\author{X. Zhao}
\affil{School of Science, Tianjin University of Technology, Tianjin, 300384, China}
\author{H. Sai}
\affil{Physics Department and Tsinghua Center for Astrophysics (THCA), Tsinghua University, Beijing, 100084, China}
\author{K. Zhang}
\affil{Physics Department and Tsinghua Center for Astrophysics (THCA), Tsinghua University, Beijing, 100084, China}
\affil{Department of Astronomy, University of Texas at Austin, Austin, TX, 78712, USA}
\author{L. Wang}
\affil{National Astronomical Observatory of China, Chinese Academy of Sciences, Beijing, 100012, China}
\affil{Chinese Academy of Sciences South America Center for Astronomy, China-Chile Joint Center for Astronomy, Camino El Observatorio 1515, Las Condes, Santiago, Chile}
\nocollaboration

\collaboration{LCO} 
\author{D. A. Howell}
\affil{Las Cumbres Observatory, 6740 Cortona Dr Ste 102, Goleta, CA 93117-5575, USA}
\affil{Department of Physics, University of California, Santa Barbara, CA 93106-9530, USA}
\author{C. McCully}
\affil{Las Cumbres Observatory, 6740 Cortona Dr Ste 102, Goleta, CA 93117-5575, USA}
\author{I. Arcavi}
\altaffiliation{Einstein Fellow}
\affiliation{Las Cumbres Observatory, 6740 Cortona Dr Ste 102, Goleta, CA 93117-5575, USA}
\affiliation{The Raymond and Beverly Sackler School of Physics and Astronomy, Tel Aviv University, Tel Aviv 69978, Israel}
\affiliation{Department of Physics, University of California, Santa Barbara, CA 93106-9530, USA}
\author{S. Valenti}
\affil{Department of Physics, University of California, Davis, CA 95616, USA}
\author{D.~Hiramatsu}
\affiliation{Las Cumbres Observatory, 6740 Cortona Dr Ste 102, Goleta, CA 93117-5575, USA}
\affiliation{Department of Physics, University of California, Santa Barbara, CA 93106-9530, USA}
\author{J.~Burke}
\affiliation{Las Cumbres Observatory, 6740 Cortona Dr Ste 102, Goleta, CA 93117-5575, USA}
\affiliation{Department of Physics, University of California, Santa Barbara, CA 93106-9530, USA}

\nocollaboration

\collaboration{KEGS} 
\author{A. Rest}
\affil{Space Telescope Science Institute, 3700 San Martin Drive, Baltimore, MD 21218, USA}
\affil{Department of Physics and Astronomy, Johns Hopkins University, Baltimore, MD 21218, USA}
\author{P. Garnavich}
\affil{Department of Physics, University of Notre Dame, 225 Nieuwland Science Hall, Notre Dame, IN, 46556-5670, USA}
\author{B.~E.~Tucker}
\affiliation{The Research School of Astronomy and Astrophysics, Mount Stromlo Observatory, Australian National University, Canberra, ACT 2611, Australia}
\affiliation{National Centre for the Public Awareness of Science, Australian National University, Canberra, ACT 2611, Australia}
\affiliation{The ARC Centre of Excellence for All-Sky Astrophysics in 3 Dimension (ASTRO 3D), Australia}
\author{G. Narayan}
\altaffiliation{Lasker Fellow}
\affil{Space Telescope Science Institute, 3700 San Martin Drive, Baltimore, MD 21218, USA}
\author{E. Shaya}
\affil{Astronomy Department, University of Maryland, College Park, Maryland 20742-2421, USA}
\author{S. Margheim}
\affil{Gemini Observatory, Southern Operations Center, c/o AURA, Casilla 603, La Serena, Chile}
\author{A. Zenteno}
\affil{Cerro Tololo Inter-American Observatory, Casilla 603, La Serena, Chile}
\author{A. Villar}
\affil{Harvard-Smithsonian Center for Astrophysics, 60 Garden Street, Cambridge, MA 02138, USA}
\nocollaboration

\collaboration{UCSC}
\author{G.~Dimitriadis}
\affil{Department of Astronomy and Astrophysics, University of California, Santa Cruz, CA 95064, USA}
\author{R.~J.~Foley}
\affil{Department of Astronomy and Astrophysics, University of California, Santa Cruz, CA 95064, USA}
\author{Y.-C.~Pan}
\affil{Department of Astronomy and Astrophysics, University of California, Santa Cruz, CA 95064, USA}
\author{D.~ A.~Coulter}
\affil{Department of Astronomy and Astrophysics, University of California, Santa Cruz, CA 95064, USA}
\author{O.~D.~Fox}
\affiliation{Space Telescope Science Institute, 3700 San Martin Drive, Baltimore, MD 21218, USA}
\author{S.~W.~Jha}
\affil{Department of Physics and Astronomy, Rutgers, The State University of New Jersey, 136 Frelinghuysen Road, Piscataway, NJ 08854, USA}
\author{D.~O.~Jones}
\affil{Department of Astronomy and Astrophysics, University of California, Santa Cruz, CA 95064, USA}
\author{D.~N.~Kasen}
\affiliation{Department of Astronomy, University of California, Berkeley, CA 94720-3411, USA}
\affiliation{Lawrence Berkeley National Laboratory, Berkeley, CA 94720, USA}
\affiliation{Lawrence Berkeley National Laboratory, Berkeley, CA, 94720, USA}
\author{C.~D.~Kilpatrick}
\affil{Department of Astronomy and Astrophysics, University of California, Santa Cruz, CA 95064, USA}
\author{A.~L.~Piro}
\affiliation{The Observatories of the Carnegie Institution for Science, 813 Santa Barbara St., Pasadena, CA 91101, USA}
\author{A.~G.~Riess}
\affiliation{Space Telescope Science Institute, 3700 San Martin Drive, Baltimore, MD 21218, USA}
\affiliation{Department of Physics and Astronomy, Johns Hopkins University, Baltimore, MD 21218, USA}
\author{C.~Rojas-Bravo}
\affil{Department of Astronomy and Astrophysics, University of California, Santa Cruz, CA 95064, USA}
\nocollaboration

\collaboration{ASAS-SN}
\author[0000-0003-4631-1149]{B.~J.~Shappee}
\affiliation{Institute for Astronomy, University of Hawai'i, 2680 Woodlawn Drive, Honolulu, HI 96822, USA}
\author[0000-0001-9206-3460]{T.~W.-S.~Holoien}
\altaffiliation{Carnegie Fellow}
\affiliation{The Observatories of the Carnegie Institution for Science, 813 Santa Barbara St., Pasadena, CA 91101, USA}
\author{K.~Z.~Stanek}
\affiliation{Center for Cosmology and AstroParticle Physics (CCAPP), The Ohio State University, 191 W.\ Woodruff Ave., Columbus, OH 43210, USA}
\affiliation{Department of Astronomy, The Ohio State University, 140 West 18th Avenue, Columbus, OH 43210, USA}
\author{M.~R.~Drout}
\altaffiliation{Hubble Fellow}
\altaffiliation{Dunlap Fellow}
\affiliation{The Observatories of the Carnegie Institution for Science, 813 Santa Barbara St., Pasadena, CA 91101, USA}
\author{K.~Auchettl}
\affiliation{Center for Cosmology and AstroParticle Physics (CCAPP), The Ohio State University, 191 W.\ Woodruff Ave., Columbus, OH 43210, USA}
\affiliation{Department of Physics, The Ohio State University, 191 W. Woodruff Avenue, Columbus, OH 43210, USA}
\author{C.~S.~Kochanek}
\affiliation{Center for Cosmology and AstroParticle Physics (CCAPP), The Ohio State University, 191 W.\ Woodruff Ave., Columbus, OH 43210, USA}
\affiliation{Department of Astronomy, The Ohio State University, 140 West 18th Avenue, Columbus, OH 43210, USA}
\author{J.~S.~Brown}
\affiliation{Department of Astronomy, The Ohio State University, 140 West 18th Avenue, Columbus, OH 43210, USA}
\author{S.~Bose}
\affiliation{Kavli Institute for Astronomy and Astrophysics, Peking University, Yi He Yuan Road 5, Hai Dian District, Beijing 100871, China}
\author{D.~Bersier}
\affiliation{Astrophysics Research Institute, Liverpool John Moores University, 146 Brownlow Hill, Liverpool L3 5RF, UK}
\author{J.~Brimacombe}
\affiliation{Coral Towers Observatory, Cairns, Queensland 4870, Australia}
\author{P.~Chen}
\affiliation{Kavli Institute for Astronomy and Astrophysics, Peking University, Yi He Yuan Road 5, Hai Dian District, Beijing 100871, China}
\author{S.~Dong}
\affiliation{Kavli Institute for Astronomy and Astrophysics, Peking University, Yi He Yuan Road 5, Hai Dian District, Beijing 100871, China}
\author{S.~Holmbo}
\affiliation{Department of Physics and Astronomy, Aarhus University, Ny Munkegade 120, DK-8000 Aarhus C, Denmark}

\author{J.~A.~Mu\~{n}oz}
\affiliation{Departamento de Astronom\'{\i}a y Astrof\'{\i}sica, Universidad de Valencia, E-46100 Burjassot, Valencia, Spain}
\affiliation{Observatorio Astron\'omico, Universidad de Valencia, E-46980 Paterna, Valencia, Spain}
\author{R.~L.~Mutel}
\affiliation{Department of Physics and Astronomy, University of Iowa, Iowa City, IA 52242, USA}
\author{R.~S.~Post}
\affiliation{Post Observatory, Lexington, MA 02421, USA}
\author{J.~L.~Prieto}
\affiliation{N\'ucleo de Astronom\'ia de la Facultad de Ingenier\'ia, Universidad Diego Portales, Av. Ej\'ercito 441, Santiago, Chile}
\affiliation{Millennium Institute of Astrophysics, Santiago, Chile}
\author{J.~Shields}
\affiliation{Department of Astronomy, The Ohio State University, 140 West 18th Avenue, Columbus, OH 43210, USA}
\author{D.~Tallon}
\affiliation{Department of Physics and Astronomy, University of Iowa, Iowa City, IA 52242, USA}
\author{T.~A.~Thompson}
\affiliation{Center for Cosmology and AstroParticle Physics (CCAPP), The Ohio State University, 191 W.\ Woodruff Ave., Columbus, OH 43210, USA}
\affiliation{Department of Astronomy, The Ohio State University, 140 West 18th Avenue, Columbus, OH 43210, USA}
\author{P.~J.~Vallely}
\affiliation{Department of Astronomy, The Ohio State University, 140 West 18th Avenue, Columbus, OH 43210, USA}
\author{S.~Villanueva~Jr.}
\affiliation{Department of Astronomy, The Ohio State University, 140 West 18th Avenue, Columbus, OH 43210, USA}
\nocollaboration

\collaboration{Pan-STARRS}
\author{S.~J.~Smartt}
\affil{Astrophysics Research Centre, School of Mathematics and Physics, Queen's University Belfast, Northern Ireland, BT7 1NN, United Kingdom}
\author{K.~W.~Smith}
\affil{Queen's University Belfast, Northern Ireland, BT7 1NN, United Kingdom}
\author{K. C. Chambers}
\affil{Institute for Astronomy, University of Hawaii at Manoa, 2680 Woodlawn Drive, Honolulu, Hawaii 96822, USA}
\author{H. A.~Flewelling}
\affil{Institute for Astronomy, University of Hawaii at Manoa, 2680 Woodlawn Drive, Honolulu, Hawaii 96822, USA}
\author{M. E.~Huber}
\affil{Institute for Astronomy, University of Hawaii at Manoa, 2680 Woodlawn Drive, Honolulu, Hawaii 96822, USA}
\author{E. A.~Magnier}
\affil{Institute for Astronomy, University of Hawaii at Manoa, 2680 Woodlawn Drive, Honolulu, Hawaii 96822, USA}
\author{C.Z.~Waters}
\affil{Institute for Astronomy, University of Hawaii at Manoa, 2680 Woodlawn Drive, Honolulu, Hawaii 96822, USA}
\author{A. S. B.~Schultz}
\affil{Institute for Astronomy, University of Hawaii at Manoa, 2680 Woodlawn Drive, Honolulu, Hawaii 96822, USA}
\author{J.~Bulger}
\affil{Institute for Astronomy, University of Hawaii at Manoa, 2680 Woodlawn Drive, Honolulu, Hawaii 96822, USA}
\author{T. B.~Lowe}
\affil{Institute for Astronomy, University of Hawaii at Manoa, 2680 Woodlawn Drive, Honolulu, Hawaii 96822, USA}
\author{M.~Willman}
\affil{Institute for Astronomy, University of Hawaii at Manoa, 2680 Woodlawn Drive, Honolulu, Hawaii 96822, USA}
\nocollaboration

\collaboration{Konkoly / Texas}
\author{K.~S\'arneczky}
\affiliation{Konkoly Observatory, MTA CSFK, Konkoly Thege M. ut 15-17, Budapest, 1121, Hungary}
\author{A.~P\'al} 
\affiliation{Konkoly Observatory, MTA CSFK, Konkoly Thege M. ut 15-17, Budapest, 1121, Hungary}
\author{J.~C.~Wheeler}
\affiliation{Department of Astronomy, University of Texas at Austin, Austin, TX 78712, USA}
\affil{George P. and Cynthia Woods Mitchell Institute for Fundamental Physics $\&$ Astronomy, Texas A. $\&$ M. University, Department of Physics and Astronomy, 4242 TAMU, College Station, TX 77843, USA}
\author{A.~B\'odi}, 
\affiliation{Konkoly Observatory, MTA CSFK, Konkoly Thege M. ut 15-17, Budapest, 1121, Hungary}
\affil{MTA CSFK Lend\"ulet Near-Field Cosmology Research Group}
\author{Zs.~Bogn\'ar}
\affiliation{Konkoly Observatory, MTA CSFK, Konkoly Thege M. ut 15-17, Budapest, 1121, Hungary}
\author{B.~Cs\'ak}
\affiliation{Konkoly Observatory, MTA CSFK, Konkoly Thege M. ut 15-17, Budapest, 1121, Hungary}
\author{B.~Cseh}
\affiliation{Konkoly Observatory, MTA CSFK, Konkoly Thege M. ut 15-17, Budapest, 1121, Hungary}
\author{G.~Cs\"ornyei}
\affiliation{Konkoly Observatory, MTA CSFK, Konkoly Thege M. ut 15-17, Budapest, 1121, Hungary}
\author{O.~Hanyecz}
\affiliation{Konkoly Observatory, MTA CSFK, Konkoly Thege M. ut 15-17, Budapest, 1121, Hungary}
\author{B.~Ign\'acz}
\affiliation{Konkoly Observatory, MTA CSFK, Konkoly Thege M. ut 15-17, Budapest, 1121, Hungary}
\author{Cs.~Kalup}
\affiliation{Konkoly Observatory, MTA CSFK, Konkoly Thege M. ut 15-17, Budapest, 1121, Hungary}
\author{R.~K\"onyves-T\'oth}
\affiliation{Konkoly Observatory, MTA CSFK, Konkoly Thege M. ut 15-17, Budapest, 1121, Hungary}
\author{L.~Kriskovics} 
\affiliation{Konkoly Observatory, MTA CSFK, Konkoly Thege M. ut 15-17, Budapest, 1121, Hungary}
\author{A.~Ordasi}
\affiliation{Konkoly Observatory, MTA CSFK, Konkoly Thege M. ut 15-17, Budapest, 1121, Hungary}
\author{I.~Rajmon}
\affiliation{Berzsenyi D\'aniel High School, K\'arp\'at utca 49-53, Budapest, 1133, Hungary}
\author{A.~S\'odor} 
\affiliation{Konkoly Observatory, MTA CSFK, Konkoly Thege M. ut 15-17, Budapest, 1121, Hungary}
\author{R.~Szab\'o}
\affiliation{Konkoly Observatory, MTA CSFK, Konkoly Thege M. ut 15-17, Budapest, 1121, Hungary}
\affiliation{MTA CSFK Lend\"ulet Near-Field Cosmology Research Group}
\author{R.~Szak\'ats} 
\affiliation{Konkoly Observatory, MTA CSFK, Konkoly Thege M. ut 15-17, Budapest, 1121, Hungary}
\author{G.~Zsidi} 
\affiliation{Konkoly Observatory, MTA CSFK, Konkoly Thege M. ut 15-17, Budapest, 1121, Hungary}
\nocollaboration

\collaboration{University of Arizona}
\author{P. Milne}
\affil{Steward Observatory, University of Arizona,  933 North Cherry Avenue, Rm. N204, Tucson, AZ 85721-0065, USA}
\author{J.~E. Andrews}
\affil{Steward Observatory, University of Arizona,  933 North Cherry Avenue, Rm. N204, Tucson, AZ 85721-0065, USA}
\author{N. Smith}
\affil{Steward Observatory, University of Arizona,  933 North Cherry Avenue, Rm. N204, Tucson, AZ 85721-0065, USA}
\author{C. Bilinski}
\affil{Steward Observatory, University of Arizona,  933 North Cherry Avenue, Rm. N204, Tucson, AZ 85721-0065, USA}

\nocollaboration

\collaboration{Swift}
\author{P.~J.~Brown}
\affiliation{ Department of Physics and Astronomy, Texas A\&M University, 4242 TAMU, College Station, TX 77843, USA}
\affil{George P. and Cynthia Woods Mitchell Institute for Fundamental Physics $\&$ Astronomy, Texas A. $\&$ M. University, Department of Physics and Astronomy, 4242 TAMU, College Station, TX 77843, USA}
\nocollaboration

\collaboration{ePESSTO}
\author{J. Nordin}
\affiliation{Institute of Physics, Humboldt-Universit\"at zu Berlin, Newtonstr. 15, 12489 Berlin, Germany}
\author{S. C. Williams}
\affiliation{Physics Department, Lancaster University, Lancaster LA1 4YB, United Kingdom}
\author{L. Galbany}
\affiliation{PITT PACC, Department of Physics and Astronomy, University of Pittsburgh, Pittsburgh, PA 15260, USA}
\author{J. Palmerio}
\affiliation{Sorbonne Universit\'es, UPMC Univ. Paris 6 et CNRS, UMR 7095, Institut d’Astrophysique de Paris, 98 bis bd Arago, 75014 Paris, France}
\author{I. M. Hook}
\affiliation{Physics Department, Lancaster University, Lancaster LA1 4YB, United Kingdom}
\author{C. Inserra}
\affiliation{Department of Physics and Astronomy, University of Southampton, Southampton, SO17 1BJ, UK}
\author{K.~Maguire}
\affiliation{Astrophysics Research Centre, School of Mathematics and Physics, Queen's University Belfast, Belfast BT7 1NN, UK}
\author{R\'egis Cartier}
\affiliation{Cerro Tololo Inter-American Observatory, National Optical Astronomy Observatory, Casilla 603, La Serena, Chile}
\author{A. Razza}
\affil{European Southern Observatory, Alonso de C\'ordova 3107, Casilla 19, Santiago, Chile}
\affil{Departamento de Astronom\'ia, Universidad de Chile, Camino El Observatorio 1515, Las Condes, Santiago, Chile}
\author{C. P. Guti\'errez}
\affil{University of Southampton, Southampton, SO17 1BJ, UK}
\nocollaboration

\collaboration{University of North Carolina}
\author{J.~J.~Hermes}
\altaffiliation{Hubble Fellow}
\affiliation{Department of Physics and Astronomy, University of North Carolina, Chapel Hill, NC 27599, USA}
\author{J.~S.~Reding}
\affiliation{Department of Physics and Astronomy, University of North Carolina, Chapel Hill, NC 27599, USA}
\author{B.~C.~Kaiser}
\affiliation{Department of Physics and Astronomy, University of North Carolina, Chapel Hill, NC 27599, USA}
\nocollaboration

\collaboration{ATLAS}
\author{J.~L.~Tonry}
\affiliation{Institute for Astronomy, University of Hawai'i, 2680 Woodlawn Drive, Honolulu, HI 96822, USA}
\author{A.~N.~Heinze}
\affiliation{Institute for Astronomy, University of Hawai'i, 2680 Woodlawn Drive, Honolulu, HI 96822, USA}
\author{L.~Denneau}
\affiliation{Institute for Astronomy, University of Hawai'i, 2680 Woodlawn Drive, Honolulu, HI 96822, USA}
\author{H.~Weiland}
\affiliation{Institute for Astronomy, University of Hawai'i, 2680 Woodlawn Drive, Honolulu, HI 96822, USA}
\author{B.~Stalder}
\affiliation{LSST, 950 North Cherry Avenue, Tucson, AZ 85719, USA}

\nocollaboration

\collaboration{{\em K2} Mission Team}
\author{G.~Barentsen}
\affil{NASA Ames Research Center, Moffett Blvd, Mountain View, CA 94035, USA}
\affiliation{Bay Area Environmental Research Institute, P.O. Box 25, Moffett Field, CA 94035, USA}
\author{J.~Dotson}
\affil{NASA Ames Research Center, Moffett Blvd, Mountain View, CA 94035, USA}
\author{T.~Barclay}
\affil{NASA Goddard Space Flight Center, 8800 Greenbelt Rd, Greenbelt, MD 20771, USA}
\affiliation{University of Maryland, Baltimore County, 1000 Hilltop Cir, Baltimore, MD 21250, USA}
\author{M.~Gully-Santiago}
\affil{NASA Ames Research Center, Moffett Blvd, Mountain View, CA 94035, USA}
\affiliation{Bay Area Environmental Research Institute, P.O. Box 25, Moffett Field, CA 94035, USA}
\author{C.~Hedges}
\affil{NASA Ames Research Center, Moffett Blvd, Mountain View, CA 94035, USA}
\affiliation{Bay Area Environmental Research Institute, P.O. Box 25, Moffett Field, CA 94035, USA}
\author{A.~M.~Cody}
\affil{NASA Ames Research Center, Moffett Blvd, Mountain View, CA 94035, USA}
\affiliation{Bay Area Environmental Research Institute, P.O. Box 25, Moffett Field, CA 94035, USA}
\author{S.~Howell}
\affiliation{NASA Ames Research Center, Moffett Field, CA 94035, USA}
\nocollaboration

\collaboration{{\em Kepler} spacecraft team}
\author{J.~Coughlin}
\author{J.~E.~Van Cleve}
\affiliation{NASA Ames Research Center, Moffett Field, CA 94035, USA}
\affiliation{SETI Institute, 189 Bernardo Avenue, Mountain View, CA 94043, USA}

\author{J.~Vin\'icius de Miranda\ Cardoso}
\affiliation{NASA Ames Research Center, Moffett Field, CA 94035, USA}
\affiliation{Universidade Federal de Campina Grande, Campina Grande, Brazil}

\author{K.~A.~Larson}
\author{K.~M.~McCalmont-Everton}
\author{C.~A.~Peterson}
\author{S.~E.~Ross}
\affiliation{Ball Aerospace and Technologies Corp., Boulder, Colorado 80301, USA}

\author{L.~H.~Reedy}
\author{D.~Osborne}
\author{C.~McGinn}
\author{L.~Kohnert}
\author{L.~Migliorini}
\author{A.~Wheaton}
\author{B.~Spencer}
\author{C.~Labonde}
\author{G.~Castillo}
\author{G.~Beerman}
\author{K.~Steward}
\author{M.~Hanley}
\author{R.~Larsen}
\author{R.~Gangopadhyay}
\author{R.~Kloetzel}
\author{T.~Weschler}
\author{V.~Nystrom}
\author{J.~Moffatt}
\author{M.~Redick}
\author{K.~Griest}
\author{M.~Packard}
\author{M.~Muszynski}
\author{J.~Kampmeier}
\author{R.~Bjella}
\author{S.~Flynn}
\author{B.~Elsaesser}
\affiliation{LASP, University of Colorado at Boulder, Boulder, CO 80303, USA}

\begin{abstract}
Supernova (SN) 2018oh (ASASSN-18bt) is the first spectroscopically-confirmed type Ia supernova (SN Ia) observed in the $Kepler$ field. The $Kepler$ data revealed an excess emission in its early light curve, allowing to place interesting constraints on its progenitor system \citep{dimitriadis-18oh,shappee-18oh}. Here, we present extensive optical, ultraviolet, and near-infrared photometry, as well as dense sampling of optical spectra, for this object. SN 2018oh is relatively normal in its photometric evolution, with a rise time of 18.3$\pm$0.3 days and $\Delta$m$_{15}(B)=0.96\pm$0.03 mag, but it seems to have bluer $B-V$ colors. We construct the ``uvoir'' bolometric light curve having peak luminosity as 1.49$\times$10$^{43}$erg s$^{-1}$, from which we derive a nickel mass as 0.55$\pm$0.04M$_{\odot}$ by fitting radiation diffusion models powered by centrally located $^{56}$Ni. Note that the moment when nickel-powered luminosity starts to emerge is +3.85 days after the first light in the $Kepler$ data, suggesting other origins of the early-time emission, e.g., mixing of $^{56}$Ni to outer layers of the ejecta or interaction between the ejecta and nearby circumstellar material or a non-degenerate companion star. The spectral evolution of SN 2018oh is similar to that of a normal SN Ia, but is characterized by prominent and persistent carbon absorption features. The C~II features can be detected from the early phases to about 3 weeks after the maximum light, representing the latest detection of carbon ever recorded in a SN Ia. This indicates that a considerable amount of unburned carbon exists in the ejecta of SN 2018oh and may mix into deeper layers.

\end{abstract}
\keywords{supernovae: general --- supernovae: individual (SN 2018oh)}

\section{Introduction}
Type Ia supernovae (SNe~Ia) have been used as standardizable candles for measuring cosmic expansion, leading to the discovery of accelerating expansion of universe and hence the ``mysterious" dark energy \citep{1998AJ....116.1009R,1999ApJ...517..565P}. However, the exact nature of their progenitor systems is still highly controversial \citep{2013Sci...340..170W,2014ARA&A..52..107M}. Two popular scenarios have been proposed so far for SN Ia progenitors. One is an explosion of a carbon-oxygen (CO) WD that accretes hydrogen-rich or helium-rich materials from a non-degenerate companion that could be a main-sequence star, a redgiant, or even a helium star \citep{1973ApJ...186.1007W,1982ApJ...253..798N,1997Sci...276.1378N}, 
this single degenerate (SD)  scenario is favored by possible detections of circumstellar material (CSM) around some SNe~Ia \citep{2003Natur.424..651H,2006ApJ...650..510A,2007Sci...317..924P,2011Sci...333..856S,2012Sci...337..942D,2013MNRAS.436..222M,2013ApJS..207....3S}. It is disfavored by the lack of narrow hydrogen emission lines in late-time spectra \citep{2005A&A...443..649M,2007ApJ...670.1275L,2013ApJ...762L...5S,2016MNRAS.457.3254M}. The other scenario involves merging explosion of two WDs, dubbed as double degenerate (DD) scenario \citep{1984ApJS...54..335I,1984ApJ...277..355W}. The DD model has recently gained more attention due to the observational findings that there are no companion signatures for some SNe Ia, including the nearby object SN 2011fe and the supernova remnant SN 1006 and SNR 0509-67.5 in LMC, down to the luminosity that is much fainter than the Sun \citep{2011Natur.480..348L,2012Natur.489..533G,2012Natur.481..164S}. Some population synthesis calculations predict delay time distributions (DTD) shapes for the birthrate of SNe Ia  in the DD scenario, which are consistent with observations \citep{2010A&A...515A..89M,2012A&A...546A..70T}.

SNe Ia also show increasing diversity in their spectroscopic and photometric properties. For instance, the so-called high velocity (HV) subclass are found to have larger ejecta velocities, redder peak $B-V$ colors, and slower late-time decline rates at bluer wavelength than those with normal ejecta velocities \citep{2008ApJ...675..626W,2009ApJ...699L.139W,2011ApJ...729...55F,2011ApJ...742...89F,2012ApJ...748..127F,2014ApJ...797...75M}. The observed differences between the HV and normal SNe~Ia have been interpreted as a geometric consequence of asymmetric explosions \citep{2010Natur.466...82M, 2010ApJ...725L.167M}. However, the fact that the HV subclass tend to be associated with more metal-rich and more luminous stellar environments indicates that SNe Ia likely arise from more than one progenitor population \citep{2013Sci...340..170W}. 

Very early observations of SNe Ia can provide clues to distinguish different progenitor models. According to the theoretical analysis by \cite{2010ApJ...708.1025K}, the collision between the material ejected by the supernova and a non-degenerate companion star will produce extra emission leading to a ``bump" feature in the early time light curves. This amount depends on the viewing angle, 
companion size and separation. Possible detections of such bump features have been reported for SNe 2012cg (\citealt{2016ApJ...820...92M} although see \citealt{2018ApJ...855....6S}), iPTF14atg \citep{2015Natur.521..328C}, iPTF16abc \citep{2018ApJ...852..100M} and SN 2017cbv (\citealt{2017ApJ...845L..11H} although see \citealt{2018arXiv180403666S}), indicating that they might have SD progenitor systems. Of these, iPTF14atg is a peculiar low luminosity supernova like SN 2002es \citep{2012ApJ...751..142G}, and is not representative of normal SNe Ia. \cite{2018ApJ...852..100M} suggested the early flux of iPTF16abc can be explained by the collision of the SN with nearby material and/or strong mixing of $^{56}$Ni in the SN ejecta. For SN 2017cbv, however, the collision of SN ejecta with a non-degenerate companion star matches well with the optical observations but overpredicts the UV flux.

The {\em Kepler} Space Telescope, observing with a time resolution of 30 minutes, can be an extremely powerful tool for finding excess early time emission \citep{2010ApJ...713L.115H}. \cite{2015Natur.521..332O} studied the {\em Kepler} light curves of 3 SNe Ia, and they found no signatures of ejecta-companion interaction in the early phase of the explosions. This is consistent with DD models. However, further studies of these SNe were limited by the lack of prompt follow up observations by other facilities. 

\begin{figure} \centering 
\subfigure[]
{ \label{fig:subfig:a} 
\includegraphics[width=3.0in]{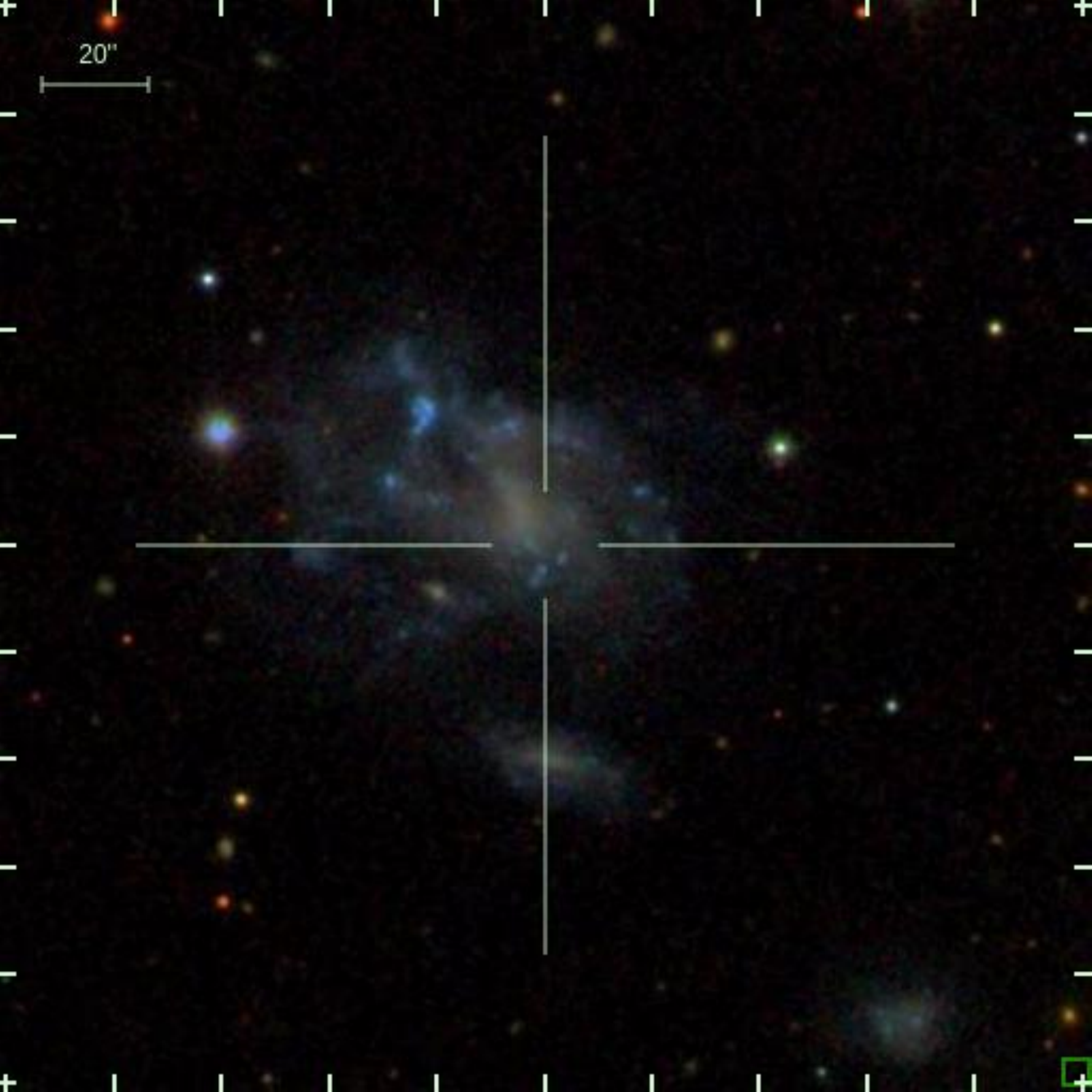}}
\subfigure[]{ \label{fig:subfig:b} 
\includegraphics[width=3in]{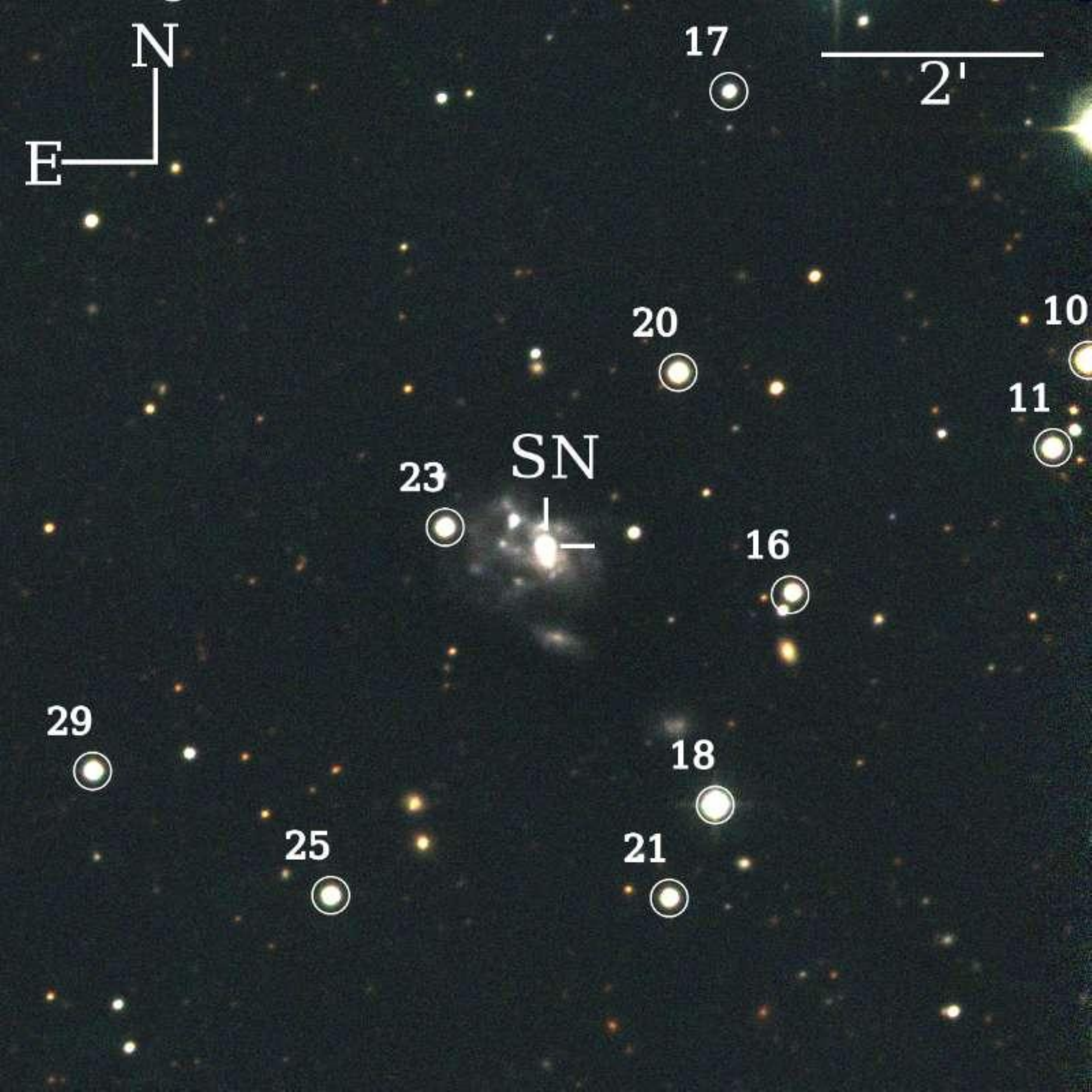}} 
\caption{(a) Pre-explosion image from the SDSS. (b) The image of SN 2018oh in UGC 4780, taken with the Tsinghua-NAOC 0.8-m telescope (TNT). Some of the reference stars listed in Table \ref{std} are marked. North is up and east is to the left (A color version of this Figure is available in the online journal).} 
\label{id} 
\end{figure}

SN 2018oh (ASASSN-18bt), a type Ia supernova in the face-on spiral galaxy UGC 4780 (see Figure \ref{id}) at a distance of about $\sim 53$ Mpc (z$\sim$0.0109), provides us a rare opportunity to examine the progenitor of a SN Ia system through the observed properties based on both continuous {\em Kepler} data and extensive follow up observations. This supernova was discovered by the All Sky Automated Survey for SuperNovae \citep[ASAS-SN; ][]{2014ApJ...788...48S} on 2018 February 4.41 (UT time is used throughout this paper) at R.A. $= 09^h06^m39^s.59$, decl. $ = +19^{\circ}20'17''.47$ \citep{2018ATel11253....1B,shappee-18oh}, located at 2$''$.0 east and 7$''$.8 north of the center of UGC 4780. It was soon identified as a normal SN Ia at about 10 days before the maximum light \citep{2018TNSCR.159....1L,2018ATel11267....1Z}. ASAS-SN monitors the {\em K2} fields at heightened cadence to help identify such SN at the earliest possible phases for detailed study.
The excess flux above a quadratic rise detected in the early rising phase of the {\em Kepler} light curve cannot be well modeled as a single power law. This is alternately explained as the collision of the SN ejecta with a non-degenerate  1-6 M$_{\odot}$ Roche-lobe-filling star at 2 $\times$ 10$^{12}$ cm (\citealt{dimitriadis-18oh} but see the caveats in \citealt{shappee-18oh}).

In this paper we present extensive follow-up observations of SN 2018oh in optical, ultraviolet (UV) and near-infrared (NIR) bands, and analyze its observational properties and explosion parameters in contrast to other well-studied SNe Ia. The observations and data reductions are described in Section 2, Section 3 presents the light/color curves, and Section 4 presents the spectral evolution. We discuss the properties of SN 2018oh and its explosion parameters in Section 5. The conclusions are summarized in Section 6.

\section{OBSERVATIONS}\label{observation}
\subsection{Photometry}
After the discovery of SN 2018oh and the recognition that it would have a {\em Kepler} light curve, follow-up photometric observations started immediately using more than a dozen of telescopes, including: (1) the 0.8~m Tsinghua-NAOC Telescope (TNT) in China \citep{2012RAA....12.1585H}; (2) the 2.4~m Lijiang Telescope (LJT) of Yunnan Astronomical Observatory (YNAO) in China \citep{2015RAA....15..918F}; (3) the Las Cumbres Observatory (LCO) 1~m telescope network \citep{2013PASP..125.1031B}; (4) Pan-STARRS1 survey (PS1) telescopes \citep{2016arXiv161205560C}; (5) Swope 1.0-m telescope at Las Campanas Observatory; (6) DEMONEXT 0.5~m telescope \citep{2018PASP..130a5001V}; (7) the 0.61-m at Post Observatory (PONM), Mayhill, NM; (8) the 60/90cm Schmidt-telescope on Piszk\'estet{\H o} Mountain Station of Konkoly Observatory in Hungary; (9) the Gemini 0.51~m telescope at the Winer Observatory; (10) CTIO 4-m Blanco telescope with DECam \citep{2008arXiv0810.3600H,2015AJ....150..150F}; (11) the 0.51-m T50 at the Astronomical Observatory of the University of Valencia in Spain and (12) the 0.6-m Super-LOTIS \citep[Livermore Optical Transient Imaging  System; ][]{2008AIPC.1000..535W} telescope at Kitt Peak Steward Observatory. Broadband $BV$- and Sloan $gri$-band photometry were obtained with all these telescopes except for the 60/90cm Schmidt-telescope of Konkoly Observatory, and the 0.6-m Super-LOTIS telescope which both used the $BVRI$ bands. Observations made with LCO 1~m telescope and Swope also used the $U$ and $u$ band, respectively.  


All CCD images were pre-processed using standard \textsc{IRAF}\footnote{IRAF is distributed by the National Optical Astronomy Observatories, which are operated by the Association of Universities for Research in Astronomy, Inc., under cooperative agreement with the National Science Foundation (NSF).} routines, including bias subtraction, flat fielding and the removal of cosmic rays. No template subtraction technique was applied in  measuring the magnitudes as the SN was still relatively bright in preparations of this work. We performed point-spread-function (PSF) photometry for both the SN and the reference stars using the pipeline $Zuruphot$ developed for automatic photometry on TNT, LJT, LCO, DEMONEXT, PONM, Gemini and T50 images (Mo et al. in prep.). This pipeline was modified to analyze the data obtained with the other telescopes involved in our study. All Swope imaging was processed using {\tt photpipe} \citep{2005ApJ...634.1103R,2014ApJ...795...44R}.

\begin{figure}[htbp]
\flushleft
\includegraphics[width=\textwidth]{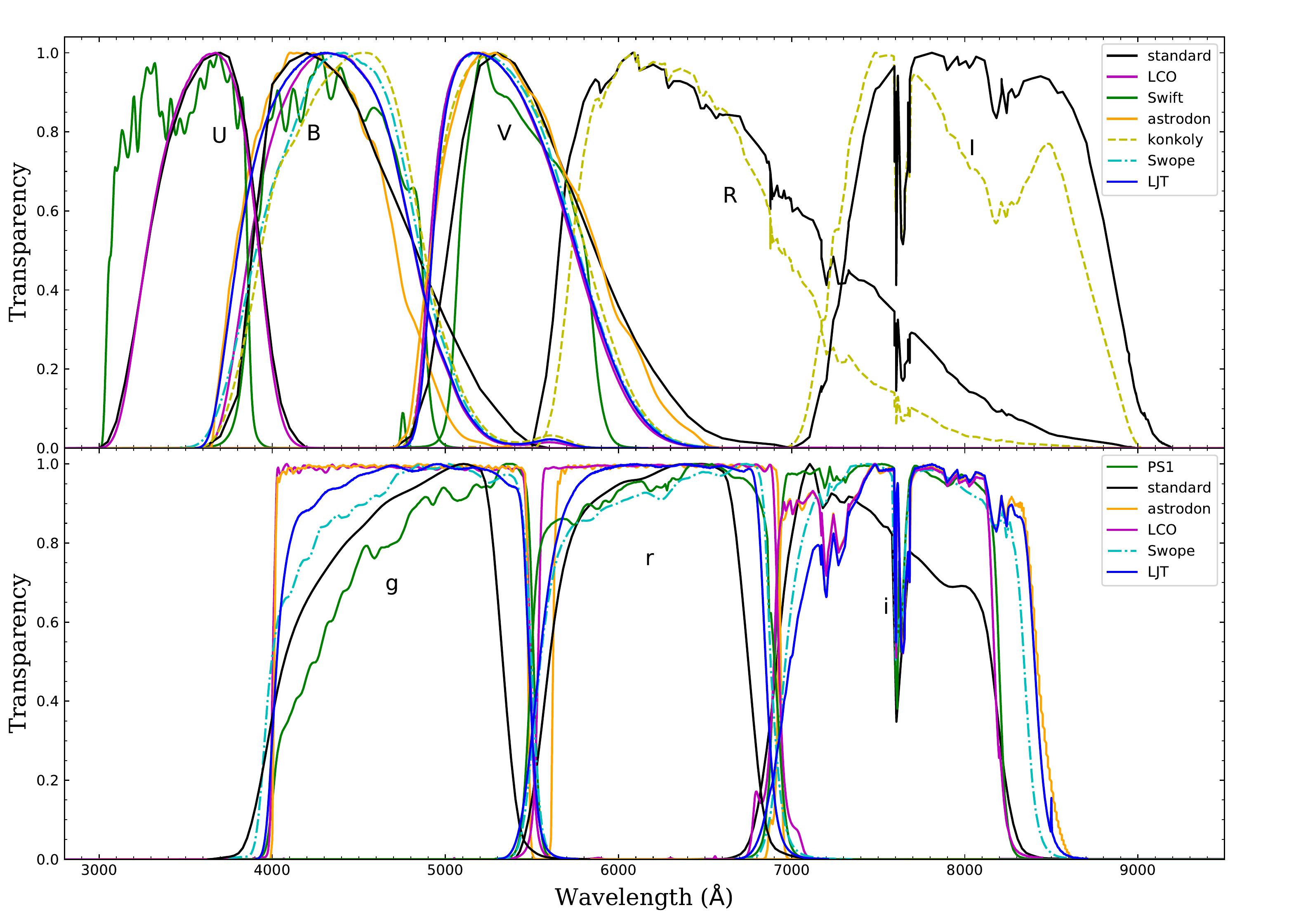}
\caption{The transmission curves of different telescopes. Curves are normalized to the peak. Black curves represent the standard filter transmission curves. 
} 
\label{filter}     
\end{figure}

The instrumental magnitudes of the supernova were converted into the standard Johnson $UBV$ \citep{1966CoLPL...4...99J}, Kron-Cousins $RI$ \citep{1981SAAOC...6....4C} and Sloan $gri$ systems using observations of a series of \cite{1992AJ....104..340L} and SDSS/PS1 \citep{2016arXiv161205560C,2016arXiv161205243F,2016arXiv161205242M,2016arXiv161205245W,2017ApJS..233...25A} standard stars on a few photometric nights. We transformed the PS1 $gri$-band magnitudes to the Swope natural system \citep[see, e.g.,][]{2010AJ....139..519C,2017AJ....154..211K} using Supercal transformations as described in \cite{2015ApJ...815..117S}. The filter transmission curves of different telescopes are displayed in Figure \ref{filter}, which are not far from the standard ones. These filter transmissions are multiplied with the CCD quantum efficiency and atmospheric transmission when the information of the latter two is available. The Astrodon filters are used by PONM and Gemini observations. Tables \ref{std} and \ref{std1} list the standard $UBVRI$ and $gri$ magnitudes of the comparison stars. The photometric results for the different photometric systems are consistent to within 0.05 mag after applying the color-term corrections. As the instrumental responses from the different photometric systems do not show noticeable differences, as shown in Figure \ref{filter}, we did not apply additional corrections (i.e., S-corrections) to the photometry due to the lack of telescope information such as CCD quantum efficiency and the mirror reflectivity for some telescopes. The final calibrated $U(u)BVRIgri$ magnitudes are presented in Table \ref{gphoto}. 

The near-infrared (NIR) photometry of SN 2018oh was obtained with two telescopes, the 3.6-m ESO New Technology Telescope (NTT) with SOFI and the 1.3-m CTIO telescope with ANDICAM. The $JHK$-band photometry from the NTT was reduced using the SOFI reduction pipeline and calibrated against the 2MASS stars in the field. The $YJH$-band images obtained with the CTIO~1.3-m telescope, were first subtracted with the sky background and then reduced with SExtractor \cite{1996A&AS..117..393B}. Magnitudes were then calibrated with the 2MASS catalogue in $JH$ bands and with the Pan-STARRS catalogue in $Y$ band.  

SN 2018oh was also observed with the Ultraviolet/Optical Telescope \citep[UVOT; ][]{2005SSRv..120...95R} onboard the $Neil~Gehrels~ Swift~Observatory$ \citep[$Swift$;][]{2004ApJ...611.1005G}. The space-based observations were obtained in uvw1, uvm2, uvw2, $U$, $B$, and $V$ filters, starting from 2018 Feb.05.4. The Swift/UVOT data reduction is based on that of the Swift Optical Ultraviolet Supernova Archive \citep[SOUSA;][]{2014Ap&SS.354...89B}. A 3\arcsec\ aperture is used to measure the source counts with an aperture correction based on an average PSF. Magnitudes are computed using the zeropoints of \cite{2011AIPC.1358..373B} for the UV and \cite{2008MNRAS.383..627P} for the optical and the 2015 redetermination of the temporal sensitivity loss. 
Table \ref{sphoto} lists the final  background-subtracted UVOT UV/optical magnitudes. The instrumental response curves of the UVOT $B$ and $V$ band are similar with standard Johnson $B$ and $V$ band. Therefore our ground-based and $Swift$ photometry of these two bands can be compared directly. Note that some differences exist between the U-band observations of $Swift$ UVOT and LCO due to different transmission curves (see Figure \ref{filter}). 

\subsection{Spectroscopy}
A total of 56 optical spectra were obtained from the Xinglong 2.16-m telescope (+BFOSC), the Lijiang 2.4-m telescope (+YFOSC), the Lick 3-m Shane telescope \citep[+KAST;][]{1993LOTRM}, the SOAR 4.1-m telescope \citep[+Goodman Spectrograph; ][]{2004SPIE.5492..331C}, the Bok 2.3-m telescope, the HET 10-m telescope \citep[+LRS2;][]{2016SPIE.9908E..4CC}, the MMT 6.5-m telescope, the Magellan 6.5-m telescope, the Las Cumbres Observatory 2.0-m telescopes (+FLOYDS), NTT \citep[+EFOSC2;][]{1984Msngr..38....9B,2015A&A...579A..40S}\footnote{NTT spectra were reduced using the PESSTO pipeline \citep{2015A&A...579A..40S}.} and the APO 3.5-m telescope (+DIS). These spectra covered the phases from $-$8.5 days to +83.8 days after the maximum light.
A log of the spectra is listed in Table \ref{log}. All spectra were reduced using standard IRAF routines. Flux calibration of the spectra was performed using spectrophotometric standard stars observed at similar airmass on the same night as the SN. The spectra were corrected for atmospheric extinction using the extinction curves of local observatories; and in most cases the telluric lines were removed. All the spectra presented in this paper will be made available via WISeREP \citep{2012PASP..124..668Y}. 

\subsection{{\em K2} photometry}\label{k2-phot}
We performed an independent photometric analysis on the {\em Kepler} long-cadence imaging 
data by involving the {\tt FITSH} package \citep{2012MNRAS.421.1825P} and using our former 
experience on photometry of stars appearing in the vicinity of background
galaxies \citep{2015ApJ...812....2M}. Astrometric jitters were derived 
using a dozen of nearby {\em K2}-stamps \citep[see also][]{2015ApJ...812....2M, 2015ApJ...804L..45P}, and the derived information is used afterwards to perform 
frame registration at sub-pixel level with an effective pixel scale of 1.0"/pixel. 
Pre-explosion images with small pointing errors were used to construct a
background reference image prior to applying image subtraction. This 
construction is based on median averaging of the first 400 frames that
were taken days before the explosion. During the subsequent differential
aperture photometry, this median-combined image was used as a template
frame. In order to correct for various systematic 
effects, including instrumental artifacts and intrinsic background-level 
variations such as the rolling band issue \citep[see e.g.][]{shappee-18oh}, we performed an additional background estimation on the subtracted images. Finally, the background-subtracted instrumental light curve was calibrated to physical units by comparing with synthetic photometry computed with the {\tt SNCOSMO} code \citep{2016ascl.soft11017B}. This was obtained using the {\em Kepler} bandpass on the extended SALT2-templates with the light curve parameters derived in Section~\ref{lc_fit}. The resulting {\em K2} light curve agreed well within the error bars of those presented in \citet{dimitriadis-18oh} and \citet{shappee-18oh}.

\section{Light Curves}

\subsection{UV/Optical Light Curves}
\begin{figure}[htbp]
\center
\includegraphics[width=\textwidth]{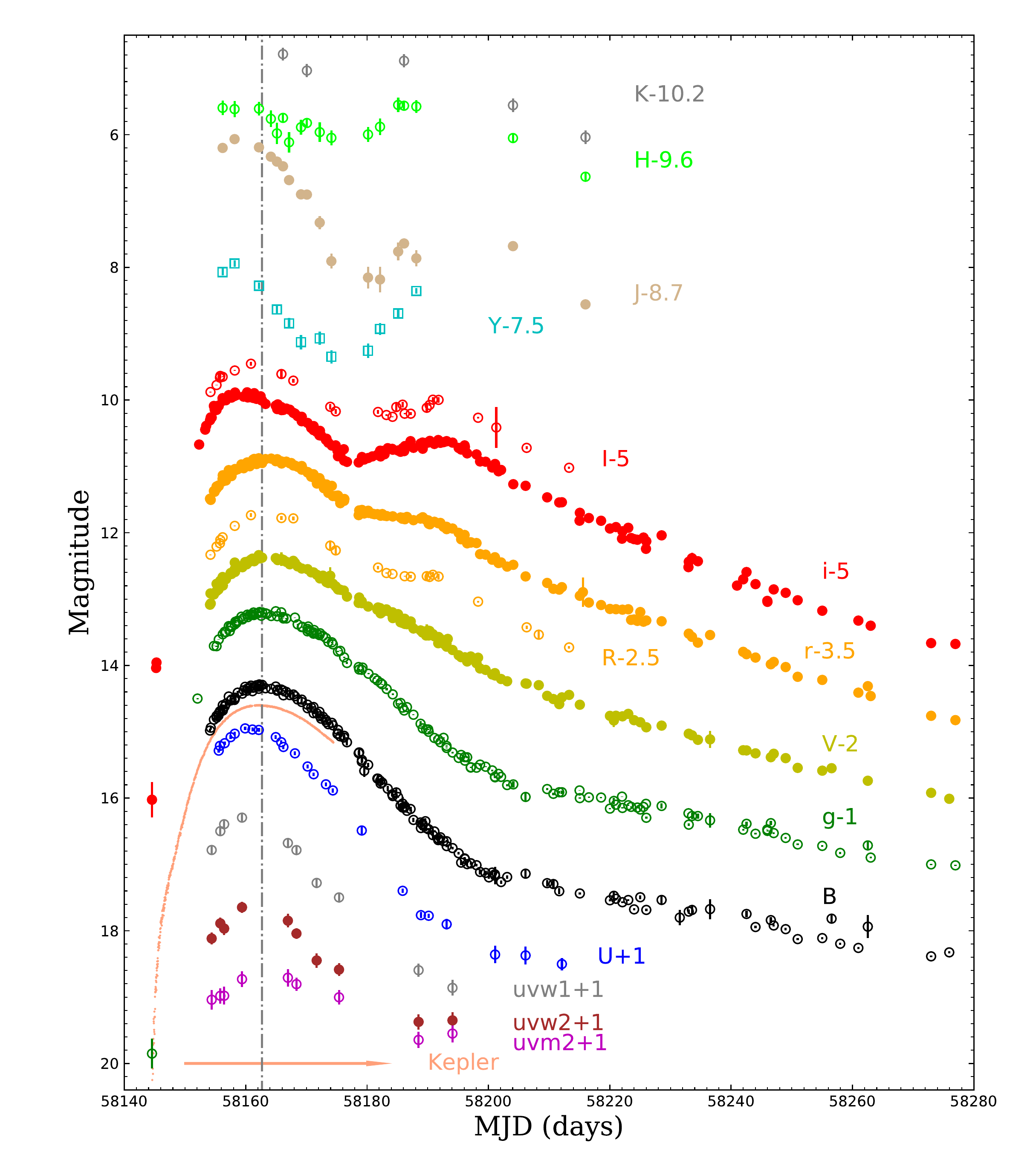}
\vspace{0.2cm}
\caption{The near-infrared, optical and ultraviolet light curves of SN 2018oh. The vertical dot-dashed line marks the date for the $B$-band maximum light $t_{Bmax} = $ MJD 58162.7$\pm$0.3 (2018 February 13.7).}
\label{uvot} \vspace{-0.0cm}
\end{figure}

Figure \ref{uvot} and \ref{uv_compare} shows the optical, UV, and NIR light curves of SN 2018oh. The optical light curves have a nearly daily cadence from $\sim$10 days before to about 100 days after maximum light of $B$ band. The earliest detections of this SN can be actually traced back to the PS1 images taken on 2018 Jan. 26.56, corresponding to $-$18.1 days relative to the peak, when the $g$ and $i$ band magnitude were 20.85$\pm$0.22 and 21.03$\pm$0.27, respectively. We take MJD 58144.37$\pm$0.04 as the explosion time, which is the average of the values adopted in \citet{dimitriadis-18oh} and \citet{shappee-18oh}. Like other normal SNe Ia, the light curves of SN 2018oh show prominent shoulders in the $R/r$ bands and secondary peaks in $I/i$ and NIR $YJHK$ bands, and they reached their peaks slightly earlier in $I/i$-, $YJHK$- and $UV$-band relative to the $B$-band. 


Using a polynomial fit to the observed light curves, we find that SN 2018oh reached peak magnitude of $B_{max}$ = 14.31$\pm$0.03 mag and $V_{max}$ = 14.37$\pm$0.03 mag on MJD 58162.7$\pm$0.3 (2018 February 13.7) and 58163.7$\pm$0.3, respectively. The post-maximum decline rate in the $B$ band, $\Delta m_{15}$(B) is 0.96$\pm$0.03 mag. The results for all the $UBVRIgriYJHK-$band light curves are reported in Table \ref{photo_info}. Results from standard light curve models like MLCS2k2 \citep{2007ApJ...659..122J}, SALT2 \citep{guy10}, and SNooPy2 \citep{2011AJ....141...19B} will be used to derive the distance to the SN and discussed in \S\ref{lc_fit}.

\begin{figure}[htbp]
\centering
\includegraphics[width=\textwidth]{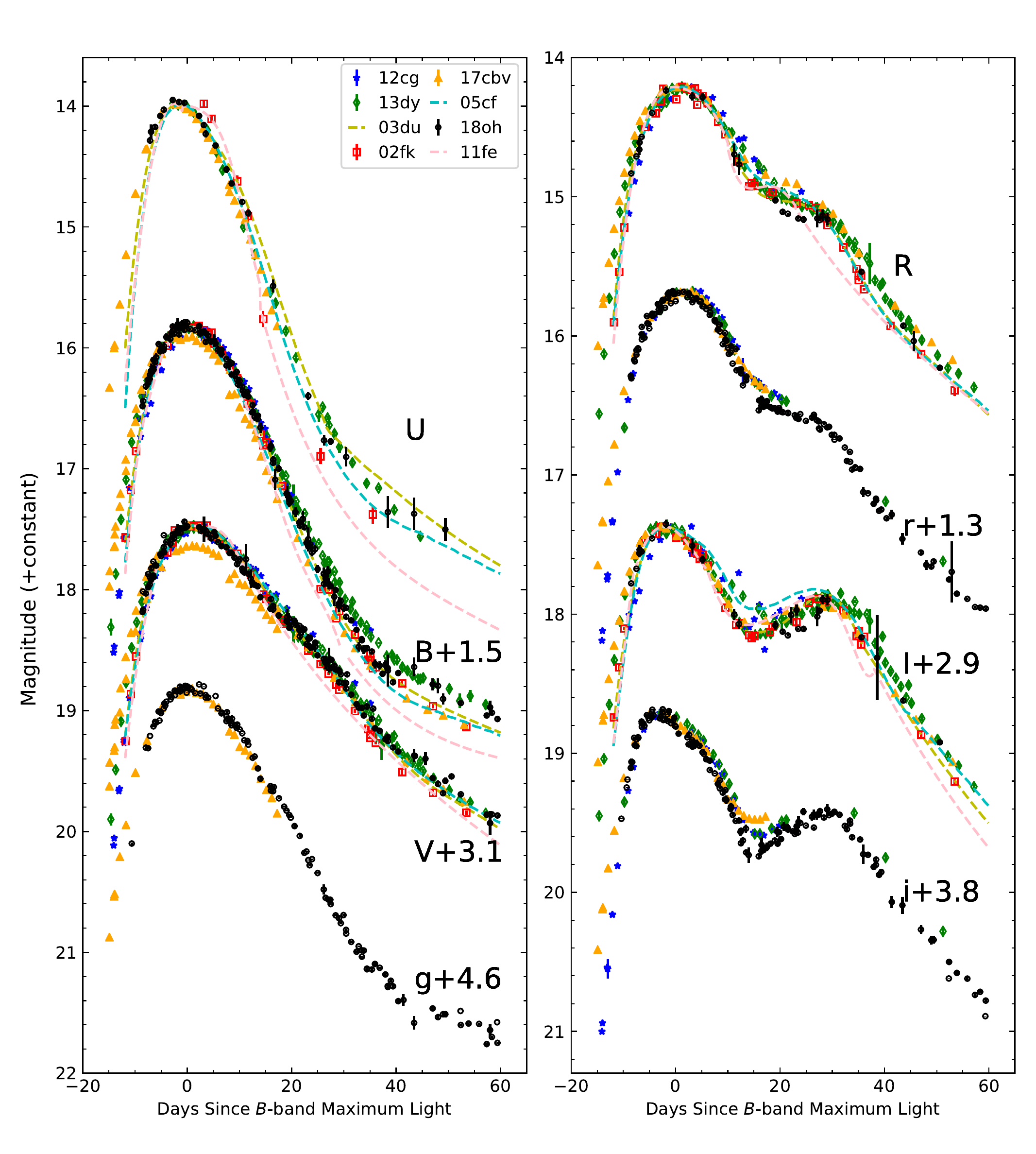}
\vspace{0.0cm}
\caption{Comparison of the optical light curves of SN 2018oh to other well-observed SN Ia with similar decline rates. The light curves of the comparison SNe Ia are normalized to match the peak magnitudes of SN 2018oh.}
\label{lc_compare} \vspace{-0.0cm}
\end{figure}

\begin{figure}[htbp]
\centering
\includegraphics[scale=.6]{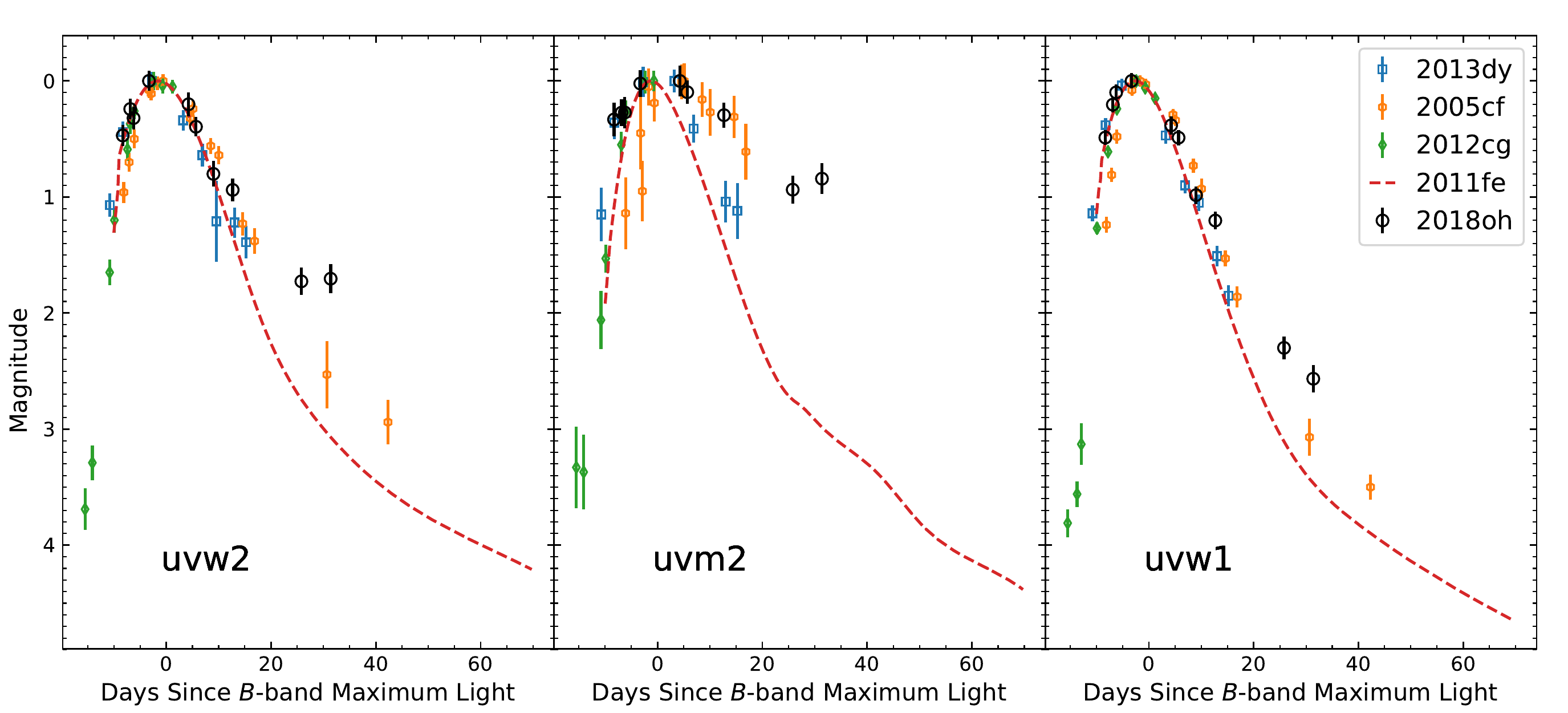}
\vspace{0.0cm}
\caption{Comparison of the $UV$ light curves of SN 2018oh with other well-observed SN Ia with similar decline rates. The magnitudes and phases of all SNe Ia are all normalized to the corresponding values at maximum light.}
\label{uv_compare} \vspace{-0.0cm}
\end{figure}

In Figure \ref{lc_compare}, we compare the light curves of SN 2018oh with other well-observed SNe Ia that have similar  $\Delta m_{15}$(B). The comparison sample includes SN 2002fk ($\Delta$m$_{15}$(B) = 1.02$\pm$0.04 mag, \cite{2014ApJ...789...89C}), SN 2003du ($\Delta$m$_{15}$(B) = 1.02$\pm$0.03 mag, \cite{2007A&A...469..645S}), SN 2005cf ($\Delta$m$_{15}$(B) = 1.07$\pm$0.03 mag, \cite{2009ApJ...697..380W}), SN 2011fe ($\Delta$m$_{15}$(B) = 1.10$\pm$0.02 mag, \cite{2013NewA...20...30M}), SN 2012cg ($\Delta m_{15}$(B) = 1.04$\pm$0.03, \cite{2013NewA...20...30M}), SN 2013dy ($\Delta m_{15}$(B) = 0.92$\pm$0.03, \cite{2015MNRAS.452.4307P}), and SN 2017cbv ($\Delta m_{15}$(B) = 1.06$\pm$0.03, \cite{2017ApJ...845L..11H}). The morphology of the light curves of SN 2018oh closely resembles to that of SN 2003du and SN 2013dy, with $\Delta m_{15}$(B) lying between these two comparison SNe Ia. 

\begin{figure}[htbp]
\center
\includegraphics[scale=.5]{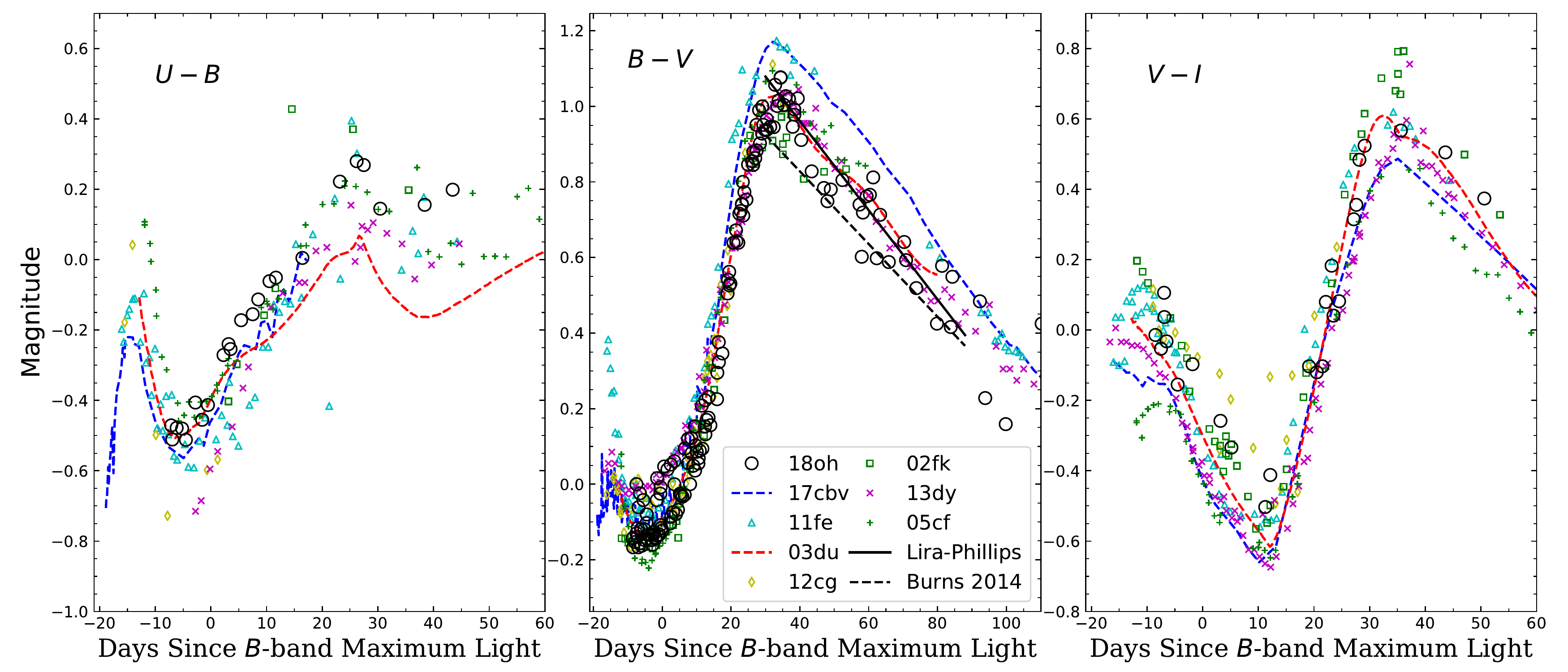}
\vspace{0.2cm}
\caption{The $U-B$, $B-V$, and $V-I$ color curves of SN 2018oh compared with those of SNe 2002fk, 2003du, 2005cf, 2011fe, 2012cg, 2013dy, and 2017cbv. All of the comparison SNe have been dereddened. The dash-dotted line in the $B - V$ panel shows the unreddened $Lira-Phillips$ loci and updated version from \cite{2014ApJ...789...32B}. The data sources are cited in the text.}
\label{color} \vspace{-0.0cm}
\end{figure}

Figure \ref{color} shows that the optical color evolution of SN 2018oh is similar to that of the comparison sample. At t $\gtrsim-$10 days, both the $U - B$ and $B - V$ colors become progressively red until t$\sim$4-5 weeks after the maximum light; the $V - I$ color initially becomes bluer until t$\sim$+10 days and it then turns redder, reaching the reddest color at t$\sim$+35 days. After t$\sim$+35 days, both the $B - V$ and $V - I$ curves colors become  bluer. In the very early phases (at t$\lesssim-$14 days), however, the color evolution of the SN is scattered. For instance, SN 2011fe evolved from very red colors towards blue ones, while SN 2017cbv (and perhaps SN 2012cg) shows the opposite trend. Bluer colors seen in the early phase of some SNe Ia have been interpreted as a result of interactions between the ejecta and a companion star, supporting SD progenitor scenario \citep{2012ApJ...753...22B, 2016ApJ...820...92M, 2017ApJ...845L..11H}. It is not clear whether SN 2018oh had such blue colors due to the lack of color information at very early times. SN 2018oh shows relatively bluer $B - V$ colors than the comparison SNe Ia, but it is redder in the $U - B$ and $V - I$ colors. The slightly redder $U - B$ color seen in SN 2018oh could be related to stronger Ca II H\&K and iron-group elements (IGEs) absorption at shorter wavelengths. We do not show the $gri$-band color evolution due to the lack of data in these bands for most of our comparison sample, but SN~2018oh shows a similar evolutionary trends to SN~2017cbv in its $g-r$ and $r-i$ colors at comparable phases. \citet{dimitriadis-18oh} show the very early $g-i $ color and conclude that before t$\sim-$10 days SN~2018oh looks bluer than SN~2011fe and is similar to SN 2017cbv. 

\begin{figure}[htbp]
\center
\includegraphics[scale=1]{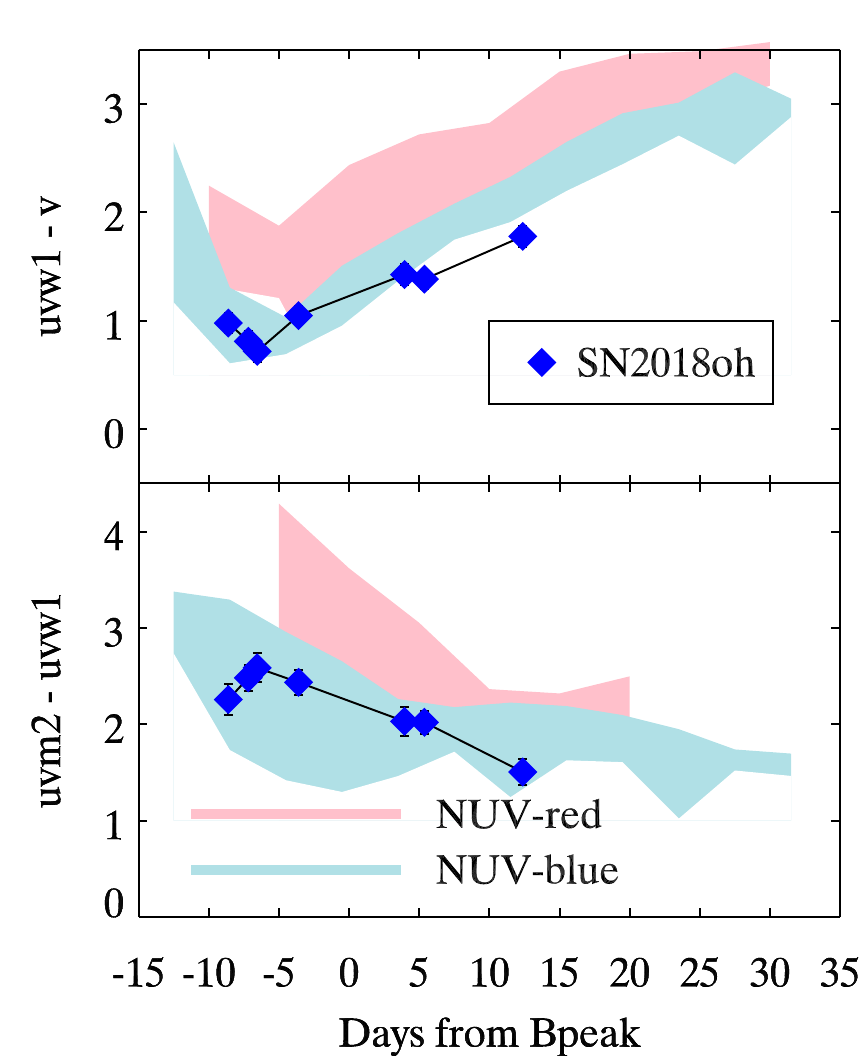}
\vspace{0.2cm}
\caption{The uvm2-uvw1 and uvw1-v colors of SN~2018oh are compared to a group of NUV-blue and NUV-red SNe \citep[see e.g. ][]{2013ApJ...779...23M}.}
\label{color_uv} \vspace{-0.0cm}
\end{figure}

\cite{2013ApJ...779...23M} found that the near-ultraviolet (NUV) colors of SNe Ia can be divided into NUV-blue and NUV-red groups. We compare SN~2018oh with these two groups in Figure \ref{color_uv}.  As shown in Figure \ref{color_uv}, SN~2018oh belongs to the NUV-blue group, consistent with the finding of \cite{2013ApJ...779...23M} that the detection of C~II (see $\S$\ref{carbon}) is common among the NUV-blue SNe Ia and rare amongst NUV-red SNe Ia. SN 2018oh has normal velocity and low velocity gradient of Si II $\lambda$6355 absorption feature which also follows the same trend as the NUV-blue group \citep{2013ApJ...779...23M}. 
These groupings (or the positions of SNe along a continuum of NUV colors) are affected by reddening but are still present for SNe Ia with low reddening \citep{2017ApJ...836..232B}.  

\begin{figure}[htbp]
\center
\includegraphics[scale=.5]{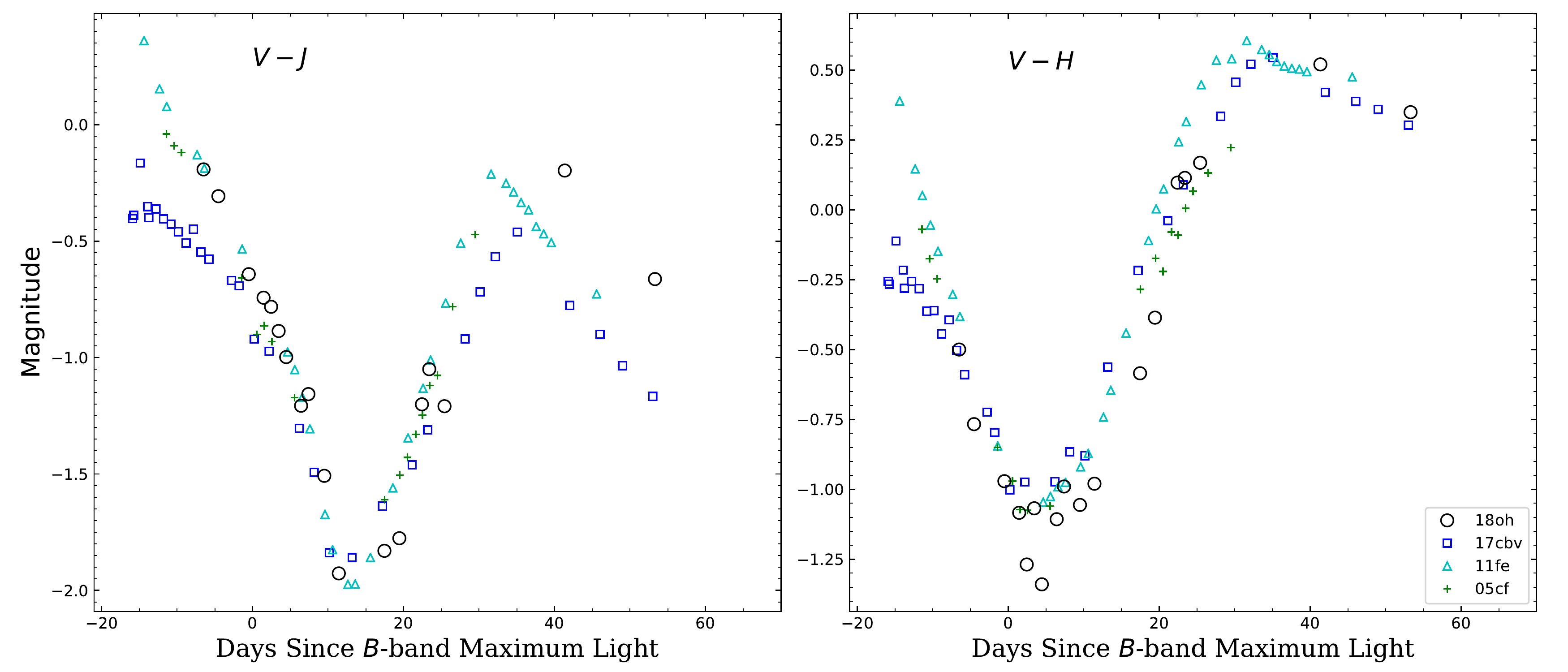}
\vspace{0.2cm}
\caption{The $V-J$ and $V-H$ color curves of SN 2018oh compared with those of SNe 2005cf, 2011fe and 2017cbv. All of the SNe have been dereddened. The data sources are cited in the text.}
\label{color_ir} \vspace{-0.0cm}
\end{figure}

We also compare the color evolution of SN 2018oh with SN 2005cf \citep{2009ApJ...697..380W}, SN 2017cbv (Wang L et al. in prep.) and SN 2011fe \citep{2012ApJ...754...19M} in the NIR bands, as shown in Figure \ref{color_ir}. SN 2017cbv is bluer in both NIR colors before maximum. SN 2018oh is bluer around maximum in $V-H$. The last two $V-J$ points of SN 2018oh are significantly redder than the others.

\subsection{Reddening Correction}\label{red}
The Galactic extinction towards SN 2018oh is estimated as $A_{V}$ (Gal) = 0.124 mag \citep{2011ApJ...737..103S}, corresponding to $E(B-V)_G = $0.040 mag for a \cite{1989ApJ...345..245C}  extinction law with $R_{V}$ = 3.1. As SN 2018oh appears close to the projected center of its host galaxy, it is necessary to examine the reddening due to the host galaxy. After corrections for the Galactic extinction, the $B - V$ colors at peak and at t = +35 days are found to be  $-$0.10 $\pm$ 0.03 mag and 1.02 $\pm$ 0.04 mag, respectively, which are consistent with typical values of unreddened SNe Ia with comparable $\Delta m_{15}$ (B) \citep{1999AJ....118.1766P, 2007ApJ...659..122J, 2009ApJ...697..380W, 2014ApJ...789...32B}. Similarly, if we fit the $B - V$ evolution over the phases from t = 30 to 90 days past the peak \citep[$Lira-Phillips$ relation,][]{1999AJ....118.1766P} using \cite{2014ApJ...789...32B}, we derive a reddening of $-$0.06$\pm$0.04 mag and 0.06$\pm$0.04 mag, respectively. Finally, we did not find any evidence for Na I D ($\lambda$5890) absorption due to the host galaxy. We thus conclude that there is no significant host-galaxy extinction, even though the SN is located near the projected center of its host galaxy.

\subsection{Light Curve Fitting} 
\label{lc_fit}


We adopt SALT 2.4 \citep{betou14} as our primary LC-fitter because it has the most flexibility in fitting multi-band light curves taken in different photometric systems, and the most recent calibrations include the dependence on the host galaxy stellar mass. We also use the SNooPy2 \citep{2011AJ....141...19B} and MLCS2k2 \citep{2007ApJ...659..122J} to verify the distances \citep[see also in][]{vinko18}. 


\begin{figure}
\centering
\includegraphics{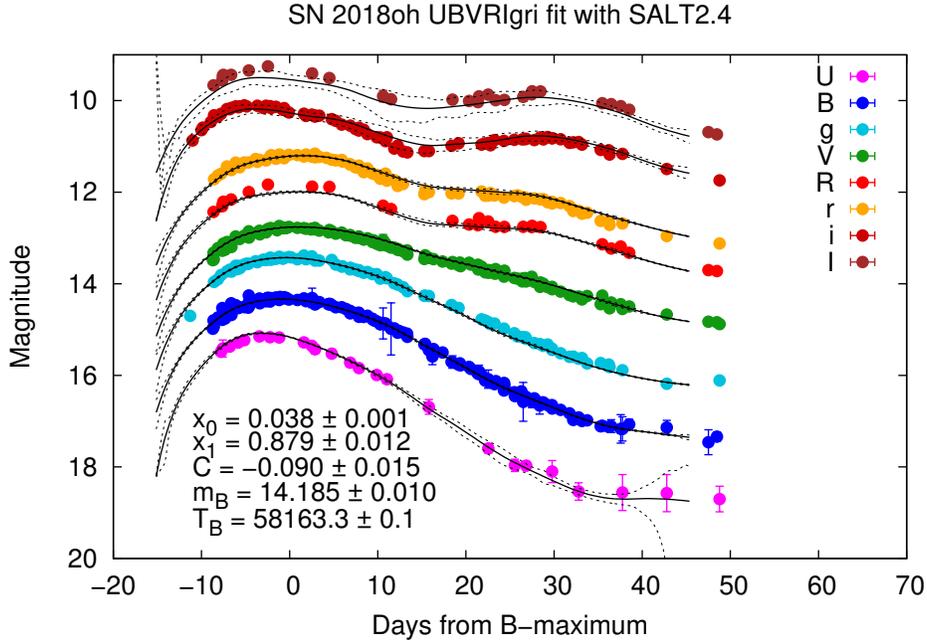}
\caption{The best-fit light curve model from SALT 2.4. The light curves are shifted vertically 
for better visibility. The dashed lines represent the 1-$\sigma$ uncertainty of the light curve templates.}
\label{fig:salt24}
\end{figure}

\begin{figure} \centering 
\subfigure[SNooPy2 template fits]
{ \label{fig:subfig:a} 
\includegraphics[width=3.0in]{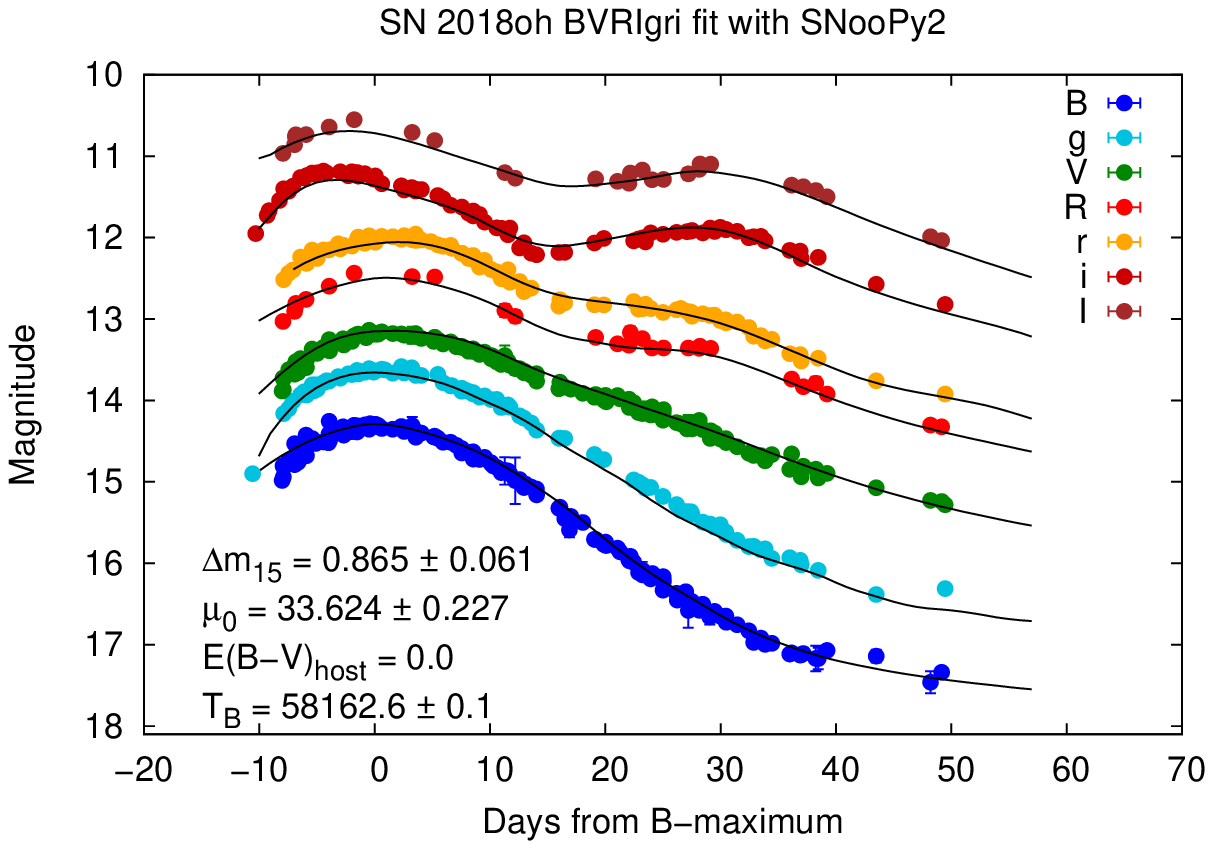}}
\subfigure[MLCS2k2 template fits]{ \label{fig:subfig:b} 
\includegraphics[width=3in]{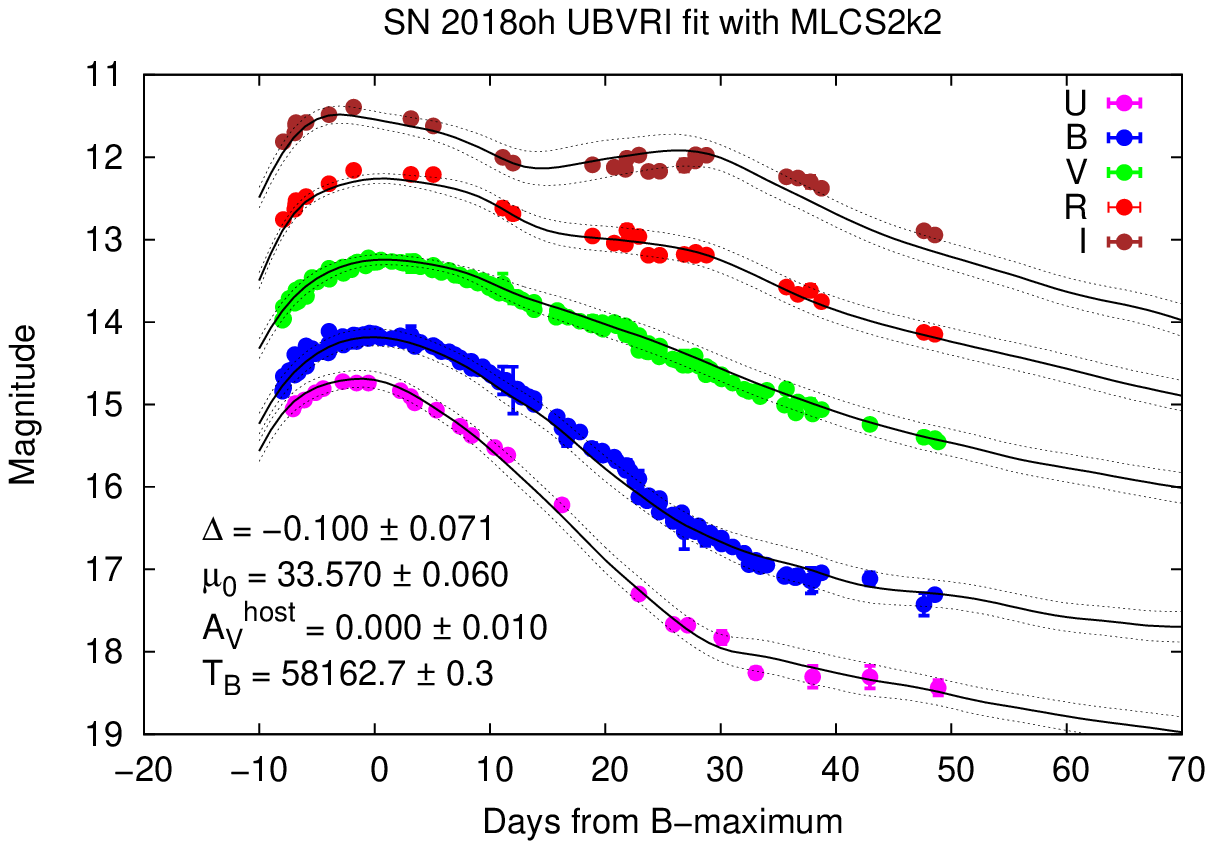}} 
\caption{The best-fit light curve models from SNooPy2 and MLCS2k2. The light curves are shifted vertically for better display. The dashed lines represent the 1-$\sigma$ uncertainty of the light curve templates.} 
\label{SNooPy} 
\end{figure}

The final, best-fit results are shown in Figure \ref{fig:salt24} and \ref{SNooPy}. Table~\ref{tab:lcfits} summarizes the LC parameters and the inferred distance moduli. The distance modulus from the SALT 2.4 best-fit parameters are derived using the calibration by \citet{betou14}. The stellar mass of the host of SN~2018oh (UGC 04780) is $\log_{10} (M_{stellar}/M_{\odot}) \sim 6.9$  (see Section \ref{c}), is taken into account as a ``mass-step'' correction of $\sim 0.06$ mag in the \citet{betou14} calibration. The distance moduli listed in the last row in Table~\ref{tab:lcfits} are brought to a common Hubble constant of $H_0 = 73$ kms$^{-1}$Mpc$^{-1}$ \citep{riess16, riess18}. 

It is readily seen that the distances from the three independent LC-fitting codes are in excellent agreement. We adopt the SALT 2.4 distance modulus of $\mu_0 = 33.61 \pm 0.05$ mag, corresponding to $52.7 \pm 1.2$ Mpc as the final result in our following analysis. 

\section{Optical spectra}
\begin{figure}[htbp]
\center
\includegraphics[scale=.85]{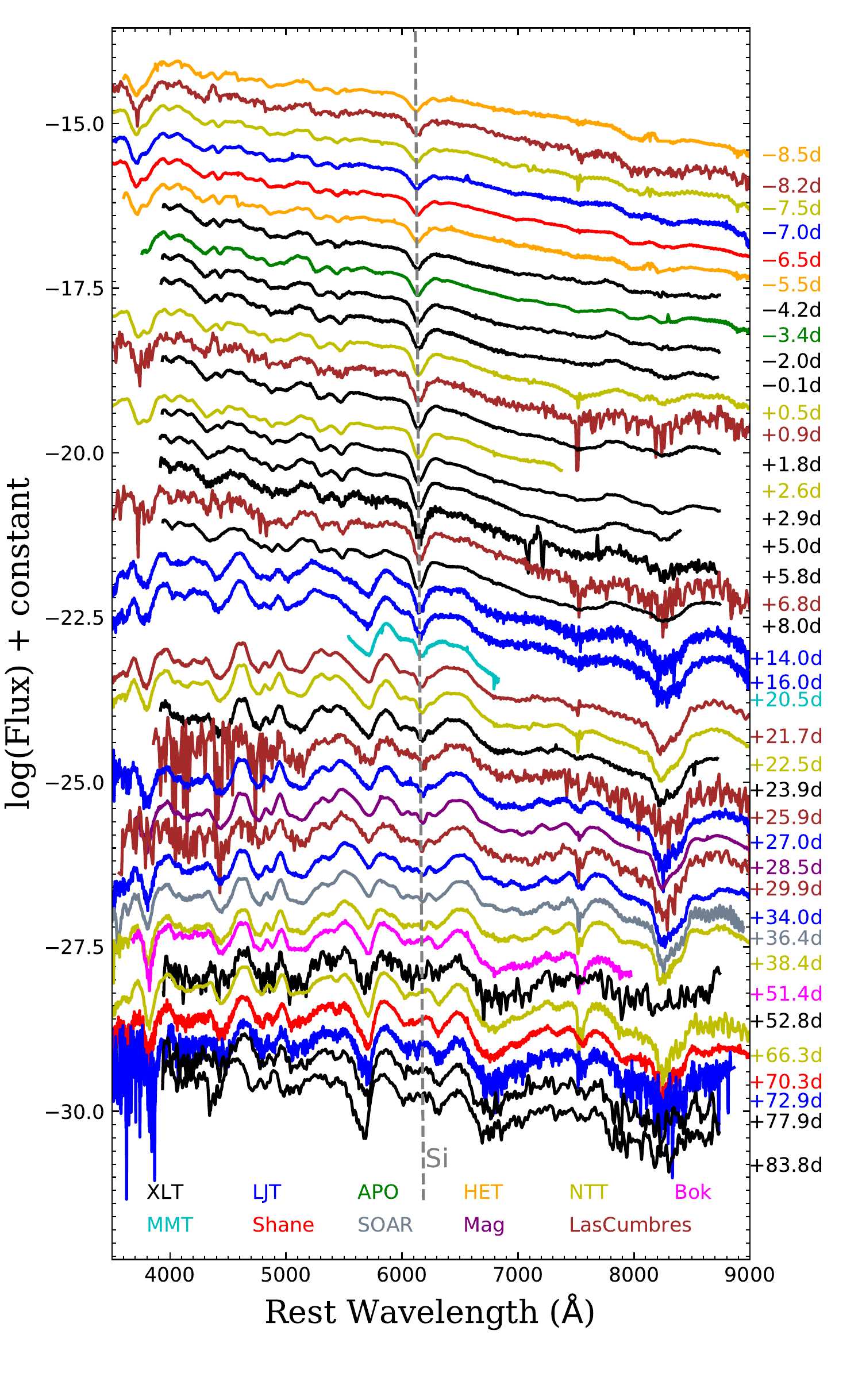}
\caption{Spectral evolution of SN 2018oh (some spectra are not displayed for limited space). The spectra have been corrected for the redshift of the host galaxy ($v_{\rm hel}$ = 3270 km s$^{-1}$) and reddening, and the slopes of the continuum are calibrated by the photometry. For better display, the spectra have been shifted vertically by arbitrary amounts. The epochs on the right side of the spectra represent the phases in days from $B$-band maximum light. The color of spectra indicates the instrument used for the observations, as shown at the bottom of the figure.}
\label{spec} \vspace{-0.0cm}
\end{figure}

Figure \ref{spec} displays the spectral evolution of SN 2018oh. The earlier spectra are dominated by absorption features of Si, Ca, S, and Fe. Near maximum light, the spectral evolution follows that of a normal SN~Ia, with the distinctive ``W"-shaped S~II lines near 5400~\AA, the blended lines of Fe~II and Si~II near 4500~\AA, and the prominent Ca~II absorption feature near 8300 \AA, respectively. A weak absorption feature that can be attributed to C~II $\lambda$6580 is seen on the red edge of Si~II $\lambda$6355 absorption feature for a long time (see discussions in Section \ref{carbon}). We discuss the spectral evolution of SN 2018oh in detail in the following subsections.

\subsection{Temporal Evolution of the Spectra}\label{spectra evo}
\begin{figure}[htbp]
\flushleft
\includegraphics[width=\textwidth]{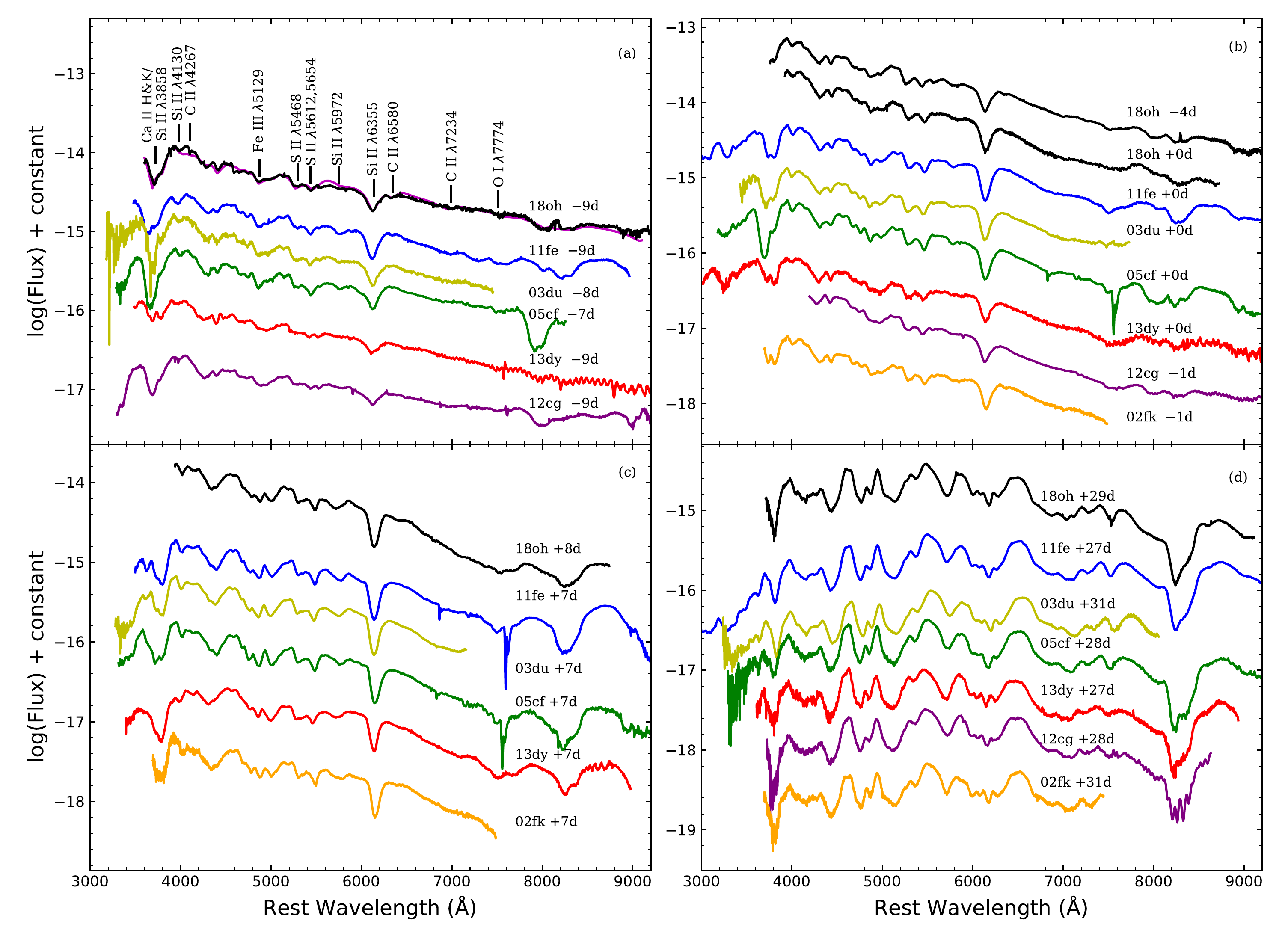}
\caption{The spectra of SN 2018oh at t$\sim$$-$9~d, $-$4 d, 0~d, +8~d, and +1 month after $B$ maximum, along with the comparable-phase spectra of SN 2002fk \citep{2012AJ....143..126B} , 2003du \citep{2007A&A...469..645S}, 2005cf \citep{2007A&A...471..527G,2009ApJ...697..380W}, 2011fe \citep{2014MNRAS.439.1959M,2016ApJ...820...67Z}, 2012cg \citep{2016ApJ...820...92M} and 2013dy \citep{2013ApJ...778L..15Z,2015MNRAS.452.4307P, 2016AJ....151..125Z}. All spectra have been corrected for reddening and the redshift of the host galaxy. For clarity, the spectra were arbitrarily shifted in the vertical direction. The SYNOW fitting result of t$\sim-$9 d spectrum of SN 2018oh is also overplotted in (a).}
\label{spec com}      
\end{figure}

In Figure \ref{spec com}, we compare the spectra of SN 2018oh with those of SNe Ia having similar decline rates at several epochs. The earliest spectrum of SN 2018oh was taken at t$\sim-$9.0 days. Figure \ref{spec com} (a) compares this spectrum with other SNe Ia at similar phases. The prominent features include Ca~II H\&K/Si~II $\lambda$3858, the ``W"-shaped S~II lines, and Si~II $\lambda$6355 absorption features. Other features include Si~II $\lambda$4130, Fe~II $\lambda$4404/ Mg~II $\lambda$4481, Si~II $\lambda$5051/Fe~II $\lambda$5018, Fe~III $\lambda$5129. The minor absorption neighbouring with Si~II $\lambda$4130 can be due to C~II $\lambda$4267. The absorption feature appearing on the right edge of S~II doublet, also visible in all of our comparison SNe Ia, is not presently identified. For SN 2018oh, the absorption due to Si~II $\lambda\lambda$5958, 5979 seems to be weaker than in SN 2011fe, SN 2003du, and SN 2005cf, but is comparable to that in SN 2012cg and SN 2013dy. The strength of Fe~III $\lambda$5129 for SN 2018oh follows the same manner as Si II $\lambda\lambda$5958, 5979 relative to the comparison SNe~Ia. A smaller line-strength ratio of Si~II $\lambda\lambda$5958, 5979 to Si~II $\lambda$6355, known as $R$(Si~II), indicates a relatively higher photospheric temperature for SN 2018oh \citep{1995ApJ...455L.147N}. Recently, \citet{2018ApJ...864L..35S} found that SNe Ia exhibiting blue colors in very early phase all belong to the shallow silicon (SS) subtype among Branch's classification scheme\citep{2006PASP..118..560B}, i.e., SNe 2012cg, 2013dy, and 2017cbv. The pseudo-equivalent widths (pEWs) of Si II$\lambda\lambda$5972, 6355 measured near the maximum light for SN 2018oh are 79\AA~ and 8\AA, respectively, suggesting that it can be also put into the SS subgroup or at least locates near the boundary between SS and core-normal subgroups. At about one week before the maximum light, absorption features of C~II 7234 and O~I 7774 are not prominent in SN 2018oh and the comparison SNe Ia except for SN 2011fe which had more unburned oxygen in the ejecta. A detached high-velocity feature (HVF) can be clearly identified in the Ca~II NIR triplet absorption features, and its relative strength is similar to that seen in SN 2013dy but weaker than SN 2005cf and SN 2012cg. A weak HVF of Si~II 6355 is also visible in SN 2018oh and the comparison SNe Ia but not in SN 2011fe. 

Figure \ref{spec com} (b) compares the near-maximum spectra. At this phase, the spectrum of SN 2018oh has evolved while maintaining most of its characteristics from the earlier epochs. The weak features (e.g., Si~II $\lambda$4130, Si~III $\lambda$4560, and the S~II ``W") become more prominent with time, as also seen in the comparison SNe Ia. The C~II absorption features are still clearly visible near 6300\AA\ and 7000\AA\ in the spectrum of SN 2018oh around maximum light, while they are barely detectable in other SNe Ia at this phase except for SN 2002fk. The O~I $\lambda$7774 line gains in strength for all the SNe, and the absorption at $\sim$7300\AA\ might be due to an O~I HVF. By t$\sim$0 days, the relative strength of the two absorption components of the Ca~II NIR triplet evolve rapidly, with the blue component (HVF) becoming weak and the red (photospheric) component becoming gradually strong and dominant. At this phase, the $R$(Si~II) parameter is measured as 0.15$\pm$0.04, which suggests a high photospheric temperature and high luminosity. This is consistent with a smaller decline rate that is characterized by an intrinsically more luminous Type Ia SN. 

At about 1 week after maximum light, most of the spectral features show no obvious evolution relative to those seen near the maximum light, as seen in Figure \ref{spec com} (c). We note that the absorption near 5700\AA\ becomes stronger in all of our sample, which is likely due to the contamination of Si II $\lambda$5972 by Na I that gradually develops after maximum light. For SN 2018oh, the most interesting spectral evolution is that the C~II 6580\AA\ absorption gains in strength during this phase, which has never been observed in other SNe Ia. Moreover, the C~II 6580 absorption can even be detected in the t $\sim$ 20.5 day spectrum, which is unusually late for a normal SN Ia. The spectral comparison at t $\approx$ 1 month is shown in Figure \ref{spec com} (d), where one can see that SN 2018oh exhibits spectral features very similar to other SNe~Ia in comparison. With the receding of the photosphere, the Fe~II features are well developed and become dominant in the wavelength range from 4700~\AA\ to 5000~\AA. By a few weeks after $B$ maximum, the region of Si~II $\lambda$5972 is dominated by Na I absorption, and the Si~II $\lambda$6355 absorption trough is affected by Fe~II $\lambda\lambda$6238, 6248 and Fe~II $\lambda\lambda$6456, 6518. Although the Ca~II NIR triplet shows the most diverse features in the earlier phases, they develop into an absorption profile that is quite smooth and similar to comparison sample at this time. 

\begin{figure}[htbp]
\center
\includegraphics[scale=.6]{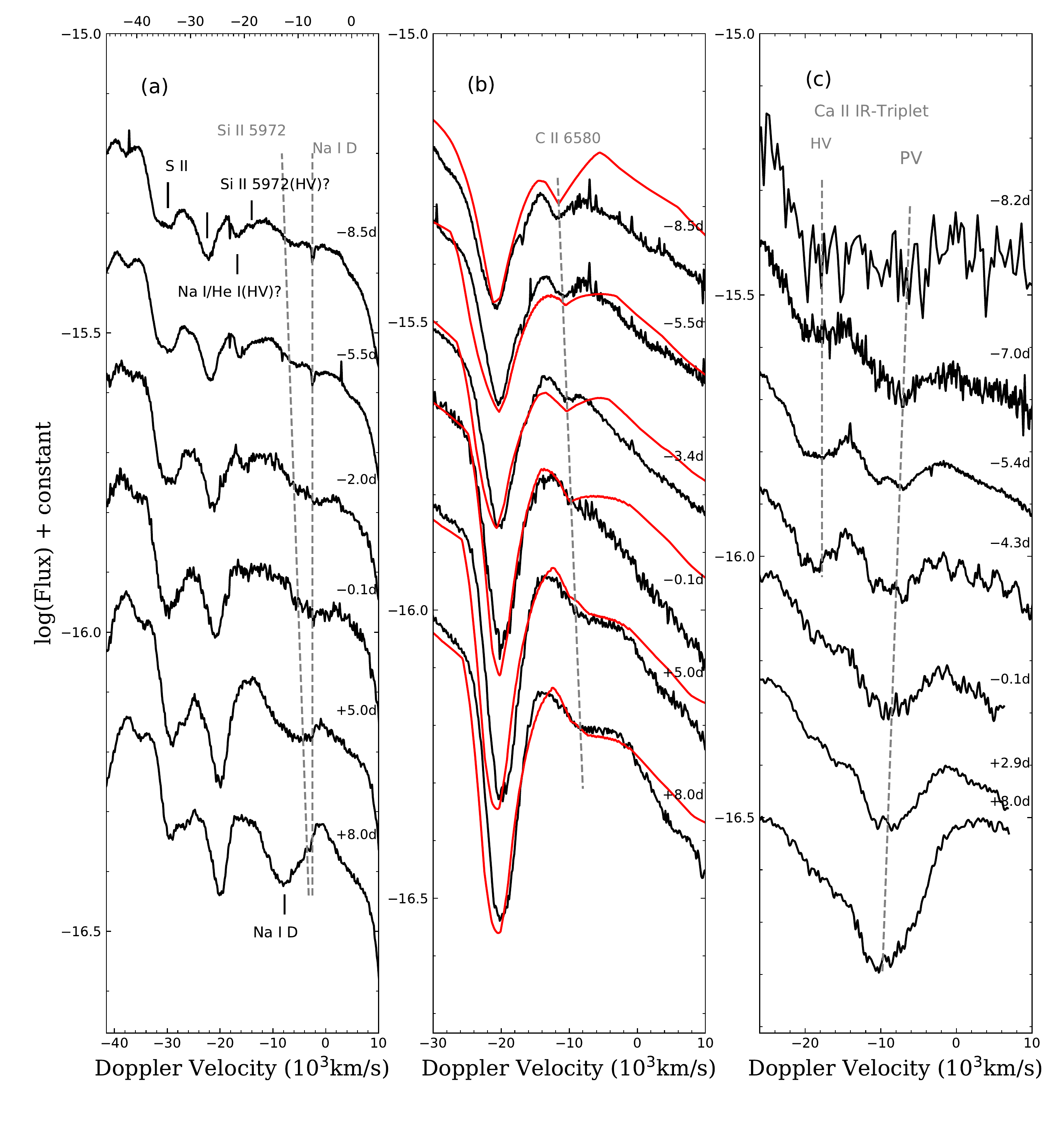}
\vspace{0.2cm}
\caption{The ``W-shaped" S~II, Si II $\lambda$5972, Si~II $\lambda$6355, C II $\lambda$6580, and Ca II IR triplet evolution of SN 2018oh. The velocity is defined relative to the rest wavelength of (a) He I $\lambda$5876 (upper axis: Si II $\lambda$5972), (b) C II $\lambda$6580, and (c) Ca II $\lambda$8542. The black solid lines label spectral features while the gray dashed lines indicate the velocity evolution trend for the corresponding lines. Overplotted red curves in (b) represent the best-fitting results from SYNOW.}
\label{c_o_ca} \vspace{-0.0cm}
\end{figure}

Figure \ref{c_o_ca} presents the detailed evolution of ``W-shaped" S~II, Si~II 5972, Si~II 6355, C~II 6580, and Ca~II NIR triplet for SN 2018oh. This evolution is shown in a velocity space. The left panel shows the line profile of S~II 5460, 5640 and Si~II 5972. One notable feature is the asymmetric absorption trough near 5500\AA\ where there is a notch on the red wing. This notch feature is likely a detached high-velocity component of Si~II 5972, since it has a velocity of $\sim$19,000 km s$^{-1}$ comparable to that of the HVF of Si~II 6355  and it became weak and disappeared in the spectra simultaneously with the Si~II 6355 HVF. The absorption feature at 5500\AA\ has not been identified but could be due to an Na I/He I HVF with a velocity at around 17,500 km s$^{-1}$. Figure \ref{c_o_ca} (b) shows the velocity evolution of Si~II 6355 and the neighboring C~II 6580 feature. The HVF of Si~II 6355 is visible in the two earliest spectra and it disappeared in the later ones. The presence of C~II 6580 is obvious, as also illustrated by the SYNOW \citep{1997ApJ...481L..89F} fit (the red curves). The C~II 6580 feature decreased in strength from t = $-$8.5 d to t = 0 d, and it then became wider and stronger in the first week after the peak. Such an evolution is unusual for a SN Ia and it is perhaps related to the interaction of the ejecta with the companion star or CSM.
The evolution of the Ca~II NIR triplet absorption feature is presented in Figure \ref{c_o_ca} (c). In the Ca~II NIR triplet, the HVF component is more separated from the photospheric component than in the Si~II line, and it dominates at earlier phases but gradually loses its strengths with time.


\subsection{High Resolution Spectra}
\begin{figure}[htbp]
\center
\includegraphics[]{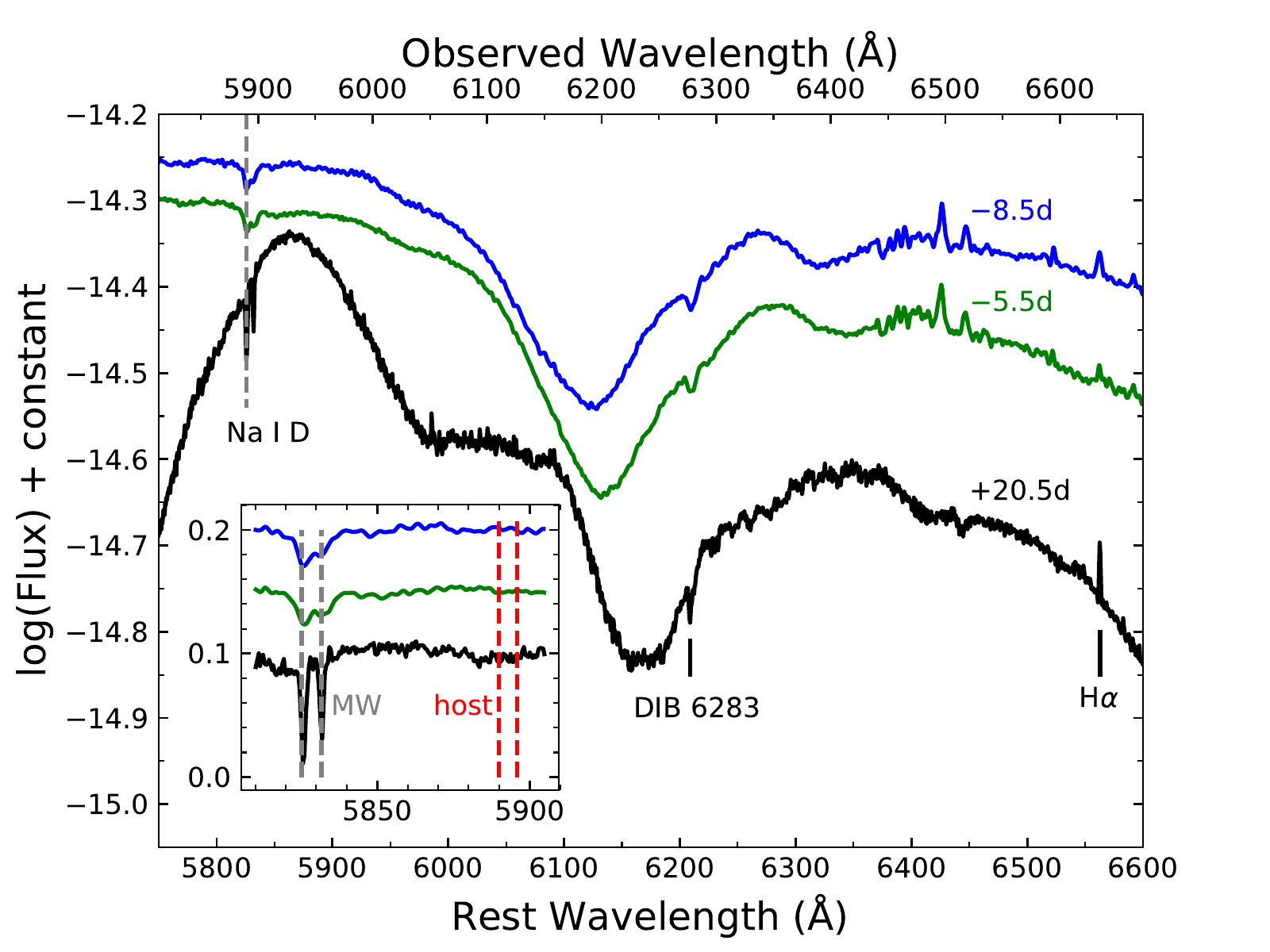}
\caption{High resolution spectra taken by the MMT and HET. Some narrow spectral features are labeled. The upper axis shows the observed wavelength. The inset shows the region of the Na~I D doublet absorption features due to Milky Way and the host galaxy.}
\label{hr3} \vspace{-0.0cm}
\end{figure}

A few spectra presented in this paper were observed with higher resolutions, i.e., the two HET spectra taken at $-$8.5d and $-$5.5d and the MMT spectrum taken at +20.5d. These spectra are shown in Figure \ref{hr3}, where we can see some narrow spectral features that are barely visible in other low-resolution spectra. One can see that the absorption by Na I D and the diffuse interstellar band (DIB) at $\lambda$6283 from the Milky Way are clearly visible in the high resolution spectra, consistent with the presence of a modest level of Galactic reddening. There are some minor absorption features in the red wing of Si II 6355, which may also be related to unidentified DIBs. A few SNe have been reported to have host-galaxy DIB detections in their spectra  \citep{1989A&A...215...21D, 2005A&A...429..559S, 2008A&A...485L...9C,2013ApJ...779...38P, 2014ApJ...792..106W}. The absence of Na~I D and DIB absorption components from UGC 4780 is consistent with that SN 2018oh suffering negligible reddening within the host galaxy. A weak, narrow $H_\alpha$ emission that is likely from the host galaxy feature can be clearly seen in both the HET and MMT spectra.

\subsection{Carbon Features}\label{carbon}

\begin{figure}[htbp]
\center
\includegraphics[scale=.8]{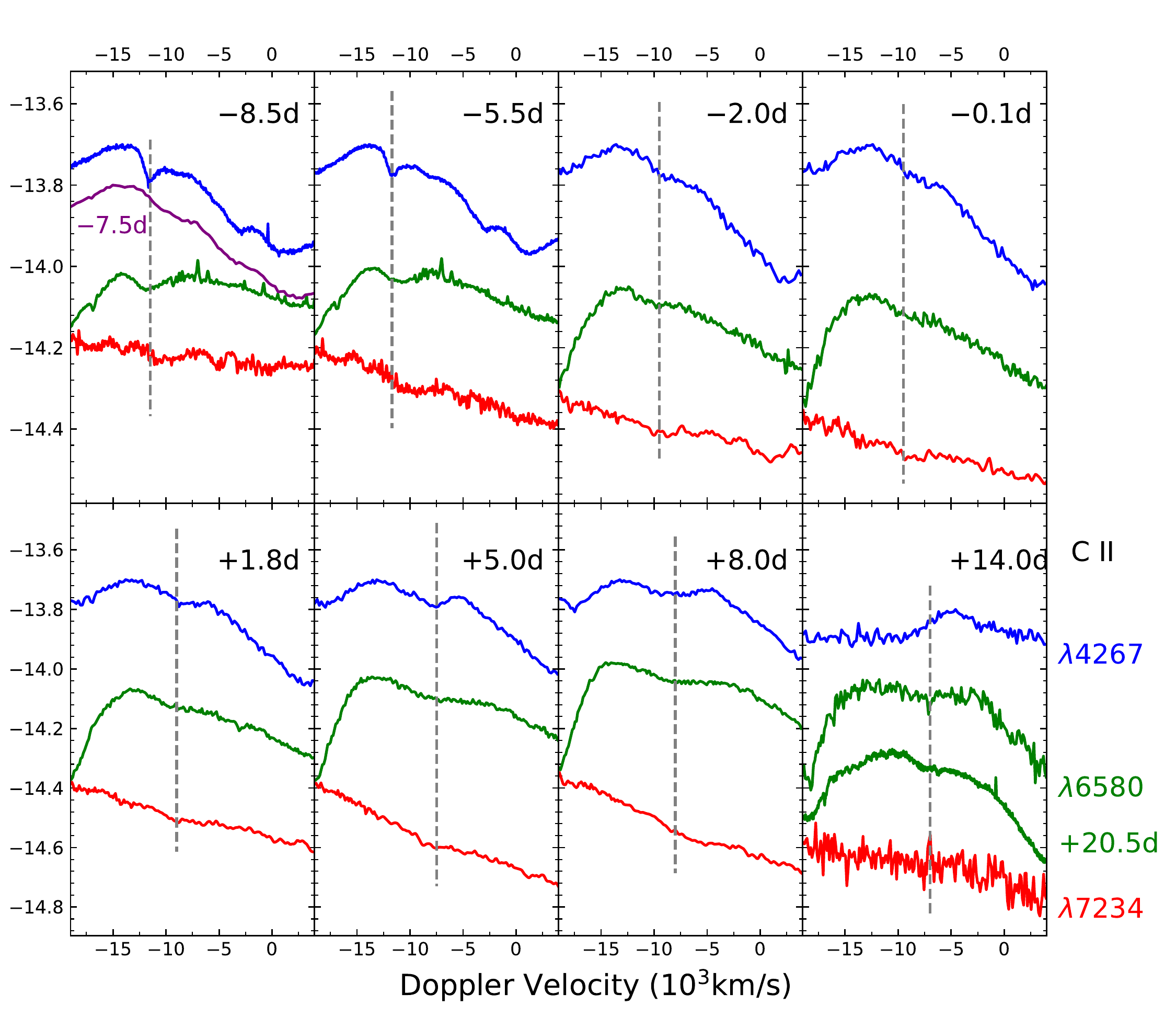}
\vspace{0.2cm}
\caption{The C II $\lambda$4267, $\lambda$6580, and $\lambda$7234 evolution of SN 2018oh in velocity space. Three lines in each subplot are from one spectrum. The purple line in the left-top panel display C II $\lambda$4267 feature from the $-$7.5d spectrum taken by NTT. The grey lines indicate the approximate velocity of the three features.}
\label{cii} \vspace{-0.0cm}
\end{figure}

C~II features are clearly detected in SN 2018oh, and they seem to persist for an unusually long time compared to other known SNe Ia. As shown in Figure \ref{cii}, the C~II 6580 absorption feature can be detected in the spectra from t = $-$8.5 to t = +20.5 days. And the C~II 4267 and C~II 7234 absorptions are also detectable in the spectra from t =$-$8.5 day to t =+8.0 days\footnote{Note that there is an instrumental trough around $\lambda$4150 in the HET spectra, which nearly coincides with the expected position
of the C II $\lambda$4267 feature and this makes it difficult to judge whether the presence of this feature is real or not. Nevertheless, a weak C II $\lambda$4267 feature can be still identified in the $-$7.5d spectrum taken by NTT as shown in the left-top panel of Figure \ref{cii}.} 
Identifying these carbon features is justified by the agreement in velocity at early phases  (see Figure \ref{cii}) and the SYNOW spectral models. It should be noted that the SYNOW velocities shown in Figure \ref{c_o_ca} are higher than the measured values by $\sim$2000 km s$^{-1}$ (see Table \ref{synow_p}). This offset is due to low optical depths at the line centers in the SYNOW fits producing a steep drop in optical depths bluewards of the best fit velocity, resulting in minimal absorption bluer than the line center, which shifts the apparent minimum in the line profile.

\cite{2012MNRAS.425.1917S} measured the velocity ratio between C II $\lambda$6580 to Si II $\lambda$6355 and found a median value of 1.05 at phases earlier than 4 days from maximum. For SN 2018oh, this ratio is 1.05$-$1.00  at t$\lesssim -$4 days, consistent with \cite{2012MNRAS.425.1917S}. However, the C/Si velocity ratio keeps decreasing after t$\gtrsim -$4d and reaches about 0.85 $\pm$ 0.06 at t$\sim$+20.5 days for SN 2018oh, which suggests that unburned carbon may be more strongly mixed than silicon and extends deep into the ejecta.

\cite{2012ApJ...745...74F} calculate the pseudo-equivalent width (pEW) evolution of C II 6580 using SYNOW synthetic spectra with different unburned carbon mass. The pEW is found to grow monotonically with the mass of carbon. For SN 2018oh, the C~II 6580 absorption has a pEW $\sim$ 4\AA ~ and 2\AA ~ around $-$4.3d and $-$2.0d, respectively, which is very similar to that of the synthetic spectra with $\approx$ 0.03 M$_\odot$ of unburned carbon in the ejecta.

\subsection{Ejecta Velocity}\label{photo v}
\begin{figure}[htbp]
\center
\includegraphics[scale=.8]{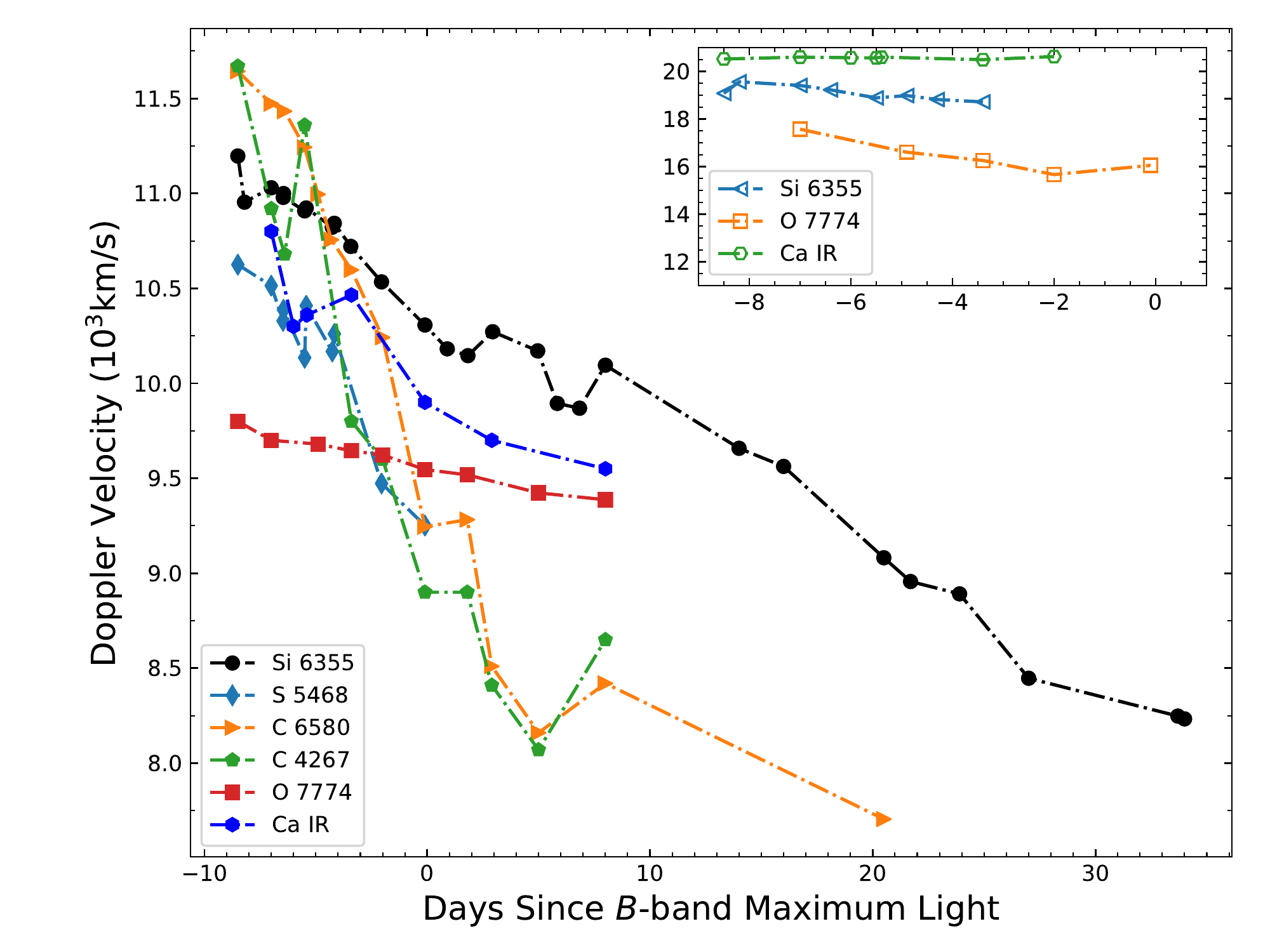}
\caption{Evolution of the expansion velocity of SN 2018oh as measured from the absorption minimum of Si~II $\lambda$6355, S~II 5640, C~II 6580, C~II 7234, O~I 7774, and Ca~II NIR triplet. Inset plot illustrates the high velocity components of three features.}
\label{vall} \vspace{-0.0cm}
\end{figure}

We measured the ejecta velocities from the blueshifted absorption features of Si~II $\lambda$6355, S~II $\lambda$5468, C~II $\lambda$6580, C~II $\lambda$7234, O~I $\lambda$7774, and the Ca~II NIR triplet lines, and the velocity evolution is shown in Figure \ref{vall}. All velocities have been corrected for the host galaxy redshift. The photospheric velocity of Si~II 6355, characterized by a linear decline from $\sim$11,000 km s$^{-1}$ to $\sim$8000 km s$^{-1}$, is comparable to that of other intermediate-mass elements at similar phases. Assuming a homologous expansion of the ejecta, this indicates a complex distribution of carbon in the ejecta. However, it is possible that the position of C~II 6580 absorption in late-time spectra might be contaminated by other unknown elements. The best-fit C~II velocities from SYNOW show an offset by $\sim$2000 km s$^{-1}$ relative to the measured values, and this suggests that carbon is detached until $\sim$+5 days from the maximum light. After that, the SYNOW velocity of C~II becomes comparable to the photospheric values, matching that of Si~II 6355.

The high-velocity features (HVF) of Si II $\lambda$6355, O I $\lambda$7774 and the Ca II IR triplet have been systematically examined in the spectra of SNe Ia \citep{2014MNRAS.437..338C, 2014MNRAS.444.3258M, 2015MNRAS.451.1973S,2015ApJS..220...20Z,2016ApJ...826..211Z}. The HVFs of both the Si II $\lambda$6355 and the Ca II IR triplet can be clearly identified in the early spectra of SN 2018oh. Since the region overlapping with the oxygen absorption has lower spectral quality for our early data, the O-HVF cannot be clear identified. The velocities measured for the HVFs identified for Si~II $\lambda$6355 and Ca~II NIR triplet can reach at about 19,000 $-$ 22,000 km s$^{-1}$, far above the photosphere. According to recent studies by \cite{2015ApJS..220...20Z,2016ApJ...826..211Z}, the HVFs cannot be explained by ionization and/or thermal processes alone, and different mechanisms are required for the creation of HVF-forming regions. 
\cite{2017MNRAS.467..778M,2018MNRAS.476.1299M} showed that a compact circumstellar shell having $\lesssim 0.01$ M$_\odot$ mass is capable of producing the observed HVF component of the Ca II NIR triplet. 
  
\begin{figure}[htbp]
\center
\includegraphics[scale=.9]{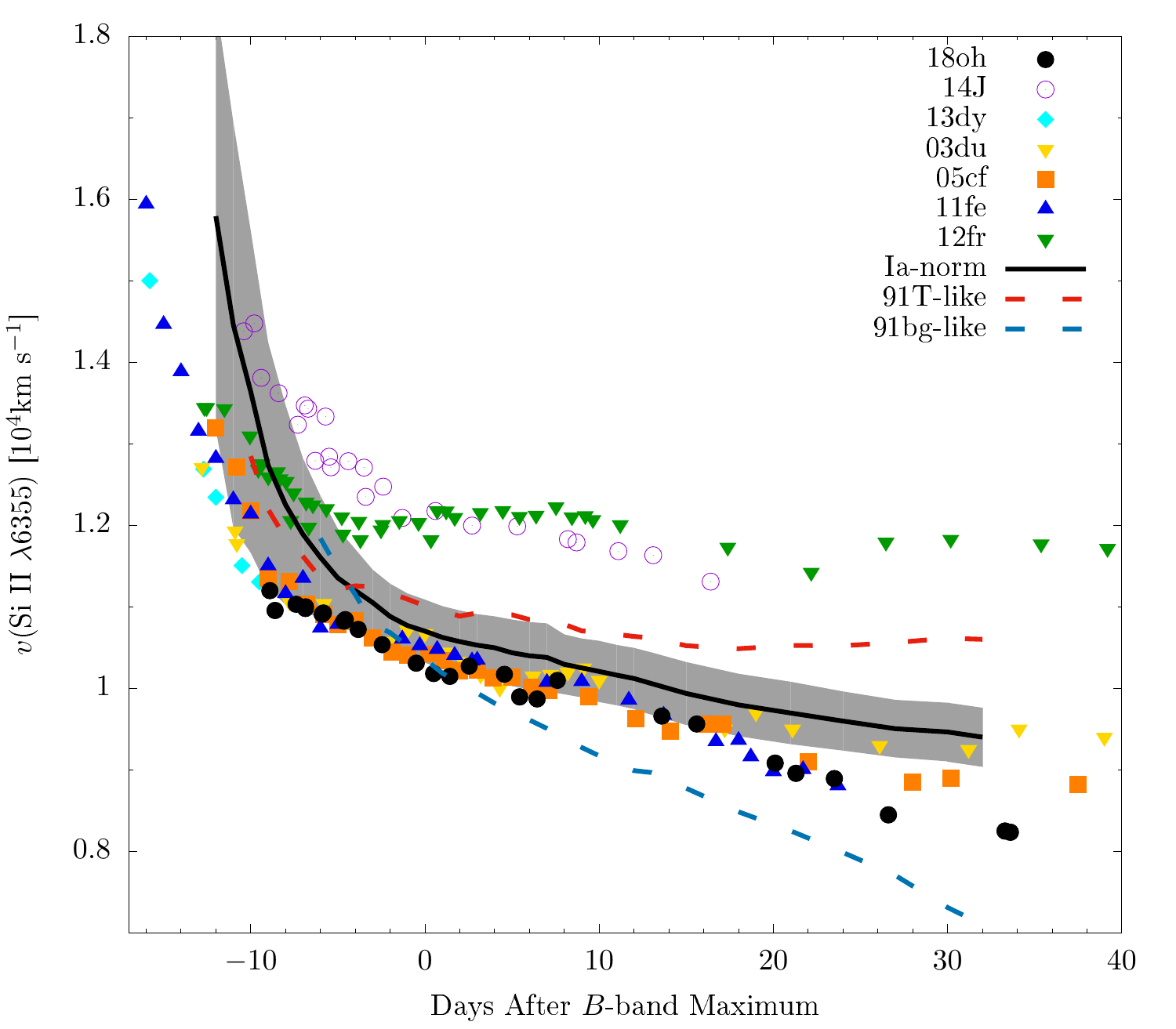}
\caption{Velocity evolution of SN 2018oh as measured from the absorption minimum of Si~II 6355, compared with SNe 2003du, 2005cf, 2011fe, 2012cg, and 2013dy (see text for the references). Overplotted are the mean curves of velocity evolution obtained SN 1991T-like (red dashed), SN 1991bg-like (blue dotted), and normal subclasses (solid black) of SNe Ia \citep{2009ApJ...699L.139W}. The shadow region represents the 1-$\sigma$ uncertainty for the mean velocity curve of normal SNe Ia.}
\label{vsi} \vspace{-0.0cm}
\end{figure}

In Figure \ref{vsi}, we compare the Si~II velocity evolution of SN 2018oh with some well-observed SNe Ia. The $v_{si}$ evolution of SN 2018oh is comparable to that of SN 2005cf and SN 2011fe, as shown in Figure \ref{vsi}. At around the $B$-band maximum light, SN 2018oh has an expansion velocity of 10,300 km s$^{-1}$, which can be clearly put into the NV group according to the classification scheme proposed by \cite{2009ApJ...699L.139W}. The velocity gradient of Si~II $\lambda$6355 during the first 10 days after $t_{Bmax}$ is measured as $\dot{v_{Si}}$ = 69 $\pm$ 4 km s$^{-1}$ d$^{-1}$, which locates just around the boundary between high-velocity gradient (HVG) and low-velocity gradient (LVG) objects \citep {2005ApJ...623.1011B}. A relatively fast velocity decline might be due to the collision of the ejecta with the nearby companion as suggested by the early light curve observed by {\em Kepler} \citet{dimitriadis-18oh} or CSM. However, \citet{shappee-18oh} found that a single power-law rise with a non-degenerate companion or CSM interaction cannot reproduce well the early $Kepler$ light curve. They derived that, at a radius of 4 $\times$ 10$^{15}$ cm from the progenitor, the CSM density $\rho_{CSM}$ is less than  4.5$\times$10$^5$ cm$^{-3}$.

\section{Discussion}

\subsection{Origin of Persistent Carbon Absorption}\label{c}
The unburned carbon features in early spectra can help to discriminate between various explosion mechanisms or progenitor models for SNe Ia. Previous studies show that the C~II signatures can be detected in 20 $-$ 30 \% of SNe Ia with ages younger than $\sim$$-$4 days from the maximum light and $>$40\% of SNe Ia have unburnt carbon before $-$10 days \citep{2011ApJ...732...30P, 2011ApJ...743...27T, 2012MNRAS.425.1917S,2014MNRAS.444.3258M}. The latest detection was at t = $-$4.4 days for SN 2008sl. In a late study, SN 2002fk showed carbon absorption lasting  $\sim+$7 days after maximum \citep{2014ApJ...789...89C} and the 2002cx-like supernova iPTF14atg showed C~II $\lambda$6580 absorption until about +2 weeks after maximum \citep{2015Natur.521..328C}.

\begin{figure}[htbp]
\center
\includegraphics[scale=.8]{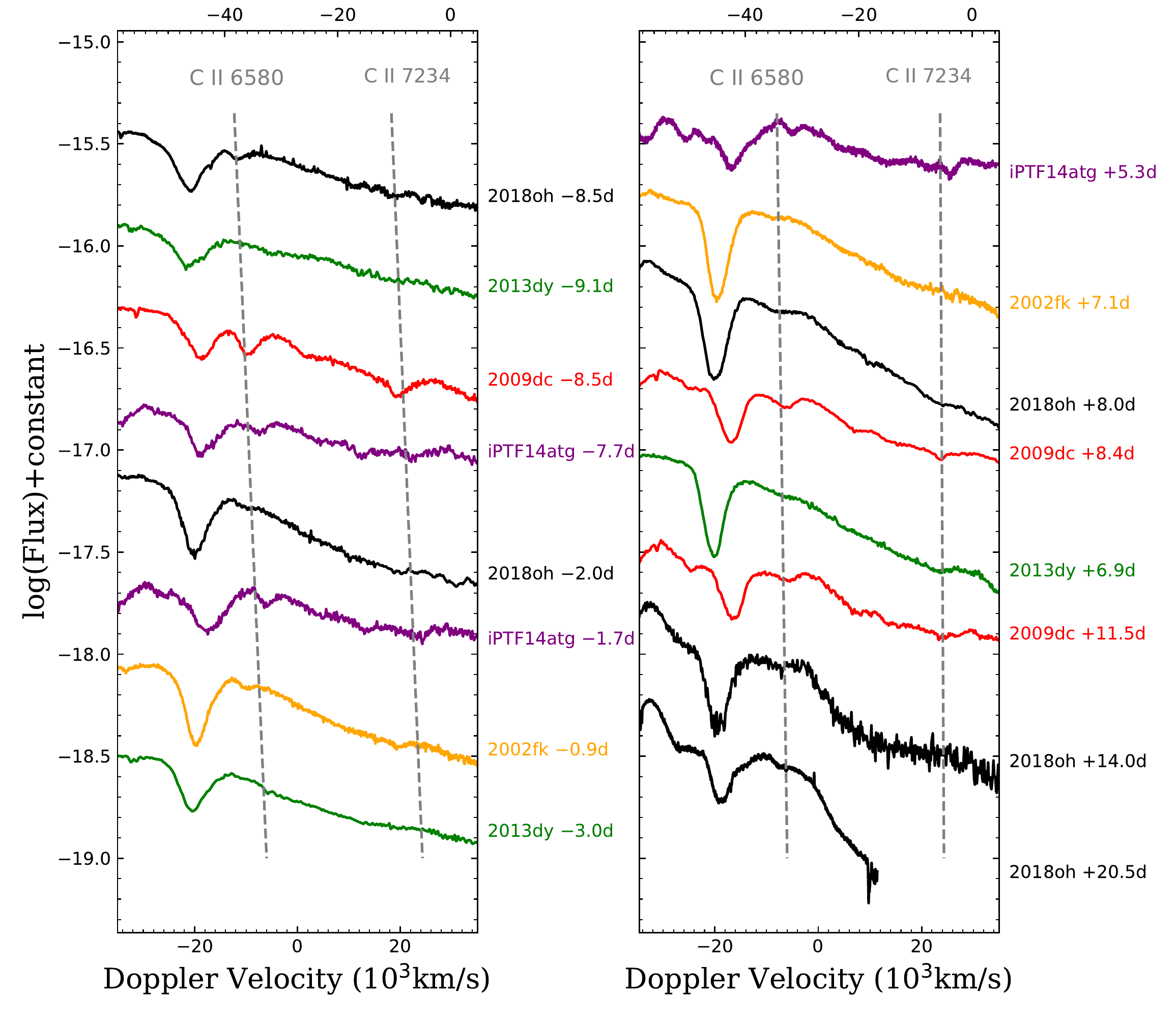}
\vspace{0.2cm}
\caption{C II $\lambda$6580 evolution of SN 2018oh compared to that of SN 2003du, SN 2011fe, iPTF14atg, and SN 2002fk. The gray dashed lines indicate the velocity evolution trend for the corresponding lines.}
\label{c_compare} \vspace{-0.0cm}
\end{figure}

The carbon absorption persists in the spectra of SN 2018oh for an unusually long time. To examine this abnormal behavior, we further compare the C~II 6580 evolution of SN 2018oh with some well-known SNe Ia with prominent carbon absorption features, including SN 2002fk, SN 2009dc, SN 2011fe, SN 2013dy, and iPTF14atg in Figure \ref{c_compare}. The C~II absorption is strong in the t=$-$8.5d and t=$-$5.5d spectra of SN 2018oh. After that, the C~II 6580 tends to become flattened, which was not seen in other normal SNe Ia. The strength of carbon absorption features is found to decrease with time (except for the period at t=$-$13 - $-$11 days from the maximum light, \cite{2012MNRAS.425.1917S}). However, the strength of C~II $\lambda$6580 absorption of SN 2018oh increases after the $B$ maximum. 

For SN 2012cg and SN 2017cbv, the C~II $\lambda$6580 of the former lasted until $-$8d \citep{2012ApJ...756L...7S}, while it disappeared in the t = $-$13d spectrum of the latter \citep{2017ApJ...845L..11H}. The super-Chandrasekhar (SC) SNe Ia like SN 2009dc are known to show prominent carbon absorptions \citep{2006Natur.443..308H, 2010ApJ...713.1073S, 2011MNRAS.410..585S, 2011MNRAS.412.2735T}. The C II 4267 absorption is difficult to identify due to several Fe-group features in this wavelength region. It was previously identified in SNLS-03D3bb and SN 2006D \citep{2006Natur.443..308H,2007ApJ...654L..53T}, while \cite{2010ApJ...713.1073S} proposed that this feature might be due to Cr~II absorption. However, this feature in SN 2018oh has similar velocity and strength evolution with C II 6580 until t$\sim$+8.0d (see Figure \ref{cii}), unlike SN 2009dc \citep{2011MNRAS.412.2735T}. This gives us more confidence in the identification of C II 4267 absorption in SN 2018oh.  

In theory, pulsating delayed-detonation (PDD) model predicts the presence of carbon in the outer ejecta during pulsation period \citep{1996ApJ...472L..81H}. \cite{2014MNRAS.441..532D} claim that PDD can leave more unburned carbon than standard delayed-detonation models and thus produce prominent C~II lines in the spectra. However, these C~II features should disappear within one week after explosion. Their models can reproduce the strong C~II lines of SN 2013dy but cannot explain the long-lasting C~II lines seen in SN 2018oh.

\cite{2017ApJ...846...15H} suggest that the emission of iron near 6100 \AA~can smear out the C~II 6580 absorption. Thus, a smaller amount of IGEs in the outer ejecta could explain for the prominent carbon feature in SN 2018oh, which could be due to stringent abundance stratification or lower metallicity for the progenitor. For example, SN 2013cv was a transitional SN Ia between normal and SC SNe Ia with persistent C~II 6580 and 7234 until one week after maximum. It has high UV luminosity and its early phase spectra were absent of Fe~II/III features, suggestive of strong stratified structure in the explosion ejecta and hence the progenitor \citep{2016ApJ...823..147C}. SN 2018oh exhibits relatively weaker Fe~III $\lambda$5129 than SN 2003du, SN 2005cf and SN 2011fe (Section \ref{spectra evo}) and has blue UV color (see Figure \ref{color_uv}), which suggests that it suffered less mixing in the explosion ejecta.

As an alternative explanation for the abundance stratification, it is possible that the progenitor of SN 2018oh has lower metallicity. In order to study the properties of host galaxy, we downloaded the spectrum from the SDSS DR14 \citep{2018ApJS..235...42A}. It corresponds to the light that falls within the 2" diameter fiber that is pointed at the center of the galaxy. Thus, to estimate the total mass of the galaxy, we scaled the synthetic broad band magnitudes measured from the spectrum to match the real photometric measurements of the integrated light of the galaxy ({\tt modelMag} parameter). However, this procedure has a caveat: it makes the assumption that the mass to light ratio (M/L) obtained from the spectrum (hence representative of the area inside the fiber) is the same as the one outside the fiber.
Then, following \cite{2014A&A...572A..38G}, we performed simple stellar population (SSP) synthesis to the spectrum with STARLIGHT \citep{2005MNRAS.358..363C} using the Granada-MILES bases \citep{2015A&A...581A.103G}, and fit all the emission lines with gaussian profiles in the subtracted gas-phase spectrum. We estimated a stellar mass of $\log _{10}(M_{stellar}/M_{\odot}) \sim 6.87$ $\pm$ 0.12, a star formation rate (SFR) of 5.54 $\pm$ 0.36 10$^{-4}$ M$_{\odot}$ yr$^{-1}$ and a subsolar oxygen abundance 12 + log$_{10}$(O/H) of 8.49 $\pm$ 0.09 dex using the O3N2 calibration from \cite{2004MNRAS.348L..59P}, confirming that UGC04780 is actually a metal-poor galaxy. These findings are in total agreement with reported numbers in the SDSS DR14 from different methods and codes\footnote{http://skyserver.sdss.org/dr14/en/tools/explore/parameters.aspx?id=1237667430628982959\&spec=2573869371524933
632\&apid=\&fieldId=0x112d13f880b60000\&ra=136.664749886541\&dec=19.3362515108894\&plateId=2573807249117964288}. In comparison, \citet{shappee-18oh} derive a larger mass of 4.68$_{-0.61}^{+0.33} \times 10^8~\rm M_{\odot}$ from GALEX and PS1 photometry, while they suggest that this value can be regarded as an upper limit, which is thus not inconsistent with our determination.

Based on the above discussions, we suggest that the outer ejecta of SN 2018oh may have few IGEs as a result of less mixing and/or having metal-poor progenitor, which could explain the presence of prominent and persistent C~II 6580 absorption feature in the spectra.

\subsection{Bolometric Light Curves and Explosion Parameters}

\begin{figure}[htbp]
\center
\includegraphics[scale=1]{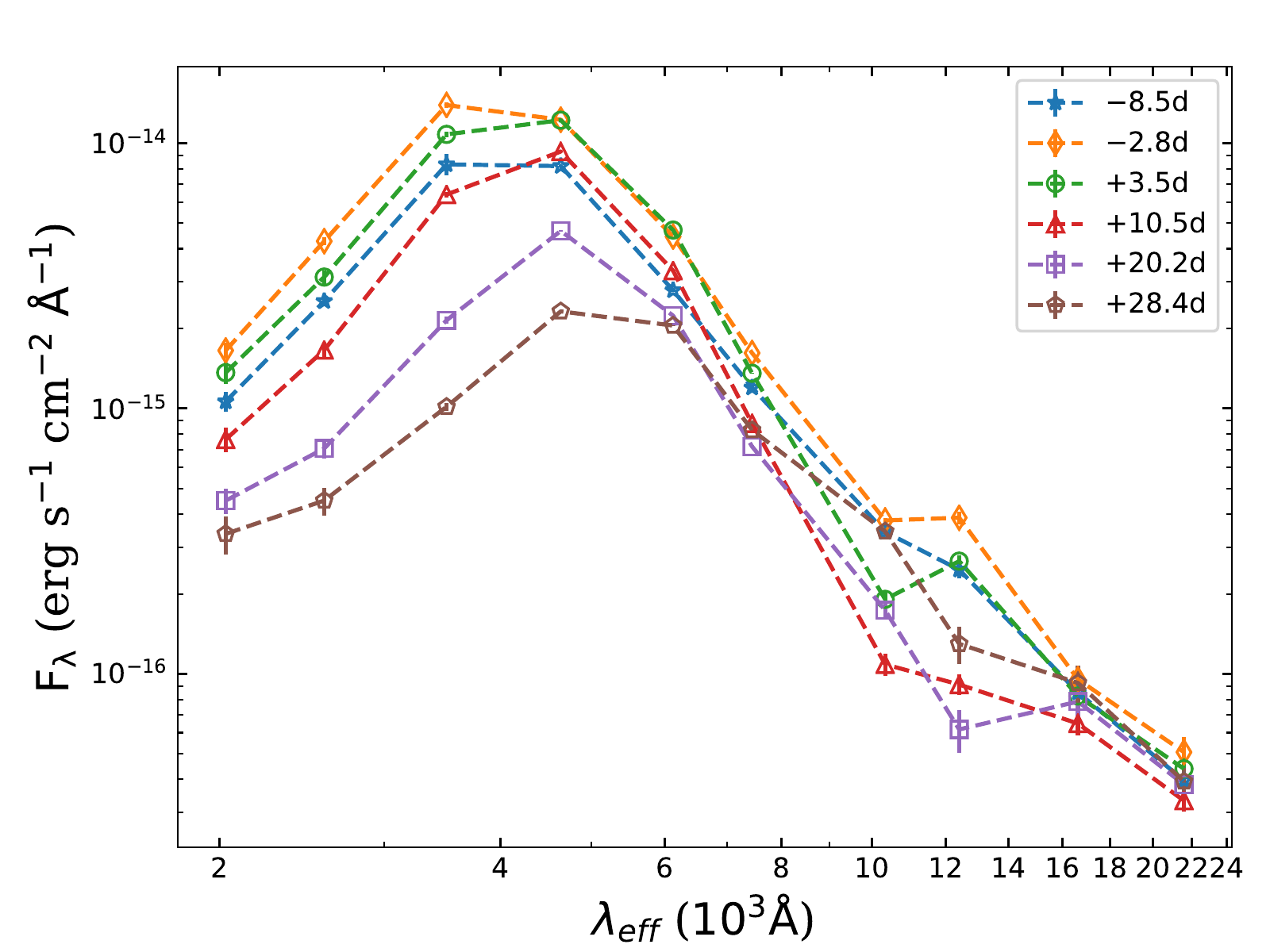}
\vspace{0.2cm}
\caption{SED evolution of SN 2018oh. The circles indicate the effective wavelength of different bands. }
\label{sed} \vspace{-0.0cm}
\end{figure}

\begin{figure}[htbp]
\center
\includegraphics[width=8.5cm]{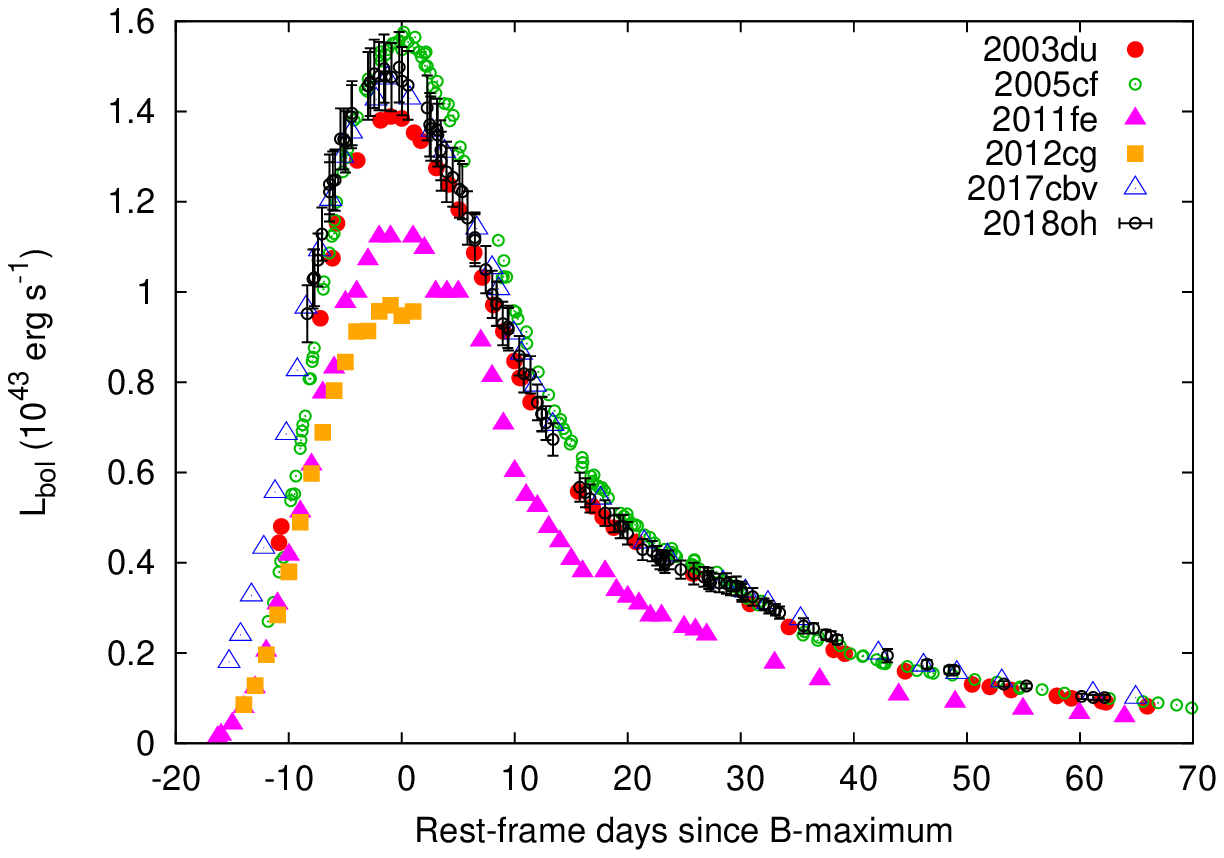}
\includegraphics[width=8.5cm]{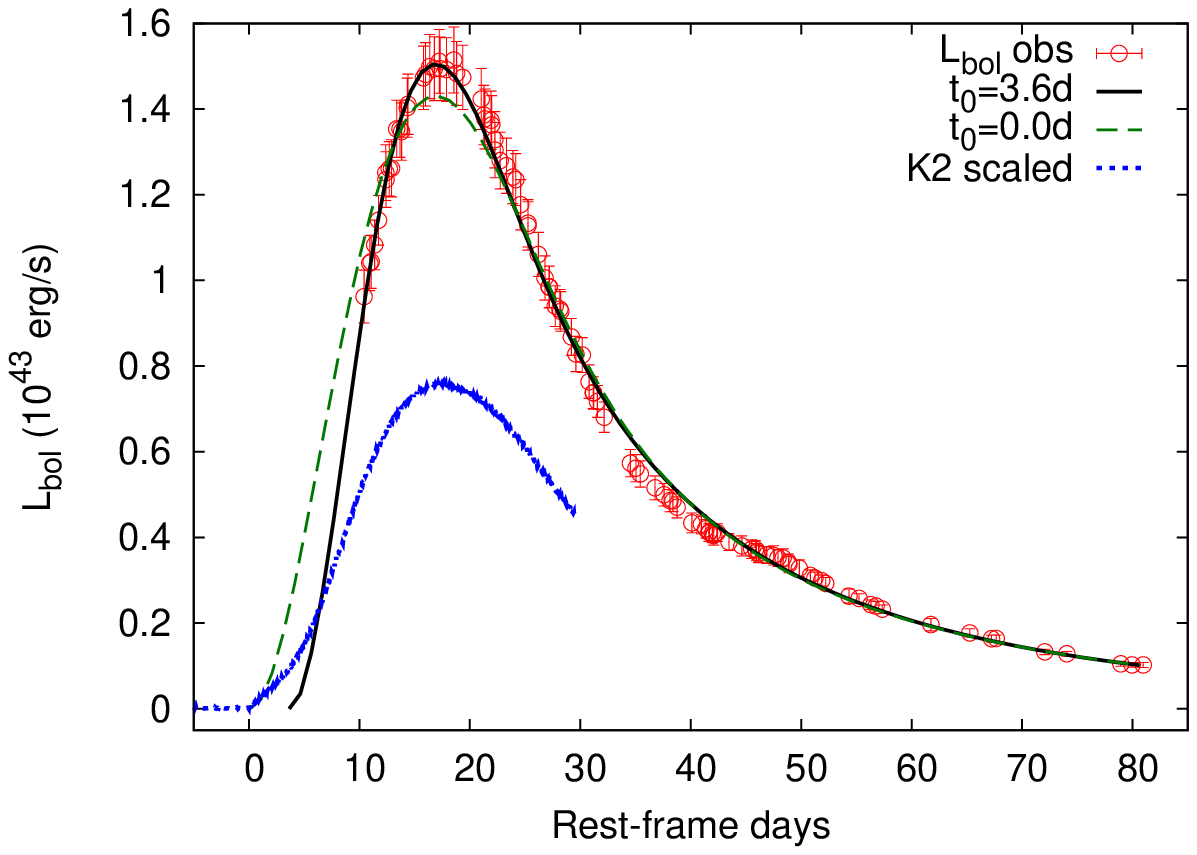}
\vspace{0.2cm}
\caption{Left panel: Luminosity evolution of SN 2018oh compared with that of SNe 2003du, 2005cf, 2011fe, 2012cg and 2017cbv. Due to the distance uncertainty of SN 2017cbv, we shift it to match the peak of SN 2018oh. Right panel: The bolometric light curve (open symbols) together with radiation diffusion Arnett-models (black curves). The continuous line shows the best-fit model, while the dashed line represents the model when the time of explosion is fixed to the appearance of the first light in the {\it {\em Kepler}} data. The scaled {\em K2} light curve (see Section~\ref{k2-phot}) is plotted with a blue dotted line. }
\label{L} \vspace{-0.0cm}
\end{figure}

The extensive photometric observations of SN 2018oh enable us to construct a $uvoir$ ``bolometric" light curve spanning the wavelength region from 0.16 to 2.3~$\mu$m. The spectral energy distribution (SED) includes the uvw2, uvm2, uvw1, $U$, $g$, $r$, $i$, $Y$, $J$, $H$, and $K$ bands. We interpolated the UV, optical and NIR photometry from their neighboring epochs or the corresponding template light curves whenever necessary.
The final SED evolution is displayed in Figure \ref{sed}. Adopting the distance d$=52.7\pm1.2$ Mpc from $\S$\ref{lc_fit}, the bolometric luminosity evolution is shown in the left panel of Figure \ref{L}. Like other comparison SNe Ia (except for SN 2005cf), SN 2018oh reached its peak about 1.5 days earlier than the $B$-band maximum. The overall shape of the light curve is quite similar to that of SN 2017cbv, and shows an apparently slower rise compared to SN 2003du. 

To estimate the nickel mass and other physical parameters of the ejecta, we apply the radiation diffusion model of \citet{1982ApJ...253..785A} \citep[see also][]{2012ApJ...746..121C}. Adopting the constant opacity approximation, we fit the bolometric light curve using the {\tt Minim} code \citep{2013ApJ...773...76C}. The fit parameters are the time of ``first light'' $t_0$ (see below), the initial mass of the radioactive nickel $M_{Ni}$, the light curve timescale $t_{lc}$ and the gamma-ray leaking timescale $t_{\gamma}$ \citep[see e.g. ][for details]{2012ApJ...746..121C}. 

If $t_0$ is constrained to the moment of first light in the {\em Kepler} data (MJD 58144.3 $\pm$ 0.1) we get $M_{Ni} = 0.662 \pm 0.003$ M$_\odot$, $t_{lc} = 14.89 \pm 0.07$ d and $t_\gamma = 39.56 \pm 0.18$ d. The model light curve is plotted as a green dashed line together with the observations in the right panel of Figure~\ref{L}. 
It is seen that this model poorly fits the light curve, because it deviates from the observed data systematically before and around maximum light: the model is too bright at $\sim +10$ days, while
it is too faint (although still within the errorbars) compared to the data around the maximum.

The fit quality improves when $t_0$ is optimized: 
bf the model having $t_0 = +3.85 \pm 0.13$ days (the black curve in Figure~\ref{L}) fits the data much better and does not show such kind of systematic deviations around maximum that the model with $t_0=0$ does. Having $t_0 > 0$ means that the radioactivity-powered light curve starts to rise $\sim 3.8$ days {\it after} the first light seen by {\em Kepler}. This is consistent with the finding by \citet{shappee-18oh}, who pointed out that the early {\em K2} light curve of SN~2018oh could be modeled with two power-laws having different starting moments ($t_1$ and $t_2$) that are separated by $t_2 - t_1 \sim 4$ days, to produce a much better fit than with a single power law starting at MJD 58144.3. Within the framework of the radiative diffusion model, their second power-law ($\sim t^{1.4}$) can be associated with the initial phase of the light curve emerging from the homologously expanding, quasi-spherical SN ejecta that is powered by the radioactive decay of $^{56}$Ni and $^{56}$Co located in the center of the ejecta. Such a delay between the moment of explosion and the emergence of the radioactivity-powered light curve is predicted in some SN Ia models as the ``dark phase'' \citep{2013ApJ...769...67P, 2014ApJ...784...85P, 2016ApJ...826...96P} caused by the location of the radioactive $^{56}$Ni within the ejecta. The duration of the dark phase is determined by the initial diffusion time of the deposited radioactive energy between the location of $^{56}$Ni and the surface of the ejecta. The Arnett model does not contain such a dark phase, because it assumes an initial temperature distribution that remains spatially constant during the SN evolution, i.e. at $t=0$ the initial diffusion wave already reached the surface. \citet{2016ApJ...826...96P} predict the length of the dark phase as $\lesssim 2$ days, while our result ($t_0 \sim 3.8$) is almost a factor of 2 longer. However, after taking into account the model-dependent uncertainties involved in such an estimate, our result of $t_0 \sim 3.8$ day could be interpreted as being this dark phase, i.e. it is the timescale of the initial diffusion wave propagating between the center and the surface of the ejecta. 

From our best-fit Arnett model we also get $t_{lc} = 10.81 \pm 0.14$ d, $t_{\gamma} = 41.36 \pm 0.18$ d and $M_{Ni} = 0.55 \pm 0.01$ M$_\odot$. The ejecta mass ($M_{ej}$) and the expansion velocity ($v_{exp}$) are related to the model timescales ($t_{lc}$ and $t_{\gamma}$) as 
\begin{equation}\label{eq:ejecta-pars}
t_{lc}^2 = {{2 \kappa M_{ej}} \over {\beta c v_{exp}}} ~,~\mathrm{and} ~~
t_{\gamma}^2 = {{3 \kappa_{\gamma} M_{ej}} \over {4 \pi v_{exp}^2}}, 
\end{equation}
\citep{1982ApJ...253..785A, 1997ApJ...491..375C,2008MNRAS.383.1485V, 2012ApJ...746..121C, 2015MNRAS.450.1295W}, 
where $\kappa$ is the effective optical opacity, $\kappa_{\gamma}$ is the opacity for $\gamma$-rays (assuming full trapping of positrons released in the cobalt decay), and $\beta \sim 13.8$ is the light curve parameter related to the density profile of the ejecta \citep{1982ApJ...253..785A}.  Combining $t_{lc}$ and $t_{\gamma}$ one can find a self-consistent solution for $M_{ej}$ and $v_{exp}$ (or the kinetic energy $E_{kin} = 0.3 M_{ej} v_{exp}^2$) depending on the chosen value of $\kappa$,
because the the $\gamma$-ray opacity is well constrained as $\kappa_{\gamma} \sim 0.03$ cm$^2$~g$^{-1}$ \citep{2015MNRAS.450.1295W}. There are additional constraints for the other parameters, as $M_{ej}$ must not exceed the Chandrasekhar mass and $v_{exp}$ must be at least as large as the observed expansion velocities (Section \ref{photo v}). For SN~2018oh, $v_{exp} > 11,000$ km~s$^{-1}$ requires $\kappa \lesssim 0.09$ cm$^2$~g$^{-1}$, while $M_{ej} \lesssim M_{Ch}$ implies $\kappa \gtrsim 0.08$ cm$^2$~g$^{-1}$. Adopting $\kappa \sim 0.085$ cm$^2$~g$^{-1}$ as a fiducial value, we 
get $M_{ej} = 1.27 \pm 0.15$ M$_{\odot}$ and $E_{kin} = 1.08 \pm 0.25 ~\times~ 10^{51}$ erg (the quoted uncertainties reflect the upper and lower value of $\kappa$ given above). These values are close to the typical ejecta masses and kinetic energies for SNe Ia \citep[e.g.][]{2014MNRAS.440.1498S, 2014MNRAS.445.2535S}. 


The uncertainty in the true explosion date has a consequence for the nickel mass estimate. Our first model having $t_0$ fixed to the moment of first light in the {\em K2} light curve gives $M_{Ni} \sim 0.66$ M$_\odot$, which is very similar to the estimate of $M_{Ni}=$0.64$\pm$0.04 M$_\odot$ based on ``Arnett's rule'' \citep{1982ApJ...253..785A, 1985Natur.314..337A,  1992ARA&A..30..359B, 2005A&A...431..423S, 2012ApJ...746..121C} Both of these estimates predict $\sim 0.1$ M$_\odot$ higher nickel mass than our best-fit Arnett model described above, due to the $\sim 3.5$ d longer rise time to maximum light.
Since this model gives a much better description of the evolution of the bolometric light curve, we adopt its final nickel mass of $M_{Ni} = 0.55 \pm 0.04$ M$_\odot$. This is very similar to the estimate of $\sim$0.57 M$_\odot$ for SN 2011fe \citep{2016ApJ...820...67Z} while smaller than the estimates of 0.77 $\pm$ 0.11 M$_\odot$ for SN 2005cf \citep{2009ApJ...697..380W} and 0.68 $\pm$ 0.14 M$_\odot$ for SN 2003du \citep{2007A&A...469..645S}.


All these are based on the assumption that the bolometric light curve of SN~2018oh is entirely powered by the Ni-Co radioactive decay located centrally within the ejecta \citep{1982ApJ...253..785A}. The early, linear rise of the flux observed by {\em Kepler} which could be due to either the interaction with a close companion star \citep{dimitriadis-18oh} or the presence of radioactive $^{56}$Ni in the outer layers of the ejecta and/or interaction with a nearby CSM \citep{shappee-18oh}, suggests that the assumptions of the Arnett model are not entirely fulfilled. For example, in the interaction model, the flux from the early shock may contribute to the full bolometric light curve non-negligibly even around and after maximum light. Subtracting the prediction of the shock-interaction model by \citet{2010ApJ...708.1025K} assuming a Roche-lobe filling companion at $A \sim 2 \times 10^{12}$ cm from the exploding white dwarf \citep{dimitriadis-18oh} and optimal viewing angle would yield $M_{Ni} = 0.54 \pm 0.01$ M$_\odot$,  $t_{lc} = 10.96 \pm 0.17$ d and $t_{\gamma} = 37.89 \pm 0.17 $ d. Thus, while $M_{Ni}$ and $t_{lc}$ are not changed significantly, the post-maximum contribution from the shock may slightly decrease the $\gamma $-ray leaking timescale. Finally, one can get $\kappa \sim 0.10 \pm 0.1$ cm$^2$~g$^{-1}$, $M_{ej} \sim 1.15 \pm 0.23$ M$_\odot$ and $E_{kin} \sim 1.06 \pm 0.4 ~\times~ 10^{51}$ erg using the same model as above. Although these parameters are somewhat less than those estimated from the pure Ni-Co model above, they are consistent within their uncertainties. The contribution of an early shock does not have a significant effect on the parameters estimated from the bolometric light curve. 

We then compare observational properties and fitting parameters of SN 2018oh with two explosion models of SNe Ia. Thermonuclear explosion near the center of the C+O WD triggered by the detonation of He near the surface of the progenitor \citep[the He detonation scenario;][]{2017MNRAS.472.2787N,2017Natur.550...80J,2018ApJ...861...78M} can produce early flux excess. Our explosion parameters are similar to model 10A/N from \cite{2018ApJ...861...78M}. One major effect of spectral evolution by the model 10A/N is the Ti trough at $\sim$4,000\AA~around maximum. However, we do not see such feature in our spectra. Therefore we disfavor this scenario for SN 2018oh. Gravitationally confined detonation \citep[the GCD model;][]{2004ApJ...612L..37P,2007ApJ...662..459K,2008ApJ...681.1448J} is another possible explosion mechanism. In the GCD, a deflagration off-center bubble ignited near the stellar core quickly rise towards the stellar surface with a lateral velocity component which will converge at the opposite side. There, a runaway detonation may be triggered. \cite{2016A&A...592A..57S} use 3D simulations to produce synthetic observables for one model, GCD200, which met their very optimistic detonation criteria. However, they yielded a nickel mass of 0.74 M$_{\odot}$ which is much larger than that of SN 2018oh. The GCD200 model also failed to reproduce the secondary peak in the $I$-band light curve. Nevertheless, the GCD model might explain the bump feature in the $Kepler$ data of SN 2018oh, as it has a strong dependence on viewing-angle caused by asymmetric deflagration ashes. The UV flux is expected to enhance if the SN was observed at a viewing angle near the detonation ignition side. However, this specific viewing angle does not  produce synthetic spectra that are consistent with the observed ones over multiple epochs. 
Thus, we conclude that the current GCD200 model cannot explain the bulk  properties of SN 2018oh.

\section{Conclusion}
We present extensive follow up photometry and spectroscopy for SN 2018oh, the first spectroscopically-confirmed SN Ia (at a distance of 52.7 Mpc) observed by {\em Kepler}. SN 2018oh reached its B-band peak on MJD = 58162.7$\pm$0.3 with an apparent magnitude of $B_{max}$ = 14.31 $\pm$ 0.03 and an absolute magnitude of M$^{B}_{max}=-$19.47$\pm$0.10. SN 2018oh has normal photometric evolution, with a rise time of 18.3$\pm$ 0.3 days and $\Delta$m$_{15}$(B) = 0.96 $\pm$0.03 mag, but it seems to have a relatively bluer $B - V$ color. 

Using three light curve models, we derive a distance to the host galaxy of UGC 4780 as d = 52.7 $\pm$ 1.2 Mpc. UGC 04780 is a star-forming dwarf galaxy with $\log_{10} (M_{stellar}/M_{\odot}) \sim 6.87 \pm 0.12$ and a low metallicity. Based on the extensive UV/optical/NIR photometry, we established the generic bolometric light curve of SN 2018oh. Fitting Arnett's radiation diffusion model powered by radioactive decay of Ni and Co to the bolometric light curve, we derived a peak luminosity of $L_{peak}$ = 1.49 $\times$ 10$^{43}$ erg s$^{-1}$ with a synthesized nickel mass $M_{Ni}$ = 0.55 $\pm$ 0.04 M$_\odot$. The moment when the luminosity begin to emerge in the radiation diffusion model, $t_0$, is found to be +3.85 days after explosion. This is consistent with the hypotheses explored by \citet{dimitriadis-18oh} and \citet{shappee-18oh} that the early flux is either due to interaction between the ejecta and some nearby material (a non-degenerate companion star or a CSM) or a non-central location of the radioactive $^{56}$Ni within the ejecta, and it does not emerge directly from the SN ejecta. In addition, we also explored two SN Ia explosion models, He-detonation and gravitationally confined detonation, while neither of them can fully explain the properties of SN 2018oh.

The overall spectral evolution of SN 2018oh is similar to normal SNe Ia like SN 2003du, but there are still some interesting features which distinguished it from other SNe Ia. For instance, the line-strength ratio of Si~II $\lambda\lambda$5958, 5979 to Si~II $\lambda$6355 ($R$(Si~II)) is found to increase from the early phase to t$=-$4 day and it then decrease towards the maximum light, suggesting a fluctuation of the photospheric temperature, consistent with the line profile change of C~II 6580. SN 2018oh can be put into the Branch shallow-silicon subtype or at the boundary between shallow-silicon and core-normal subtypes based on the pEWs of Si II $\lambda\lambda$5972, 6355, similar to other few SNe Ia showing excess emissions in the early phase in \citet{2018ApJ...864L..35S}. The velocity of Si~II 6355 (i.e., $\sim$10,300$\pm$200 km s$^{-1}$ at t$\sim$ 0 day) suggests that SN 2018oh belongs to the normal subclass but it shows a somewhat larger velocity gradient (near the boundary between LVG and HVG groups) after the maximum light. 

The most striking spectral feature identified for SN 2018oh is the long-lasting C II absorptions. We can identify C II 4267, 6580 and 7234 in early spectra, which all have similar velocity and strength evolution from t$\sim$$-$9 days to t$\sim$+8 days. During this phase, the velocity of C~II 6580 and 4267 decreases from $\sim$ 11,700 km s$^{-1}$ to $\sim$ 8,000 km s$^{-1}$, suggesting a strong mixing of carbon in the exploding ejecta. The C~II 6580 absorption can be even detected in the t=+20.5d spectrum, which is never seen in other SNe Ia. The origin of the persistent carbon in SN 2018oh is unclear but may be related to nature of progenitor systems such as lower metallicity. Detailed modeling is needed to clarify this issue.

\acknowledgments

We acknowledge the support of the staff of the Lijiang 2.4m, Xinglong 2.16m telescope. Funding for the LJT has been provided by Chinese Academy of Sciences and the People's Government of Yunnan Province. The LJT is jointly operated and administrated by Yunnan Observatories and Center for Astronomical Mega-Science, CAS. This work is supported by the National Natural Science Foundation of China (NSFC grants 11325313, 11633002, and 11761141001), the National Program on Key Research and Development Project (grant no. 2016YFA0400803), and Tsinghua University Initiative Scientific Research Program (20161080144). This work was also partially supported by the Collaborating research Program (OP201702) of the Key Laboratory of the Structure and Evolution of Celestial Objects, Chinese Academy of Sciences. This work is sponsored (in part) by the Chinese Academy of Sciences (CAS), through a grant to the CAS South America Center for Astronomy (CASSACA) in Santiago, Chile.  J.-J. Zhang is supported by the NSFC (grants 11403096, 11773067), the Key Research Program of the CAS (Grant NO. KJZD-EW- M06), the Youth Innovation Promotion Association of the CAS (grants 2018081), and the CAS ``Light of West China" Program. F. Huang is supported by the Collaborating Research Program (OP201702) of the Key Laboratory of the Structure and Evolution of Celestial Objects, Chinese Academy of Sciences.

This work makes use of observations from Las Cumbres Observatory. DAH, CM, and GH are supported by the US National Science Foundation grant 1313484. Support for IA was provided by NASA through the Einstein Fellowship Program, grant PF6-170148. The work made use of Swift/UVOT data reduced by P. J. Brown and released in the Swift Optical/Ultraviolet Supernova Archive (SOUSA). SOUSA is supported by NASA's Astrophysics Data Analysis Program through grant NNX13AF35G. 
Research by DJS is supported by NSF grants AST-1821967, 1821987, 1813708 and 1813466. 

This work includes data obtained with the Swope Telescope at Las Campanas Observatory, Chile, as part of the Swope Time Domain Key Project (PI Piro, Co-PIs Shappee, Drout, Madore, Phillips, Foley, and Hsiao). We thank I. Thompson and the Carnegie Observatory Time Allocation Committee for approving the Swope project and scheduling this program. Parts of this research were supported by the Australian Research Council Centre of Excellence for All Sky Astrophysics in 3 Dimensions (ASTRO 3D), through project number CE170100013. EB and JD acknowledge partial support from NASA Grant NNX16AB5G. 

J.V. and his group at Konkoly Observatory is supported by the project ``Transient Astrophysical Objects'' GINOP 2.3.2-15-2016-00033 of the National Research, Development and Innovation Office (NKFIH), Hungary, funded by the European Union, and by the ``Lend\"ulet''Program of the Hungarian Academy of Sciences, project Nos. LP2012-31 and LP2018-7/2018. This project has been supported by the NKFIH grant K-115709. ZsB acknowledges the support provided from the National Research, Development and Innovation Fund of Hungary, financed under the PD\_17 funding scheme, project No. PD123910. Support for J.J.H. was provided by NASA through Hubble Fellowship grant \#HST-HF2-51357.001-A, awarded by the Space Telescope Science Institute, which is operated by the Association of Universities for Research in Astronomy, Incorporated, under NASA contract NAS5-26555, as well as NASA {\em {\em K2}} Cycle 4 Grant NNX17AE92G.

Based on observations obtained at the Southern Astrophysical Research (SOAR) telescope, which is a joint project of the Minist\'{e}rio da Ci\^{e}ncia, Tecnologia, e Inova\c{c}\~{a}o da Rep\'{u}blica Federativa do Brasil, the U.S. National Optical Astronomy Observatory, the University of North Carolina at Chapel Hill, and Michigan State University.

Based in part on observations collected at the European Organisation for Astronomical Research in the Southern Hemisphere under ESO programme 199.D-0143. L.G. was supported in part by the US National Science Foundation under Grant AST-1311862. KM acknowledges support from the STFC through an Ernest Rutherford Fellowship and the EU Horizon 2020 ERC grant no. 758638. C.P.G. acknowledges support from EU/FP7-ERC grant No. [615929]

We thank the Las Cumbres Observatory and its staff for its continuing support of the ASAS-SN project.  ASAS-SN is supported by the Gordon and Betty Moore Foundation through grant GBMF5490 to the Ohio State University and NSF grant AST-1515927. Development of ASAS-SN has been supported by NSF grant AST-0908816, the Mt. Cuba Astronomical Foundation, the Center for Cosmology and AstroParticle Physics at the Ohio State University, the Chinese Academy of Sciences South America Center for Astronomy (CASSACA), the Villum Foundation, and George Skestos.

S.B., P.C. and S.D. acknowledge Project 11573003 supported by NSFC. This research uses data obtained through the Telescope Access Program (TAP), which has been funded by “the Strategic Priority Research Program-The Emergence of Cosmological Structures” of the Chinese Academy of Sciences (Grant No.11 XDB09000000) and the Special Fund for Astronomy from the Ministry of Finance.
The UCSC group is supported in part by NASA grants NNG17PX03C and 80NSSC18K0303, NSF grant AST-1518052, the Gordon \& Betty Moore Foundation, the Heising-Simons Foundation, and by fellowships from the Alfred P.\ Sloan Foundation and the David and Lucile Packard Foundation to R.J.F.

This paper includes data collected by the {\em K2} mission. Funding for the {\em K2} mission is provided by the NASA Science Mission directorate. KEGS is supported in part by NASA {\em K2} cycle 4 and 5 grants NNX17AI64G and 80NSSC18K0302, respectively. AR and his groups is supported in part by HST grants GO-12577 and HST AR-12851.

\software{IRAF \citep{1986SPIE..627..733T,1993ASPC...52..173T}, Minim \citep{2013ApJ...773...76C}, FITSH \citep{2012MNRAS.421.1825P}, SNCOSMO \citep{2016ascl.soft11017B}, PESSTO pipeline \citep{2015A&A...579A..40S}, SALT2.4 \citep{betou14}, SNooPy2 \citep{2011AJ....141...19B}, MLCS2k2 \citep{2007ApJ...659..122J}, photpipe \citep{2005ApJ...634.1103R,2014ApJ...795...44R}, SYNOW \citep{1997ApJ...481L..89F}, STARLIGHT \citep{2005MNRAS.358..363C}, SExtractor \cite{1996A&AS..117..393B}    }
{}

\clearpage


\clearpage

\startlongtable
\begin{deluxetable}{ccccccccc}
\tablecolumns{6} \tablewidth{0pc} \tabletypesize{\scriptsize}
\tablecaption{Photometric Standards in the SN 2018oh Field 1\tablenotemark{a}\label{std} }
\tablehead{\colhead{Num.} &\colhead{$\alpha$(J2000)} &
\colhead{$\delta$(J2000)} & \colhead{\textit{U} (mag)}&\colhead{\textit{B} (mag)} & \colhead{\textit{V} (mag)}  & \colhead{\textit{g} (mag)} & \colhead{\textit{r} (mag)} & \colhead{\textit{i} (mag)}} \startdata			
1	&	$09^h05^m59^s.52$	&	+19$^{\circ}$15'08''.52	&	17.207(045)	&	16.391(156)	&	15.528(022)	&	15.839(004)	&	15.111(003)	&	14.839(001)	\\
2	&	$09^h05^m59^s.95$	&	+19$^{\circ}$20'47''.95	&	17.495(234)	&	17.294(136)	&	16.594(018)	&	16.787(004)	&	16.278(003)	&	16.091(004)	\\
3	&	$09^h06^m02^s.44$	&	+19$^{\circ}$25'11''.44	&	14.466(128)	&	14.265(012)	&	13.785(036)	&	\nd	&	\nd	&	\nd	\\
4	&	$09^h06^m02^s.74$	&	+19$^{\circ}$22'59''.74	&	16.876(225)	&	16.411(136)	&	15.682(063)	&	15.889(002)	&	15.377(002)	&	15.189(003)	\\
5	&	$09^h06^m05^s.30$	&	+19$^{\circ}$15'21''.30	&	18.374(121)	&	17.322(161)	&	16.094(030)	&	16.637(005)	&	15.538(001)	&	14.899(003)	\\
6	&	$09^h06^m08^s.24$	&	+19$^{\circ}$23'45''.24	&	14.683(130)	&	14.562(093)	&	14.089(051)	&	14.177(003)	&	13.877(001)	&	13.754(007)	\\
7	&	$09^h06^m09^s.43$	&	+19$^{\circ}$19'47''.43	&	16.944(205)	&	16.190(108)	&	15.371(027)	&	15.612(002)	&	14.998(003)	&	14.777(002)	\\
8	&	$09^h06^m09^s.72$	&	+19$^{\circ}$26'37''.72	&	14.249(157)	&	14.085(119)	&	13.530(051)	&	\nd	&	\nd	&	\nd	\\
9	&	$09^h06^m11^s.47$	&	+19$^{\circ}$23'57''.47	&	14.049(117)	&	13.876(117)	&	13.336(045)	&	\nd	&	\nd	&	\nd	\\
10	&	$09^h06^m18^s.41$	&	+19$^{\circ}$21'59''.41	&	15.101(122)	&	14.207(142)	&	13.265(037)	&	\nd	&	\nd	&	\nd	\\
11	&	$09^h06^m19^s.78$	&	+19$^{\circ}$21'11''.78	&	15.624(116)	&	15.034(140)	&	14.300(037)	&	14.504(001)	&	13.989(004)	&	13.794(002)	\\
12	&	$09^h06^m22^s.84$	&	+19$^{\circ}$11'53''.84	&	\nd	&	\nd	&	\nd	&	17.049(003)	&	15.857(003)	&	15.206(003)	\\
13	&	$09^h06^m23^s.26$	&	+19$^{\circ}$27'45''.26	&	\nd	&	\nd	&	11.120(005)	&	\nd	&	\nd	&	\nd	\\
14	&	$09^h06^m25^s.39$	&	+19$^{\circ}$26'07''.39	&	\nd	&	\nd	&	\nd	&	13.991(002)	&	13.547(003)	&	13.401(006)	\\
15	&	$09^h06^m28^s.27$	&	+19$^{\circ}$13'37''.27	&	16.763(165)	&	16.324(098)	&	15.654(029)	&	15.818(004)	&	15.371(001)	&	15.213(001)	\\
16	&	$09^h06^m30^s.03$	&	+19$^{\circ}$19'50''.03	&	\nd	&	\nd	&	\nd	&	15.225(002)	&	14.559(003)	&	14.305(001)	\\
17	&	$09^h06^m32^s.41$	&	+19$^{\circ}$24'27''.41	&	16.777(175)	&	16.659(117)	&	16.120(051)	&	16.227(003)	&	15.886(003)	&	15.760(005)	\\
18	&	$09^h06^m32^s.94$	&	+19$^{\circ}$17'54''.94	&	\nd	&	\nd	&	12.832(012)	&	\nd	&	\nd	&	\nd	\\
19	&	$09^h06^m34^s.32$	&	+19$^{\circ}$28'33''.32	&	17.113(179)	&	16.248(103)	&	15.445(076)	&	15.663(001)	&	15.028(003)	&	14.773(003)	\\
20	&	$09^h06^m34^s.39$	&	+19$^{\circ}$21'52''.39	&	15.806(063)	&	15.085(130)	&	14.215(028)	&	14.485(003)	&	13.851(004)	&	13.586(003)	\\
21	&	$09^h06^m34^s.74$	&	+19$^{\circ}$17'03''.74	&	16.377(162)	&	15.671(110)	&	14.867(012)	&	15.112(002)	&	14.537(003)	&	14.315(003)	\\
22	&	$09^h06^m36^s.26$	&	+19$^{\circ}$29'46''.26	&	14.544(113)	&	14.470(059)	&	13.949(071)	&	14.067(002)	&	13.711(005)	&	13.562(001)	\\
23	&	$09^h06^m43^s.46$	&	+19$^{\circ}$20'27''.46	&	15.631(135)	&	15.244(108)	&	14.606(016)	&	14.765(005)	&	14.329(001)	&	14.183(001)	\\
24	&	$09^h06^m47^s.84$	&	+19$^{\circ}$25'33''.84	&	15.742(079)	&	15.395(123)	&	14.760(056)	&	14.929(002)	&	14.502(002)	&	14.373(001)	\\
25	&	$09^h06^m47^s.92$	&	+19$^{\circ}$17'04''.92	&	15.781(157)	&	15.318(070)	&	14.615(037)	&	14.813(001)	&	14.324(004)	&	14.141(002)	\\
26	&	$09^h06^m48^s.17$	&	+19$^{\circ}$13'56''.17	&	14.501(06)	&	14.504(082)	&	13.928(016)	&	14.065(002)	&	13.715(005)	&	13.563(002)	\\
27	&	$09^h06^m52^s.18$	&	+19$^{\circ}$11'57''.18	&	\nd	&	\nd	&	16.723(122)	&	16.773(002)	&	16.411(003)	&	16.269(003)	\\
28	&	$09^h06^m54^s.07$	&	+19$^{\circ}$25'28''.07	&	14.718(118)	&	14.603(097)	&	14.087(052)	&	14.204(002)	&	13.886(002)	&	13.766(008)	\\
29	&	$09^h06^m57^s.19$	&	+19$^{\circ}$18'13''.19	&	15.555(151)	&	15.226(058)	&	14.562(022)	&	14.766(003)	&	14.322(003)	&	14.139(004)	\\
30	&	$09^h06^m58^s.25$	&	+19$^{\circ}$13'56''.25	&	16.208(156)	&	16.047(051)	&	15.389(018)	&	15.557(001)	&	15.158(001)	&	15.009(004)	\\
31	&	$09^h07^m02^s.35$	&	+19$^{\circ}$17'23''.35	&	14.296(083)	&	14.237(081)	&	13.623(007)	&	\nd	&	\nd	&	\nd	\\
32	&	$09^h07^m02^s.62$	&	+19$^{\circ}$13'50''.62	&	14.784(190)	&	14.403(054)	&	13.763(021)	&	13.936(001)	&	13.561(002)	&	13.432(007)	\\
33	&	$09^h07^m03^s.14$	&	+19$^{\circ}$15'58''.14	&	16.385(141)	&	16.251(074)	&	15.469(014)	&	15.713(001)	&	15.173(002)	&	14.914(003)	\\
34	&	$09^h07^m03^s.82$	&	+19$^{\circ}$17'49''.82	&	15.334(171)	&	14.481(074)	&	13.531(009)	&	\nd	&	\nd	&	\nd	\\
35	&	$09^h07^m04^s.07$	&	+19$^{\circ}$26'20''.07	&	\nd	&	\nd	&	12.647(039)	&	\nd	&	\nd	&	\nd	\\
36	&	$09^h07^m16^s.62$	&	+19$^{\circ}$21'05''.62	&	14.992(097)	&	14.878(071)	&	14.288(034)	&	14.441(003)	&	14.103(006)	&	13.984(008)	\\
37	&	$09^h07^m20^s.56$	&	+19$^{\circ}$21'50''.56	&	\nd	&	\nd	&	\nd	&	15.357(002)	&	14.823(003)	&	14.621(005)	\\
38	&	$09^h07^m20^s.99$	&	+19$^{\circ}$23'49''.99	&	16.216(221)	&	16.153(062)	&	15.558(018)	&	15.742(003)	&	15.394(003)	&	15.264(003)	\\
39	&	$09^h07^m21^s.73$	&	+19$^{\circ}$15'09''.73	&	15.603(164)	&	15.536(026)	&	14.904(038)	&	15.106(001)	&	14.756(002)	&	14.639(002)	\\
\enddata
\tablenotetext{a}{See Figure 1 for a finder chart of SN 2018oh and part of the comparison stars.}
\tablenotetext{}{Note: Uncertainties, in units of 0.001 mag, are $1\sigma$.}
\end{deluxetable}

\startlongtable
\begin{deluxetable}{ccccccc}
\tablecolumns{6} \tablewidth{0pc} \tabletypesize{\scriptsize}
\tablecaption{Photometric Standards in the SN 2018oh Field 2\tablenotemark{a}\label{std1} }
\tablehead{\colhead{Num.} &\colhead{$\alpha$(J2000)} &
\colhead{$\delta$(J2000)} & \colhead{\textit{B} (mag)} & \colhead{\textit{V} (mag)} & \colhead{\textit{R} (mag)}  & \colhead{\textit{I} (mag)} } \startdata	
1	&	$09^h06^m53^s.43$	&	+19$^{\circ}$18'22''.43	&	16.553(032)	&	15.932(012)	&	15.567(015)	&	15.189(017)	\\
2	&	$09^h06^m36^s.12$	&	+19$^{\circ}$20'24''.12	&	19.686(033)	&	18.143(014)	&	17.237(015)	&	16.027(018)	\\
3	&	$09^h06^m54^s.98$	&	+19$^{\circ}$21'32''.98	&	18.460(032)	&	17.350(012)	&	16.706(015)	&	16.151(018)	\\
4	&	$09^h06^m58^s.91$	&	+19$^{\circ}$20'26''.91	&	18.413(032)	&	17.563(013)	&	17.070(016)	&	16.572(017)	\\
5	&	$09^h06^m30^s.32$	&	+19$^{\circ}$19'41''.32	&	17.908(032)	&	17.173(012)	&	16.746(015)	&	16.309(017)	\\
6	&	$09^h06^m55^s.78$	&	+19$^{\circ}$15'40''.78	&	17.785(032)	&	17.189(013)	&	16.837(015)	&	16.464(017)	\\
7	&	$09^h06^m55^s.71$	&	+19$^{\circ}$14'56''.71	&	17.990(032)	&	17.170(012)	&	16.693(015)	&	16.204(017)	\\
8	&	$09^h06^m57^s.27$	&	+19$^{\circ}$23'16''.27	&	17.654(032)	&	16.772(012)	&	16.261(015)	&	15.777(017)	\\
9	&	$09^h07^m07^s.05$	&	+19$^{\circ}$18'52''.05	&	19.857(033)	&	18.292(014)	&	17.372(015)	&	16.333(018)	\\
10	&	$09^h06^m29^s.08$	&	+19$^{\circ}$22'45''.08	&	19.367(033)	&	18.039(013)	&	17.265(016)	&	16.581(018)	\\
11	&	$09^h06^m36^s.11$	&	+19$^{\circ}$14'10''.11	&	19.765(032)	&	18.286(013)	&	17.418(015)	&	16.608(018)	\\
12	&	$09^h07^m09^s.67$	&	+19$^{\circ}$20'53''.67	&	17.617(032)	&	16.848(012)	&	16.400(015)	&	15.973(017)	\\
13	&	$09^h06^m50^s.90$	&	+19$^{\circ}$13'22''.90	&	18.832(033)	&	17.281(013)	&	16.370(015)	&	15.384(017)	\\
14	&	$09^h07^m07^s.41$	&	+19$^{\circ}$22'17''.41	&	17.490(032)	&	16.434(012)	&	15.822(015)	&	15.235(017)	\\
\enddata
\tablenotetext{a}{Standards for Konkoly and Super-lotis observations.}
\tablenotetext{}{Note: Uncertainties, in units of 0.001 mag, are $1\sigma$.}
\end{deluxetable}

\startlongtable
\begin{deluxetable}{ccccccccccc}
\tablecolumns{6} \tablewidth{0pc} \tabletypesize{\scriptsize}
\tablecaption{Ground-based Optical Photometry of SN 2018oh  \label{gphoto} }
\tablehead{\colhead{Date} &\colhead{\tablenotemark{a}Epoch} &
\colhead{\textit{U} (mag)} & \colhead{\textit{B} (mag)} & \colhead{\textit{V} (mag)} & \colhead{\textit{R} (mag)}& \colhead{\textit{I} (mag)} & \colhead{\textit{g} (mag)} & \colhead{\textit{r} (mag)} & \colhead{\textit{i} (mag)} & \colhead{Telescope} } \startdata
2018-01-26.6 & -18.1 & \nd & \nd & \nd & \nd & \nd & 20.852(223) & \nd & 21.025(269) & PS1\\
2018-01-27.2 & -17.5 & \nd & \nd & \nd & \nd & \nd & \nd & \nd & 19.039(009) & Decam\\
2018-01-27.3 & -17.4 & \nd & \nd & \nd & \nd & \nd & \nd & \nd & 18.957(008) & Decam\\
2018-02-03.1 & -10.6 & \nd & \nd & \nd & \nd & \nd & 15.500(010) & \nd & \nd & ASAS-SN\\
2018-02-03.3 & -10.4 & \nd & \nd & \nd & \nd & \nd & \nd & \nd & 15.671(004) & PS1\\
2018-02-04.3 & -9.4 & \nd & \nd & \nd & \nd & \nd & \nd & \nd & 15.446(004) & PS1\\
2018-02-04.5 & -9.2 & \nd & \nd & \nd & \nd & \nd & \nd & \nd & 15.389(003) & PS1\\
2018-02-05.1 & -8.6 & \nd & 14.982(025) & 15.085(025) & \nd & \nd & \nd & 14.988(017) & 15.305(025) & DEMONEXT\\
2018-02-05.2 & -8.5 & \nd & 14.940(026) & 15.067(025) & \nd & \nd & \nd & 15.005(013) & 15.262(024) & PONM\\
2018-02-05.2 & -8.5 & \nd & \nd & 14.915(005) & 14.832(006) & 14.879(007) & \nd & \nd & \nd & slotis\\
2018-02-05.4 & -8.3 & \nd & \nd & \nd & \nd & \nd & \nd & \nd & 15.264(010) & PS1\\
2018-02-05.8 & -7.9 & \nd & 14.818(031) & 14.931(011) & \nd & \nd & 14.707(005) & 14.872(007) & 15.089(009) & TNT\\
2018-02-05.9 & -7.8 & \nd & \nd & \nd & \nd & \nd & \nd & 14.883(006) & 15.142(008) & LCO\\
2018-02-06.2 & -7.5 & \nd & 14.789(031) & 14.879(013) & \nd & \nd & \nd & 14.803(007) & 15.101(006) & PONM\\
2018-02-06.2 & -7.5 & 15.615(013) & 14.762(012) & 14.722(010) & \nd & \nd & 14.711(010) & 14.835(008) & 15.149(009) & Swope\\
2018-02-06.2 & -7.5 & \nd & \nd & 14.771(005) & 14.709(006) & 14.774(006) & \nd & \nd & \nd & slotis\\
2018-02-06.5 & -7.2 & \nd & 14.758(027) & 14.857(024) & \nd & \nd & \nd & 14.789(016) & 15.085(022) & DEMONEXT\\
2018-02-06.6 & -7.1 & 14.285(027) & 14.723(016) & 14.799(014) & \nd & \nd & 14.610(003) & 14.798(003) & 15.089(006) & LCO\\
2018-02-06.8 & -6.9 & 14.213(062) & 14.691(043) & 14.818(015) & 14.660(041) & 14.662(029) & \nd & \nd & \nd & LJT\\
2018-02-06.8 & -6.9 & \nd & 14.703(072) & 14.724(027) & 14.609(032) & 14.637(035) & \nd & \nd & \nd & Konkoly\\
2018-02-07.2 & -6.5 & \nd & 14.666(022) & 14.783(016) & \nd & \nd & \nd & 14.688(013) & 14.972(019) & DEMONEXT\\
2018-02-07.2 & -6.5 & \nd & 14.681(034) & 14.797(014) & \nd & \nd & \nd & 14.725(004) & 15.010(007) & PONM\\
2018-02-07.2 & -6.5 & 15.345(035) & 14.600(012) & 14.584(011) & \nd & \nd & 14.531(010) & 14.635(008) & 14.985(009) & Swope\\
2018-02-07.2 & -6.5 & \nd & \nd & 14.668(006) & 14.569(006) & 14.650(006) & \nd & \nd & \nd & slotis\\
2018-02-07.5 & -6.2 & 14.173(026) & 14.618(015) & 14.697(014) & \nd & \nd & 14.494(002) & 14.702(003) & 14.987(006) & LCO\\
2018-02-07.7 & -6.0 & \nd & 14.578(034) & 14.701(020) & \nd & \nd & 14.499(016) & 14.713(018) & 15.003(013) & TNT\\
2018-02-08.2 & -5.5 & 15.291(032) & 14.469(011) & 14.471(010) & \nd & \nd & 14.408(010) & 14.557(008) & 14.926(008) & Swope\\
2018-02-08.3 & -5.4 & \nd & \nd & \nd & \nd & \nd & \nd & 14.609(009) & \nd & Gemini\\
2018-02-08.4 & -5.3 & \nd & \nd & \nd & \nd & \nd & 14.483(002) & 14.613(002) & \nd & PS1\\
2018-02-08.5 & -5.2 & 14.082(027) & 14.529(015) & 14.608(014) & \nd & \nd & 14.409(002) & 14.614(003) & 14.947(006) & LCO\\
2018-02-08.7 & -5.0 & \nd & 14.514(030) & 14.622(010) & \nd & \nd & 14.421(003) & 14.646(004) & 14.963(006) & TNT\\
2018-02-09.1 & -4.6 & \nd & 14.521(025) & 14.584(022) & \nd & \nd & \nd & 14.550(014) & 14.940(026) & DEMONEXT\\
2018-02-09.2 & -4.5 & 14.030(027) & 14.509(015) & 14.549(013) & \nd & \nd & 14.382(001) & 14.551(002) & 14.904(003) & LCO\\
2018-02-09.2 & -4.5 & \nd & 14.487(033) & 14.591(012) & \nd & \nd & \nd & 14.540(004) & 14.887(005) & PONM\\
2018-02-09.2 & -4.5 & \nd & \nd & 14.450(004) & 14.397(005) & 14.555(006) & \nd & \nd & \nd & slotis\\
2018-02-09.3 & -4.4 & \nd & \nd & \nd & \nd & \nd & 14.342(018) & 14.556(019) & \nd & Gemini\\
2018-02-09.5 & -4.2 & \nd & \nd & \nd & \nd & \nd & 14.359(002) & \nd & 14.926(003) & PS1\\
2018-02-09.7 & -4.0 & \nd & 14.411(030) & 14.505(010) & \nd & \nd & 14.327(003) & 14.543(004) & \nd & TNT\\
2018-02-10.3 & -3.4 & \nd & \nd & \nd & \nd & \nd & 14.266(008) & 14.471(012) & \nd & Gemini\\
2018-02-10.5 & -3.2 & \nd & \nd & \nd & \nd & \nd & 14.307(002) & 14.476(002) & \nd & PS1\\
2018-02-10.5 & -3.2 & \nd & 14.426(032) & 14.520(032) & \nd & \nd & \nd & 14.511(027) & \nd & DEMONEXT\\
2018-02-10.7 & -3.0 & \nd & 14.375(031) & 14.483(010) & \nd & \nd & 14.287(005) & 14.528(008) & 14.949(010) & TNT\\
2018-02-10.9 & -2.8 & 13.950(026) & 14.323(015) & 14.443(013) & \nd & \nd & \nd & 14.493(010) & \nd & LCO\\
2018-02-10.9 & -2.8 & \nd & \nd & \nd & \nd & \nd & \nd & 14.499(005) & 14.936(007) & LCO\\
2018-02-11.2 & -2.5 & \nd & 14.383(022) & 14.476(022) & \nd & \nd & \nd & 14.449(013) & 14.905(024) & DEMONEXT\\
2018-02-11.2 & -2.5 & \nd & 14.352(028) & 14.438(010) & \nd & \nd & \nd & 14.441(005) & 14.884(007) & PONM\\
2018-02-11.3 & -2.3 & \nd & \nd & \nd & \nd & \nd & 14.273(002) & \nd & 14.961(003) & PS1\\
2018-02-11.3 & -2.4 & \nd & \nd & \nd & \nd & \nd & 14.235(014) & 14.446(013) & \nd & Gemini\\
2018-02-11.7 & -2.0 & \nd & 14.331(031) & 14.435(010) & \nd & \nd & 14.241(004) & 14.498(005) & 14.957(008) & TNT\\
2018-02-11.9 & -1.8 & \nd & 14.305(050) & 14.406(045) & 14.236(028) & 14.453(030) & \nd & \nd & \nd & Konkoly\\
2018-02-12.1 & -1.6 & 13.967(028) & 14.389(022) & 14.391(014) & \nd & \nd & 14.231(001) & 14.427(002) & 14.920(004) & LCO\\
2018-02-12.2 & -1.5 & 15.215(010) & 14.337(011) & 14.281(010) & \nd & \nd & 14.572(018) & 14.398(008) & 14.968(009) & Swope\\
2018-02-12.3 & -1.4 & \nd & \nd & \nd & \nd & \nd & 14.242(002) & 14.402(002) & \nd & PS1\\
2018-02-12.3 & -1.4 & \nd & \nd & \nd & \nd & \nd & 14.203(013) & 14.424(024) & \nd & Gemini\\
2018-02-12.4 & -1.3 & \nd & 14.333(022) & 14.430(023) & \nd & \nd & \nd & 14.388(013) & 14.897(020) & DEMONEXT\\
2018-02-12.8 & -0.9 & \nd & 14.300(030) & 14.393(012) & \nd & \nd & 14.220(003) & 14.465(004) & 14.982(004) & TNT\\
2018-02-13.2 & -0.5 & 13.973(028) & 14.353(015) & 14.338(014) & \nd & \nd & 14.210(002) & 14.378(002) & \nd & LCO\\
2018-02-13.2 & -0.5 & \nd & 14.320(025) & 14.388(018) & \nd & \nd & \nd & 14.383(016) & 14.967(019) & DEMONEXT\\
2018-02-13.2 & -0.4 & 15.243(025) & 14.287(011) & 14.255(010) & \nd & \nd & \nd & \nd & \nd & Swope\\
2018-02-13.3 & -0.4 & \nd & \nd & \nd & \nd & \nd & 14.196(012) & 14.398(009) & \nd & Gemini\\
2018-02-13.5 & -0.2 & \nd & 14.309(025) & 14.385(012) & \nd & \nd & \nd & 14.400(008) & 14.944(010) & PONM\\
2018-02-13.6 & -0.1 & \nd & \nd & \nd & \nd & \nd & 14.252(002) & \nd & 15.002(003) & PS1\\
2018-02-13.7 & +0.0 & \nd & 14.293(030) & 14.376(009) & \nd & \nd & 14.201(003) & 14.450(003) & 15.008(005) & TNT\\
2018-02-14.3 & +0.6 & 15.296(061) & 14.345(011) & 14.258(010) & \nd & \nd & 14.215(012) & 14.390(008) & 15.059(010) & Swope\\
2018-02-14.5 & +0.8 & \nd & \nd & \nd & \nd & \nd & 14.225(002) & \nd & \nd & PS1\\
2018-02-15.0 & +2.3 & 14.081(027) & 14.319(015) & 14.383(014) & \nd & \nd & 14.184(002) & 14.422(003) & 15.085(007) & LCO\\
2018-02-15.2 & +1.5 & 15.317(010) & 14.354(011) & 14.275(009) & \nd & \nd & 14.256(008) & 14.383(008) & \nd & Swope\\
2018-02-16.2 & +2.5 & 15.410(023) & 14.382(011) & 14.290(009) & \nd & \nd & 14.259(009) & 14.392(008) & 15.136(009) & Swope\\
2018-02-16.3 & +2.6 & \nd & 14.367(025) & 14.413(010) & \nd & \nd & \nd & 14.400(007) & 15.066(010) & PONM\\
2018-02-16.9 & +3.2 & 14.157(027) & 14.364(016) & 14.398(013) & \nd & \nd & 14.200(002) & 14.425(004) & 15.106(009) & LCO\\
2018-02-16.9 & +3.2 & \nd & \nd & \nd & \nd & \nd & \nd & 14.457(009) & 15.152(009) & LCO\\
2018-02-16.9 & +3.2 & \nd & \nd & 14.400(104) & 14.279(030) & 14.608(054) & \nd & \nd & \nd & Konkoly\\
2018-02-17.2 & +3.5 & 14.230(031) & 14.451(016) & 14.411(013) & \nd & \nd & 14.293(002) & 14.437(002) & 15.127(006) & LCO\\
2018-02-17.2 & +3.5 & 15.455(013) & 14.380(011) & 14.279(009) & \nd & \nd & 14.272(009) & 14.367(008) & 15.149(010) & Swope\\
2018-02-17.7 & +4.0 & \nd & 14.400(042) & 14.439(013) & \nd & \nd & 14.293(009) & 14.429(013) & 15.129(027) & LJT\\
2018-02-18.3 & +4.6 & \nd & 14.448(029) & 14.476(013) & \nd & \nd & \nd & 14.453(007) & 15.146(016) & PONM\\
2018-02-18.5 & +4.8 & \nd & \nd & \nd & \nd & \nd & \nd & 14.450(002) & \nd & PS1\\
2018-02-18.8 & +5.2 & \nd & 14.439(052) & 14.424(025) & 14.284(029) & 14.708(030) & \nd & \nd & \nd & Konkoly\\
2018-02-18.9 & +5.2 & \nd & \nd & \nd & \nd & \nd & \nd & 14.487(004) & 15.193(009) & LCO\\
2018-02-19.1 & +5.4 & 14.326(027) & 14.465(015) & 14.455(013) & \nd & \nd & 14.280(002) & 14.489(003) & 15.205(008) & LCO\\
2018-02-19.6 & +5.9 & \nd & 14.514(035) & 14.507(011) & \nd & \nd & 14.382(005) & 14.506(004) & 15.243(023) & LJT\\
2018-02-20.2 & +6.5 & \nd & 14.552(029) & 14.540(014) & \nd & \nd & \nd & 14.549(015) & 15.253(022) & PONM\\
2018-02-20.2 & +6.5 & 15.720(023) & 14.516(011) & 14.382(009) & \nd & \nd & 14.423(008) & 14.495(007) & 15.320(008) & Swope\\
2018-02-20.3 & +6.6 & \nd & \nd & \nd & \nd & \nd & 14.418(002) & 14.532(002) & \nd & PS1\\
2018-02-21.0 & +8.3 & \nd & \nd & \nd & \nd & \nd & \nd & 14.619(004) & \nd & T50\\
2018-02-21.2 & +7.5 & 14.524(029) & 14.646(016) & 14.550(014) & \nd & \nd & 14.450(002) & 14.592(002) & 15.345(005) & LCO\\
2018-02-21.2 & +7.5 & 15.862(126) & 14.592(010) & 14.432(008) & \nd & \nd & 14.487(008) & 14.585(007) & \nd & Swope\\
2018-02-21.3 & +7.6 & \nd & \nd & \nd & \nd & \nd & 14.414(012) & 14.594(026) & \nd & Gemini\\
2018-02-21.5 & +7.8 & \nd & \nd & \nd & \nd & \nd & 14.472(002) & \nd & \nd & PS1\\
2018-02-21.8 & +8.1 & \nd & 14.637(049) & 14.597(015) & \nd & \nd & 14.490(009) & 14.655(014) & 15.420(028) & LJT\\
2018-02-22.2 & +8.5 & 14.641(027) & 14.721(015) & 14.599(014) & \nd & \nd & 14.503(002) & 14.668(002) & 15.393(007) & LCO\\
2018-02-22.2 & +8.5 & 15.959(017) & 14.627(011) & 14.465(009) & \nd & \nd & 14.522(008) & 14.619(007) & 15.453(009) & Swope\\
2018-02-22.3 & +8.6 & \nd & \nd & \nd & \nd & \nd & 14.467(028) & \nd & \nd & Gemini\\
2018-02-22.7 & +9.0 & \nd & 14.724(032) & 14.635(016) & \nd & \nd & 14.524(006) & 14.759(008) & 15.483(012) & TNT\\
2018-02-23.2 & +9.5 & \nd & 14.765(032) & 14.689(012) & \nd & \nd & \nd & 14.727(004) & 15.465(009) & PONM\\
2018-02-23.2 & +9.5 & 16.051(017) & 14.700(011) & 14.507(009) & \nd & \nd & 14.544(008) & 14.678(008) & 15.532(009) & Swope\\
2018-02-23.3 & +9.6 & \nd & \nd & \nd & \nd & \nd & 14.516(011) & 14.727(033) & \nd & Gemini\\
2018-02-23.4 & +9.7 & \nd & 14.803(025) & 14.660(026) & \nd & \nd & \nd & 14.735(019) & 15.535(032) & DEMONEXT\\
2018-02-23.7 & +10.0 & \nd & 14.774(031) & 14.648(010) & \nd & \nd & 14.556(003) & 14.785(005) & \nd & TNT\\
2018-02-23.9 & +10.2 & \nd & \nd & \nd & \nd & \nd & \nd & 14.829(009) & \nd & T50\\
2018-02-24.2 & +10.5 & 14.793(027) & 14.821(016) & 14.726(015) & \nd & \nd & 14.588(003) & 14.820(007) & 15.600(014) & LCO\\
2018-02-24.3 & +10.6 & \nd & 14.854(024) & 14.745(019) & \nd & \nd & \nd & 14.773(016) & 15.583(029) & DEMONEXT\\
2018-02-24.6 & +10.9 & \nd & 14.887(032) & 14.758(012) & \nd & \nd & 14.646(004) & 14.899(006) & 15.643(010) & TNT\\
2018-02-24.9 & +11.2 & \nd & \nd & 14.649(126) & 14.695(081) & 15.102(039) & \nd & \nd & \nd & Konkoly\\
2018-02-25.2 & +11.5 & 16.307(027) & 14.866(012) & 14.618(010) & \nd & \nd & 14.652(009) & 14.794(008) & 15.686(009) & Swope\\
2018-02-25.4 & +11.7 & 14.885(032) & 14.903(018) & 14.769(017) & \nd & \nd & 14.683(004) & 14.947(008) & \nd & LCO\\
2018-02-25.8 & +12.1 & \nd & \nd & \nd & \nd & \nd & \nd & 14.944(010) & 15.684(017) & LCO\\
2018-02-25.8 & +12.1 & \nd & \nd & 14.809(076) & 14.767(076) & 15.171(035) & \nd & \nd & \nd & Konkoly\\
2018-02-26.1 & +12.4 & \nd & 15.034(024) & 14.841(026) & \nd & \nd & \nd & 14.956(023) & 15.765(028) & DEMONEXT\\
2018-02-26.2 & +12.5 & \nd & 15.028(031) & 14.856(013) & \nd & \nd & \nd & 14.969(006) & 15.757(011) & PONM\\
2018-02-26.2 & +12.5 & 16.489(036) & 14.973(013) & 14.711(011) & \nd & \nd & 14.791(010) & 14.940(009) & 15.847(011) & Swope\\
2018-02-26.6 & +12.9 & \nd & 15.070(032) & 14.860(011) & \nd & \nd & 14.780(006) & 15.055(007) & 15.827(012) & TNT\\
2018-02-27.2 & +13.5 & \nd & 15.088(027) & 14.872(026) & \nd & \nd & \nd & 14.998(018) & 15.742(030) & DEMONEXT\\
2018-02-27.2 & +13.5 & 16.564(035) & 15.061(016) & 14.790(013) & \nd & \nd & 14.879(012) & 15.019(012) & 15.913(016) & Swope\\
2018-02-27.3 & +13.6 & \nd & \nd & \nd & \nd & \nd & \nd & 14.992(013) & \nd & Gemini\\
2018-02-27.7 & +14.0 & \nd & 15.159(045) & 14.963(022) & \nd & \nd & 14.964(014) & \nd & 15.932(056) & LJT\\
2018-03-01.6 & +15.9 & \nd & 15.323(038) & 15.058(016) & \nd & \nd & 15.025(012) & 15.235(012) & 15.941(020) & TNT\\
2018-03-01.7 & +16.0 & \nd & 15.312(043) & 14.977(015) & \nd & \nd & 15.062(013) & 15.164(010) & \nd & LJT\\
2018-03-02.1 & +16.4 & 15.489(054) & 15.451(028) & 15.035(021) & \nd & \nd & 15.067(013) & 15.201(015) & 15.902(029) & LCO\\
2018-03-02.2 & +16.5 & \nd & 15.425(034) & 15.062(034) & \nd & \nd & \nd & 15.157(028) & 15.859(052) & DEMONEXT\\
2018-03-02.3 & +16.6 & \nd & \nd & \nd & \nd & \nd & 15.033(011) & 15.189(017) & \nd & Gemini\\
2018-03-02.5 & +16.8 & \nd & 15.590(089) & \nd & \nd & \nd & \nd & \nd & \nd & TNT\\
2018-03-02.5 & +16.8 & \nd & \nd & \nd & \nd & \nd & \nd & 15.210(012) & 15.897(016) & LCO\\
2018-03-03.2 & +17.5 & \nd & 15.499(032) & 15.113(031) & \nd & \nd & \nd & 15.169(021) & 15.874(031) & DEMONEXT\\
2018-03-03.2 & +17.5 & \nd & \nd & \nd & \nd & \nd & 15.126(009) & 15.202(018) & \nd & Gemini\\
2018-03-03.9 & +18.2 & \nd & \nd & \nd & \nd & \nd & \nd & 15.213(008) & 15.850(013) & LCO\\
2018-03-04.2 & +18.5 & \nd & \nd & \nd & \nd & \nd & 15.195(009) & 15.221(012) & \nd & Gemini\\
2018-03-04.7 & +19.0 & \nd & 15.706(036) & 15.161(013) & \nd & \nd & 15.228(008) & 15.220(007) & 15.827(013) & TNT\\
2018-03-04.8 & +19.1 & \nd & 15.708(068) & 15.127(068) & 15.025(025) & 15.180(031) & \nd & \nd & \nd & Konkoly\\
2018-03-05.2 & +19.5 & \nd & 15.739(026) & 15.137(043) & \nd & \nd & \nd & 15.185(023) & 15.764(039) & DEMONEXT\\
2018-03-05.2 & +19.5 & \nd & 15.782(033) & 15.216(023) & \nd & \nd & \nd & 15.244(010) & 15.850(020) & PONM\\
2018-03-05.2 & +19.5 & \nd & \nd & \nd & \nd & \nd & 15.265(012) & 15.232(020) & \nd & Gemini\\
2018-03-05.5 & +19.8 & \nd & 15.767(032) & 15.196(011) & \nd & \nd & 15.286(005) & 15.219(007) & 15.768(011) & TNT\\
2018-03-05.9 & +20.2 & \nd & \nd & \nd & \nd & \nd & \nd & 15.244(007) & 15.817(012) & LCO\\
2018-03-06.2 & +20.5 & \nd & \nd & \nd & \nd & \nd & 15.356(020) & 15.249(022) & \nd & Gemini\\
2018-03-06.2 & +20.5 & \nd & \nd & 15.156(008) & 15.108(009) & 15.226(013) & \nd & \nd & \nd & slotis\\
2018-03-06.4 & +20.7 & \nd & 15.859(040) & 15.180(039) & \nd & \nd & \nd & 15.150(017) & 15.726(031) & DEMONEXT\\
2018-03-07.2 & +21.5 & \nd & 15.945(032) & 15.233(041) & \nd & \nd & \nd & 15.200(015) & 15.736(031) & DEMONEXT\\
2018-03-07.2 & +21.5 & \nd & 15.970(029) & 15.287(033) & \nd & \nd & \nd & 15.255(012) & 15.749(017) & PONM\\
2018-03-07.2 & +21.5 & \nd & \nd & \nd & \nd & \nd & 15.437(012) & 15.253(019) & \nd & Gemini\\
2018-03-07.2 & +21.5 & \nd & \nd & 15.200(006) & 15.122(007) & 15.253(008) & \nd & \nd & \nd & slotis\\
2018-03-07.8 & +22.1 & \nd & 15.915(069) & 15.236(041) & 14.967(075) & 15.106(059) & \nd & \nd & \nd & Konkoly\\
2018-03-08.1 & +22.4 & 17.766(037) & 15.984(014) & 15.228(010) & \nd & \nd & 15.577(010) & 15.184(009) & 15.761(010) & Swope\\
2018-03-08.5 & +22.8 & \nd & 16.113(032) & 15.350(011) & \nd & \nd & 15.566(004) & 15.270(005) & 15.780(007) & TNT\\
2018-03-08.8 & +23.1 & \nd & \nd & \nd & \nd & \nd & \nd & 15.289(007) & 15.753(010) & LCO\\
2018-03-08.9 & +23.2 & 16.398(032) & 16.143(019) & 15.361(015) & \nd & \nd & 15.637(005) & 15.275(006) & 15.725(019) & LCO\\
2018-03-08.9 & +23.2 & \nd & 16.083(100) & 15.303(035) & 15.041(033) & 15.070(041) & \nd & \nd & \nd & Konkoly\\
2018-03-09.1 & +23.4 & 17.813(033) & 16.144(014) & 15.304(010) & \nd & \nd & 15.680(010) & 15.216(008) & 15.770(009) & Swope\\
2018-03-09.2 & +23.5 & \nd & 16.126(031) & 15.375(035) & \nd & \nd & \nd & 15.297(024) & 15.694(043) & PONM\\
2018-03-09.2 & +23.5 & \nd & \nd & 15.297(008) & 15.156(009) & 15.206(009) & \nd & \nd & \nd & slotis\\
2018-03-09.6 & +23.9 & \nd & 16.192(032) & 15.379(011) & \nd & \nd & 15.629(006) & 15.262(005) & 15.701(008) & TNT\\
2018-03-10.2 & +24.5 & \nd & 16.165(036) & 15.372(045) & \nd & \nd & \nd & 15.242(024) & 15.620(026) & DEMONEXT\\
2018-03-10.2 & +24.5 & \nd & \nd & 15.338(016) & 15.163(024) & 15.206(023) & \nd & \nd & \nd & slotis\\
2018-03-10.7 & +25.0 & \nd & 16.330(033) & 15.444(011) & \nd & \nd & 15.744(005) & 15.311(005) & 15.720(007) & TNT\\
2018-03-11.8 & +26.1 & \nd & 16.368(051) & 15.473(016) & \nd & \nd & 15.880(043) & 15.287(010) & 15.656(027) & LJT\\
2018-03-11.9 & +26.2 & 16.765(045) & 16.452(031) & \nd & \nd & \nd & \nd & \nd & \nd & LCO\\
2018-03-12.1 & +26.4 & \nd & 16.389(037) & 15.504(055) & \nd & \nd & \nd & 15.282(020) & 15.641(036) & DEMONEXT\\
2018-03-12.2 & +26.5 & 17.965(116) & 16.406(016) & 15.486(010) & \nd & \nd & 15.949(010) & 15.267(008) & 15.734(009) & Swope\\
2018-03-12.6 & +26.9 & \nd & 16.349(045) & \nd & \nd & \nd & \nd & \nd & \nd & TNT\\
2018-03-12.7 & +27.0 & \nd & 16.452(046) & 15.549(016) & \nd & \nd & 15.964(011) & 15.306(013) & 15.647(029) & LJT\\
2018-03-12.8 & +27.1 & \nd & \nd & 15.482(103) & 15.155(064) & 15.117(072) & \nd & \nd & \nd & Konkoly\\
2018-03-13.1 & +27.4 & 18.174(058) & 16.468(018) & 15.477(011) & \nd & \nd & 15.965(011) & 15.314(009) & 15.641(010) & Swope\\
2018-03-13.2 & +27.5 & 16.774(031) & 16.472(018) & 15.549(015) & \nd & \nd & 16.001(005) & 15.365(005) & 15.633(011) & LCO\\
2018-03-13.3 & +27.6 & \nd & \nd & 15.481(009) & 15.167(009) & 15.075(009) & \nd & \nd & \nd & slotis\\
2018-03-13.4 & +27.7 & \nd & \nd & \nd & \nd & \nd & \nd & 15.373(008) & 15.616(014) & LCO\\
2018-03-13.8 & +28.2 & \nd & 16.559(075) & 15.529(023) & 15.134(044) & 14.995(036) & \nd & \nd & \nd & Konkoly\\
2018-03-14.1 & +28.4 & 18.259(055) & 16.503(016) & 15.561(011) & \nd & \nd & 16.094(010) & 15.335(008) & 15.662(009) & Swope\\
2018-03-14.2 & +28.5 & \nd & \nd & \nd & \nd & \nd & \nd & \nd & \nd & slotis\\
2018-03-14.7 & +29.0 & \nd & 16.634(054) & 15.666(026) & \nd & \nd & 16.118(014) & 15.357(020) & 15.603(036) & LJT\\
2018-03-14.8 & +29.1 & \nd & 16.612(051) & 15.572(035) & 15.161(045) & 14.998(041) & \nd & \nd & \nd & Konkoly\\
2018-03-15.1 & +29.4 & 18.335(059) & 16.591(018) & 15.600(011) & \nd & \nd & 16.159(010) & 15.353(009) & 15.645(009) & Swope\\
2018-03-15.6 & +29.9 & \nd & 16.650(034) & 15.673(015) & \nd & \nd & 16.092(009) & 15.411(011) & 15.635(017) & TNT\\
2018-03-16.1 & +30.4 & 16.902(081) & 16.724(018) & 15.718(015) & \nd & \nd & 16.245(005) & 15.446(004) & 15.621(007) & LCO\\
2018-03-16.1 & +30.4 & 18.300(054) & 16.650(016) & 15.672(012) & \nd & \nd & 16.214(011) & 15.411(009) & 15.621(009) & Swope\\
2018-03-16.3 & +30.6 & \nd & \nd & 15.601(013) & \nd & \nd & \nd & \nd & \nd & slotis\\
2018-03-17.1 & +31.4 & 18.478(064) & 16.754(018) & 15.768(012) & \nd & \nd & 16.314(011) & 15.437(009) & 15.642(009) & Swope\\
2018-03-18.1 & +32.4 & 18.632(107) & 16.832(018) & 15.847(012) & \nd & \nd & 16.395(011) & 15.502(008) & 15.722(009) & Swope\\
2018-03-18.5 & +32.8 & \nd & 16.973(039) & 15.876(015) & \nd & \nd & 16.350(008) & 15.599(013) & 15.746(014) & TNT\\
2018-03-19.1 & +33.4 & \nd & 16.919(022) & 15.878(015) & \nd & \nd & 16.436(007) & 15.605(006) & 15.706(009) & LCO\\
2018-03-19.1 & +33.4 & 18.615(148) & 16.910(018) & \nd & \nd & \nd & 16.383(011) & 15.534(008) & 15.684(009) & Swope\\
2018-03-19.5 & +33.8 & \nd & 16.996(036) & 15.941(014) & \nd & \nd & 16.385(012) & 15.660(009) & 15.804(016) & TNT\\
2018-03-20.1 & +34.4 & 18.770(153) & 16.984(018) & 15.868(013) & \nd & \nd & 16.535(011) & 15.644(008) & \nd & Swope\\
2018-03-20.1 & +34.4 & \nd & 16.983(015) & 15.866(010) & \nd & \nd & 16.540(008) & 15.649(005) & \nd & Konkoly\\
2018-03-21.0 & +35.3 & 18.759(154) & 17.012(018) & 15.968(012) & \nd & \nd & 16.542(012) & 15.656(009) & 15.820(009) & Swope\\
2018-03-21.3 & +35.6 & \nd & \nd & 15.884(011) & 15.539(012) & 15.267(013) & \nd & \nd & \nd & slotis\\
2018-03-21.6 & +35.9 & \nd & 17.114(039) & 16.047(021) & \nd & \nd & 16.496(018) & 15.823(037) & 15.926(069) & TNT\\
2018-03-22.5 & +36.8 & \nd & 17.128(035) & 16.068(013) & \nd & \nd & 16.528(008) & 15.831(008) & 15.930(011) & TNT\\
2018-03-23.1 & +37.4 & 19.104(161) & 17.196(020) & \nd & \nd & \nd & \nd & \nd & \nd & Swope\\
2018-03-23.6 & +37.9 & \nd & 17.123(037) & 16.137(018) & \nd & \nd & 16.583(013) & 15.907(012) & 16.017(016) & TNT\\
2018-03-24.1 & +38.4 & 17.359(131) & 17.170(133) & 16.152(021) & \nd & \nd & 16.687(016) & 15.882(011) & 15.964(017) & LCO\\
2018-03-24.1 & +38.4 & 19.102(168) & 17.149(023) & 16.117(013) & \nd & \nd & 16.679(015) & 15.870(010) & 16.002(011) & Swope\\
2018-03-24.3 & +38.6 & \nd & \nd & \nd & \nd & 15.414(306) & \nd & \nd & \nd & slotis\\
2018-03-24.7 & +39.0 & \nd & \nd & \nd & \nd & \nd & 16.635(012) & 15.957(010) & 16.074(015) & TNT\\
2018-03-25.1 & +39.4 & \nd & 17.266(025) & 16.205(015) & \nd & \nd & 16.681(016) & 15.948(011) & 16.054(012) & Swope\\
2018-03-26.1 & +40.4 & 19.129(157) & 17.190(025) & 16.239(015) & \nd & \nd & 16.804(016) & 16.011(011) & \nd & Swope\\
2018-03-27.1 & +41.4 & \nd & \nd & \nd & \nd & \nd & 16.794(048) & 15.984(034) & 16.269(044) & Swope\\
2018-03-29.1 & +43.4 & 17.372(135) & 17.140(079) & 16.272(051) & \nd & \nd & 16.983(055) & 16.159(041) & 16.293(061) & LCO\\
2018-03-29.3 & +43.6 & \nd & \nd & 16.276(021) & 15.926(023) & 15.721(022) & \nd & \nd & \nd & slotis\\
2018-03-31.3 & +45.6 & \nd & \nd & 16.298(054) & 16.037(073) & \nd & \nd & \nd & \nd & slotis\\
2018-04-01.7 & +47.0 & \nd & 17.283(049) & 16.459(019) & \nd & \nd & 16.864(015) & 16.257(015) & 16.467(030) & LJT\\
2018-04-02.7 & +48.0 & \nd & 17.296(063) & 16.507(028) & \nd & \nd & 16.936(020) & 16.346(035) & \nd & LJT\\
2018-04-03.7 & +49.0 & \nd & 17.404(047) & 16.584(028) & \nd & \nd & 16.912(017) & 16.358(016) & 16.544(032) & LJT\\
2018-04-04.1 & +49.4 & 17.502(093) & \nd & 16.482(028) & \nd & \nd & 16.912(026) & 16.321(019) & 16.541(028) & LCO\\
2018-04-04.2 & +49.5 & \nd & \nd & \nd & \nd & \nd & \nd & \nd & \nd & slotis\\
2018-04-05.3 & +50.6 & \nd & \nd & 16.445(016) & 16.229(017) & 16.021(018) & \nd & \nd & \nd & slotis\\
2018-04-07.0 & +52.3 & \nd & \nd & \nd & \nd & \nd & 16.885(002) & \nd & 16.819(002) & Decam\\
2018-04-07.1 & +52.4 & 19.186(284) & 17.438(018) & 16.593(013) & \nd & \nd & 17.001(011) & 16.451(010) & 16.700(013) & Swope\\
2018-04-07.6 & +52.9 & \nd & \nd & \nd & \nd & \nd & \nd & 16.396(221) & \nd & TNT\\
2018-04-08.6 & +53.9 & \nd & \nd & \nd & \nd & \nd & 16.989(009) & 16.553(010) & 16.779(013) & TNT\\
2018-04-10.6 & +55.9 & \nd & \nd & \nd & \nd & \nd & 16.992(008) & 16.590(008) & 16.820(010) & TNT\\
2018-04-12.0 & +57.3 & 19.274(126) & 17.541(014) & 16.761(011) & \nd & \nd & 17.159(011) & 16.649(009) & 16.937(012) & Swope\\
2018-04-12.7 & +58.0 & \nd & 17.473(056) & 16.831(101) & \nd & \nd & 17.043(048) & 16.861(340) & \nd & LJT\\
2018-04-13.0 & +58.3 & 19.290(240) & 17.518(015) & 16.759(012) & \nd & \nd & 17.100(010) & 16.653(009) & 16.915(011) & Swope\\
2018-04-14.0 & +59.3 & \nd & \nd & \nd & \nd & \nd & 16.978(002) & \nd & 17.09(001) & Decam\\
2018-04-14.1 & +59.4 & \nd & 17.570(015) & 16.768(011) & \nd & \nd & 17.149(010) & 16.660(009) & 16.978(011) & Swope\\
2018-04-15.1 & +60.4 & 19.439(167) & 17.537(015) & 16.731(011) & \nd & \nd & 17.108(010) & 16.654(009) & 16.926(011) & Swope\\
2018-04-15.5 & +60.8 & \nd & \nd & \nd & \nd & \nd & 17.133(011) & 16.814(016) & 17.080(017) & TNT\\
2018-04-16.0 & +61.3 & 19.431(274) & 17.678(016) & 16.826(013) & \nd & \nd & 17.513(031) & 16.808(011) & 17.097(016) & Swope\\
2018-04-16.6 & +61.9 & \nd & \nd & \nd & \nd & \nd & 17.143(015) & 16.829(017) & 17.106(015) & TNT\\
2018-04-17.0 & +62.3 & 19.540(264) & 17.495(028) & 16.857(024) & \nd & \nd & 17.173(015) & 16.696(016) & \nd & Swope\\
2018-04-17.0 & +63.3 & \nd & \nd & \nd & \nd & \nd & 17.087(002) & \nd & 17.243(002) & Decam\\
2018-04-17.6 & +62.9 & \nd & \nd & \nd & \nd & \nd & 17.133(017) & 16.841(014) & 17.073(018) & TNT\\
2018-04-18.0 & +63.3 & 19.553(246) & 17.686(022) & 16.933(016) & \nd & \nd & 17.297(015) & 16.823(015) & 17.129(027) & Swope\\
2018-04-20.5 & +65.8 & \nd & 17.536(069) & 16.908(039) & \nd & \nd & 17.119(040) & 16.835(023) & 17.040(043) & TNT\\
2018-04-23.5 & +68.8 & \nd & 17.802(114) & \nd & \nd & \nd & \nd & \nd & \nd & TNT\\
2018-04-24.0 & +70.3 & \nd & \nd & \nd & \nd & \nd & 17.229(002) & \nd & 17.519(002) & Decam\\
2018-04-25.0 & +70.3 & 19.518(266) & 17.711(030) & 17.030(019) & \nd & \nd & 17.401(019) & 17.022(015) & 17.438(020) & Swope\\
2018-04-25.6 & +70.9 & \nd & 17.687(079) & 17.054(041) & \nd & \nd & 17.271(049) & 17.070(042) & 17.384(084) & TNT\\
2018-04-26.5 & +71.8 & \nd & \nd & 17.122(039) & \nd & \nd & 17.271(040) & 17.157(035) & 17.430(063) & TNT\\
2018-04-28.5 & +73.8 & \nd & 17.674(150) & 17.115(127) & \nd & \nd & 17.336(108) & 17.042(076) & 17.268(114) & TNT\\
2018-05-02.0 & +78.3 & \nd & \nd & \nd & \nd & \nd & \nd & \nd & 17.797(004) & Decam\\
2018-05-03.0 & +79.3 & \nd & \nd & 17.279(011) & \nd & \nd & 17.477(011) & 17.296(010) & 17.703(016) & Swope\\
2018-05-04.5 & +79.8 & \nd & 17.747(051) & 17.282(039) & \nd & \nd & 17.388(034) & 17.331(029) & 17.593(055) & TNT\\
2018-05-06.0 & +81.3 & 19.837(241) & 17.944(019) & 17.326(016) & \nd & \nd & 17.538(013) & 17.382(013) & 17.775(022) & Swope\\
2018-05-07.0 & +83.3 & \nd & \nd & \nd & \nd & \nd & 17.473(004) & \nd & 18.028(01) & Decam\\
2018-05-07.0 & +83.3 & \nd & \nd & \nd & \nd & \nd & 17.495(009) & \nd & 18.042(016) & Decam\\
2018-05-08.5 & +83.8 & \nd & 17.840(047) & 17.384(042) & \nd & \nd & 17.378(032) & 17.483(066) & 17.497(045) & TNT\\
2018-05-09.0 & +84.3 & \nd & 17.921(017) & 17.332(014) & \nd & \nd & 17.528(012) & 17.448(015) & 17.855(020) & Swope\\
2018-05-10.0 & +86.3 & 19.585(252) & 17.976(021) & 17.398(016) & \nd & \nd & 17.601(014) & 17.525(019) & 17.906(029) & Swope\\
2018-05-12.0 & +88.3 & 19.440(359) & 18.126(019) & 17.545(017) & \nd & \nd & 17.699(013) & 17.672(016) & 18.018(027) & Swope\\
2018-05-17.0 & +92.3 & 19.717(242) & 18.111(016) & 17.587(016) & \nd & \nd & 17.721(011) & 17.719(012) & 18.176(022) & Swope\\
2018-05-18.6 & +93.9 & \nd & 17.818(063) & 17.550(056) & \nd & \nd & \nd & \nd & \nd & TNT\\
2018-05-19.0 & +95.3 & \nd & 18.196(018) & \nd & \nd & \nd & 17.827(014) & \nd & \nd & Swope\\
2018-05-20.0 & +95.3 & 19.922(271) & \nd & \nd & \nd & \nd & \nd & \nd & \nd & Swope\\
2018-05-22.0 & +98.3 & 20.149(316) & 18.259(017) & \nd & \nd & \nd & \nd & 17.911(013) & 18.325(019) & Swope\\
2018-05-24.5 & +99.8 & \nd & 17.938(174) & 17.739(080) & \nd & \nd & 17.714(060) & 17.813(050) & 18.131(108) & TNT\\
2018-05-25.0 & +100.3 & \nd & \nd & \nd & \nd & \nd & 17.897(013) & 17.962(013) & 18.402(017) & Swope\\
2018-06-03.0 & +110.3 & 19.680(241) & 18.386(012) & 17.921(010) & \nd & \nd & 18.000(010) & 18.260(018) & 18.664(030) & Swope\\
2018-06-06.9 & +113.3 & 20.098(306) & 18.326(016) & 18.011(015) & \nd & \nd & \nd & \nd & \nd & Swope\\
2018-06-07.0 & +114.3 & \nd & \nd & \nd & \nd & \nd & 18.014(010) & 18.325(013) & 18.677(025) & Swope\\
\enddata
\tablenotetext{a}{Days relative to the B-banx maximum on 2018-02-13.7 (JD 2458163.2).}
\tablenotetext{}{Note: Uncertainties, in units of 0.001 mag, are $1\sigma$.}
\end{deluxetable}

\startlongtable
\begin{deluxetable}{cccccc}
\tablecolumns{6} \tablewidth{0pc} \tabletypesize{\scriptsize}
\tablecaption{NIR Photometry of SN 2018oh  \label{sphoto} }
\tablehead{\colhead{Date} &\colhead{\tablenotemark{a}Epoch} &
\colhead{$Y$ (mag)} & \colhead{$J$ (mag)} & \colhead{$H$ (mag)} & \colhead{$K$ (mag)} } \startdata
2018-02-07.2 & $-$6.5 & 15.571(051) & 14.900(061) & 15.196(111) & \nd \\
2018-02-09.2 & $-$4.5 & 15.442(049) & 14.767(061) & 15.215(121) & \nd \\
2018-02-13.2 & $-$0.5 & 15.778(047) & 14.891(058) & 15.208(100) & \nd \\
2018-02-15.2 & +1.5 & \nd & 15.032(055) & 15.361(124) & \nd \\
2018-02-16.2 & +2.5 & 16.136(069) & 15.106(067) & 15.581(159) & \nd \\
2018-02-17.2 & +3.5 & \nd & 15.177(051) & 15.347(070) & 14.986(098) \\
2018-02-18.2 & +4.5 & 16.344(079) & 15.385(074) & 15.715(156) & \nd \\
2018-02-20.1 & +6.4 & 16.627(108) & 15.600(079) & 15.488(114) & \nd \\
2018-02-21.1 & +7.4 & \nd & 15.603(052) & 15.424(070) & 15.234(098) \\
2018-02-23.2 & +9.5 & 16.570(099) & 16.026(100) & 15.562(147) & \nd \\
2018-02-25.1 & +11.4 & 16.848(099) & 16.607(112) & 15.648(113) & \nd \\
2018-03-03.2 & +17.5 & 16.757(110) & 16.854(163) & 15.597(111) & \nd \\
2018-03-05.1 & +19.4 & 16.428(090) & 16.883(194) & 15.481(125) & \nd \\
2018-03-08.1 & +22.4 & 16.195(054) & 16.462(134) & 15.152(108) & \nd \\
2018-03-09.1 & +23.4 & \nd & 16.340(054) & 15.164(071) & 15.086(098) \\
2018-03-11.1 & +25.4 & 15.856(041) & 16.564(120) & 15.175(098) & \nd \\
2018-03-27.1 & +41.4 & \nd & 16.380(057) & 15.651(071) & 15.757(100) \\
2018-04-08.0 & +53.3 & \nd & 17.258(059) & 16.234(073) & 16.238(102) \\
\enddata
\tablenotetext{a}{Days relative to B-band maximum on 2018-02-13.7 (JD 2458163.2).}
\tablenotetext{}{Note: Uncertainties, in units of 0.001 mag, are $1\sigma$.}
\end{deluxetable}

\startlongtable
\begin{deluxetable}{cccccccc}
\tablecolumns{6} \tablewidth{0pc} \tabletypesize{\scriptsize}
\tablecaption{$Swift$ Photometry of SN 2018oh  \label{sphoto} }
\tablehead{\colhead{Date} &\colhead{\tablenotemark{a}Epoch} &
\colhead{uvw2 (mag)} & \colhead{uvm2 (mag)} & \colhead{uvw1 (mag)} & \colhead{\textit{U} (mag)}& \colhead{\textit{B} (mag)} & \colhead{\textit{V} (mag)}} \startdata
2018-02-05.4 & $-$8.3 & 17.117(092) & 18.04(146) & 15.783(065) & 14.228(045) & 14.784(045) & 14.804(061) \\
2018-02-06.8 & $-$6.9 & 16.889(087) & 17.981(116) & 15.499(066) & 13.955(044) & 14.611(044) & 14.688(063) \\
2018-02-07.5 & $-$6.2 & 16.967(096) & 17.98(135) & 15.394(069) & 13.912(044) & 14.501(044) & 14.675(067) \\
2018-02-10.4 & $-$3.3 & 16.647(086) & 17.73(116) & 15.295(066) & 13.714(043) & 14.305(043) & 14.247(054) \\
2018-02-17.0 & +4.3 & 16.848(102) & 17.708(132) & 15.678(078) & 14.147(047) & 14.36(043) & 14.251(056) \\
2018-02-19.4 & +5.7 & 17.042(083) & 17.805(1) & 15.784(066) & 14.291(046) & 14.4(043) & 14.4(057) \\
2018-02-22.7 & +9.0 & 17.448(11) & \nd & 16.279(076) & 14.735(052) & 14.66(044) & \nd \\
2018-02-26.4 & +12.7 & 17.586(098) & 18.003(107) & 16.498(077) & 15.153(058) & 14.911(046) & 14.719(06) \\
2018-03-11.5 & +25.8 & 18.374(117) & 18.645(121) & 17.595(099) & 16.792(082) & 16.332(063) & 15.416(064) \\
2018-03-17.1 & +31.4 & 18.351(125) & 18.55(133) & 17.861(119) & 16.887(091) & 16.719(07) & 15.663(071) \\
\enddata
\tablenotetext{a}{Days relative to B-band maximum on 2018-02-13.7 (JD 2458163.2).}
\tablenotetext{}{Note: Uncertainties, in units of 0.001 mag.}
\end{deluxetable}

\startlongtable
\begin{deluxetable}{cccccc}
\tablecolumns{6} \tablewidth{0pc} \tabletypesize{\scriptsize}
\tablecaption{Log of Spectroscopic Observations of SN 2018oh  \label{log} }
\tablehead{\colhead{UT Date} &\colhead{MJD} &
\colhead{Epoch$^{a}$} & \colhead{Range (\AA)} & \colhead{Res. (\AA) } & \colhead{Inst.}}\startdata
2018-02-05.2	&	58154.2	&	$-$8.5	&	3640-10298	&	4.0　　　&	HET	\\
2018-02-05.2	&	58154.2	&	$-$8.5	&	3640-5220	&	2.0	　　&	SOAR\\
2018-02-05.5	&	58154.5	&	$-$8.2	&	3300-10000	&	10.0	&	Las Cumbres\\
2018-02-06.2	&	58155.2	&	$-$7.5	&	3380-10320	&	15.8	&NTT\\
2018-02-06.7	&	58155.7	&	$-$7.0	&	3498-9173	&	25.0	&	LJT	\\
2018-02-07.2	&	58156.2	&	$-$6.5	&	3190-10914	&	7.0	　　&	Shane	\\
2018-02-07.3	&	58156.3	&	$-$6.4	&	3640-7977	&	10.0	&	Bok	\\
2018-02-07.3	&	58156.3	&	$-$6.4	&	3685-9315	&	21.2	&NTT\\
2018-02-08.2	&	58157.2	&	$-$5.5	&	3640-10298	&	4.0	　　&	HET	\\
2018-02-08.3	&	58157.3	&	$-$5.4	&	3180-11252	&	7.0	　　　&	Shane	\\
2018-02-09.4	&	58158.4	&	$-$4.3	&	3250-10000	&	10.0	&	Las Cumbres	\\
2018-02-09.5	&	58158.5	&	$-$4.2	&	3986-8834	&	15.0	&	XLT	\\
2018-02-10.1	&	58159.1	&	$-$3.6	&	3640-5220	&	2.8　	&	SOAR	\\
2018-02-10.3	&	58159.3	&	$-$3.4	&	3799-9627	&	15.0	&	APO	\\
2018-02-11.7	&	58160.7	&	$-$2.0	&	3976-8830	&	15.0	&	XLT	\\
2018-02-13.6	&	58162.6	&	$-$0.1	&	3966-8816	&	15.0	&	XLT	\\
2018-02-14.2	&	58163.2	&	+0.5	&	3380-10320	&	15.8	&NTT\\
2018-02-14.2	&	58163.2	&	+0.5	&	3640-5220	&	2.8	　　&	SOAR	\\
2018-02-14.6	&	58163.6	&	+0.9	&	3249-10000	&	10.0	&	Las Cumbres	\\
2018-02-15.5	&	58164.5	&	+1.8	&	3976-8831	&	2.8	　　&	XLT	\\
2018-02-16.3	&	58165.3	&	+2.6	&	3380-7520	&	15.8	&　NTT　\\
2018-02-16.6	&	58165.6	&	+2.9	&	3975-8831	&	15.0	&	XLT	\\
2018-02-18.7	&	58167.7	&	+5.0	&	3958-8812	&	15.0	&	XLT	\\
2018-02-19.2	&	58168.2	&	+5.5	&	3380-7520	&	15.8	&　NTT　\\
2018-02-19.5	&	58168.5	&	+5.8	&	3959-8816	&	15.0	&	XLT	\\
2018-02-20.5	&	58169.5	&	+6.8	&	3400-10000	&	10.0	&	Las Cumbres	\\
2018-02-21.7	&	58170.7	&	+8.0	&	3981-8835	&	15.0	&	XLT	\\
2018-02-22.2	&	58171.2	&	+8.5	&	3380-7520	&	15.8	&　NTT　\\
2018-02-27.7	&	58176.7	&	+14.0	&	3501-9166	&	25.0	&	LJT	\\
2018-03-01.7	&	58178.7	&	+16.0	&	3501-9155	&	25.0	&	LJT	\\
2018-03-06.2	&  	58183.2 & 	+20.5   &	5601-6905 	&	1.5 	&	MMT	\\ 
2018-03-07.4	&	58184.4	&	+21.7	&	3250-10000	&	10.0	&	Las Cumbres	\\
2018-03-08.2	&	58185.2	&	+22.5	&	3380-10320	&	15.8	&　NTT　\\
2018-03-09.6	&	58186.6	&	+23.9	&	3961-8815	&	15.0	&	XLT	\\
2018-03-11.6	&	58188.6	&	+25.9	&	3899-9299	&	10.0	&	Las Cumbres	\\
2018-03-12.7	&	58189.7	&	+27.0	&	3497-9166	&	25.0	&	LJT	\\
2018-03-14.2	&	58191.2	&	+28.5	&	3752-9208	&	2.0　　　&	Magellan	\\
2018-03-15.6	&	58192.6	&	+29.9	&	3600-9999	&	10.0	&	Las Cumbres	\\
2018-03-19.4	&	58196.4	&	+33.7	&	3249-9999	&	10.0	&	Las Cumbres	\\
2018-03-19.7	&	58196.7	&	+34.0	&	3503-9165	&	25.0	&	LJT	\\
2018-03-22.1	&	58199.1	&	+36.4	&	3500-9040	&	6.0	&	SOAR	\\
2018-03-23.5	&	58200.5	&	+37.8	&	3965-8822	&	15.0	&	XLT	\\
2018-03-23.7	&	58200.7	&	+38.0	&	3492-9160	&	25.0	&	LJT	\\
2018-03-24.1	&	58201.1	&	+38.4	&	3380-10320	&	15.8	&NTT\\
2018-03-25.0	&	58202.0	&	+39.3	&	3966-8822	&	15.0	&	XLT	\\
2018-04-06.1	&	58214.1	&	+51.4	&	3715-8061	&	10.0	&	Bok	\\
2018-04-06.1	&	58214.1	&	+51.4	&	3380-10320	&	15.8	&NTT\\
2018-04-07.5	&	58215.5	&	+52.8	&	3966-8822	&	15.0	&	XLT	\\
2018-04-21.0	&	58229.0	&	+66.3	&	3560-8948	&	6.0	&	SOAR	\\
2018-04-21.0	&	58229.0	&	+66.3	&	3380-10320	&	15.8	&NTT\\
2018-04-25.0	&	58233.0	&	+70.3	&	3180-11252	&	7.0	&	Shane	\\
2018-04-27.6	&	58235.6	&	+72.9	&	3966-8822	&	15.0	&	XLT	\\
2018-04-27.6	&	58235.6	&	+72.9	&	3492-9160	&	25.0	&	LJT	\\
2018-05-02.6	&	58240.6	&	+77.9	&	3966-8822	&	15.0	&	XLT	\\
2018-05-08.0	&	58246.0	&	+83.3	&	3180-11252	&	7.0　　　&	Shane	\\
2018-05-08.5	&	58246.5	&	+83.8	&	3966-8822	&	15.0	&	XLT	\\
\enddata
\tablenotetext{a}{Days relative to B-band maximum on 2018-02-13.7 (JD 2458163.2).}
\end{deluxetable}

\startlongtable
\begin{deluxetable}{cccccc}
\tablecolumns{5} \tablewidth{0pc} \tabletypesize{\scriptsize}
\tablecaption{Photometry parameters of SN 2018oh  \label{photo_info} }
\tablehead{\colhead{Band} &\colhead{$\lambda_{eff}$ (\AA)} &
\colhead{$t_{max}$ (MJD)} & \colhead{$m_{peak}$ (mag)} & \colhead{$\Delta m15$  (mag)} } \startdata
uvw2	&	2030	&	58161.2 $\pm$ 0.2	&	16.67 $\pm$ 0.07	&	1.08 $\pm$ 0.49	\\
uvm2	&	2228	&	58164.1	$\pm$ 0.8	&	17.71 $\pm$ 0.05	&	0.49 $\pm$ 0.46\\
uvw1	&	2589	&	58160.5 $\pm$ 0.1	&	15.31 $\pm$ 0.07	&	1.32 $\pm$ 0.45	\\
$U$ 	&	3663	&	58161.1 $\pm$ 0.1	&	13.98 $\pm$ 0.01	&	1.19 $\pm$ 0.12	\\
$B$ 	&	4360	&	58162.9 $\pm$ 0.1	&	14.32 $\pm$ 0.01	&	0.96 $\pm$ 0.02	\\
$V$	&	5446	&	58164.1 $\pm$ 0.1	&	14.37 $\pm$ 0.01	&	0.63 $\pm$ 0.06	\\
$R$	&	6414	&	58163.7 $\pm$ 0.2	&	14.21 $\pm$ 0.01	&	0.69 $\pm$ 0.09	\\
$I$	&	7979	&	58161.7 $\pm$ 0.2	&	14.47 $\pm$ 0.02	&	0.64 $\pm$ 0.15	\\
$g$	&	4640	&	58163.6 $\pm$ 0.2	&	14.22 $\pm$ 0.01	&	0.82 $\pm$ 0.07	\\
$r$	&	6122	&	58163.3 $\pm$ 0.1	&	14.38 $\pm$ 0.01	&	0.70 $\pm$ 0.08	\\
$i$ &	7440	&	58160.4 $\pm$ 0.1	&	14.91 $\pm$ 0.01	&	0.85 $\pm$ 0.07	\\
\enddata
\end{deluxetable}

\begin{table}
\caption{Best-fit parameters from the applied LC-fitters}
\label{tab:lcfits}
\centering
\begin{tabular}{llll}
\hline
\hline
Parameter & SALT2.4 & SNooPy2 & MLCS2k2 \\
\hline
$T_{max}(B)$ (MJD) & 58163.34 (0.02) & 58162.67 (0.05) &  58162.70 (0.02) \\
$x_0$ & 0.038 (0.001) & -- & -- \\
$x_1$ & 0.879 (0.012) & -- & -- \\
$C$ & $-0.09$ (0.010) & -- & -- \\
$E(B-V)_{host}$ & -- & 0.00 (0.01) & 0.00 (0.01) \\
$\Delta m_{15}$ & -- & 0.865 (0.060) & -- \\
$\Delta_{MLCS}$ & -- & -- & -0.100 (0.08) \\
$\mu_0$ (mag) & 33.614 (0.05) & 33.62 (0.22) & 33.57 (0.06) \\
\hline
\end{tabular}
\end{table}

\begin{table}[h]
\tiny
\centering
\caption{SYNOW fitting parameters of SN 2018oh  \label{synow_p} }
\begin{tabular}{ccccccccccc}
\hline\hline
Phase	& T$_{bb}$ 	& Si~II $\lambda$6355 (HV)	& Si~II $\lambda$6355	& C~II $\lambda$6580 	& Ca~II $\lambda$8498 (HV)	& Ca~II $\lambda$8498	& S~II $\lambda$5454	& O~I $\lambda$7774	& Fe~III $\lambda$5129	& Fe~III $\lambda$4404\\
(d) & (kK) & ($10^3$ km/s) & ($10^3$ km/s) & ($10^3$ km/s)  & ($10^3$ km/s)  & ($10^3$ km/s)  & ($10^3$ km/s)  & ($10^3$ km/s)  & ($10^3$ km/s)  & ($10^3$ km/s) \\
\hline
$-$8.0	& 10.20	& 14.13	& 11.90	& 14.00	& 19.25	& 12.56	& 10.40	& 10.69	& 10.29	& 9.81\\
$-$5.5	& 11.20	& 14.49	& 11.80	& 14.00	& 18.36	& 11.12	& 10.34	& 11.16	& 10.18	& 9.81\\
$-$3.0	& 11.81	& 13.77	& 11.95	& 14.50	& 18.65	& 11.86	& 9.63	& 9.77	& 9.88	& 10.15\\
+0.0	& 10.45	& 12.28	& 11.10	& 12.00	& 18.76	& 10.64	& 9.52	& 9.37	& 10.06	& 10.08\\
+5.0	& 9.79	& 12.60	& 9.75	& 10.00 	& 19.59	& 12.17	& 9.57	& 10.68	& 10.18	& 9.67\\
+8.0	& 9.41	& 12.61	& 9.42	& 9.67	& 19.48	& 11.32	& 9,79	& 10.58	& 10.12	& 10.15\\
+14.0	& 9.43	& \nd	& 10.15	& 8.16	& \nd	& 12.43	& 9.74	& 10.69	& 10.06	& 9.54\\
\hline
\end{tabular}
\end{table}

\begin{table}[h]
\centering
\caption{Parameters of SN 2018oh}
\label{pars}
\begin{tabular*}{4in}{@{\extracolsep{1in}}ll}
\hline\hline
\multicolumn{1}{c}{Parameter} & \multicolumn{1}{c}{Value} \\\hline
\multicolumn{2}{c}{Photometric}                            \\ 
$B\rm _{max}$             & 14.32$\pm$0.01 mag  \\
$B \rm_{max}$$-V \rm _{max}$ & $-$0.09$\pm$0.02 mag \\
$E(B-V)\rm _{host}$  & 0.00$\pm$0.04 mag   \\
$\Delta m_{15}(B)$ & 0.96$\pm$0.03 mag \\
$t\rm _{max}$($B$) 	& 58162.7$\pm$0.3 day \\
$t_0$	&	58144.37$\pm$0.04 day \\
$\tau_{rise}$	& 18.3$\pm$0.3 day \\
$L_{bol}^{max}$	& 1.49$\times10^{43}$ erg s$^{-1}$\\
$M_{^{56}Ni}$	& 0.55$\pm$0.04 $\rm M\odot$\\
\multicolumn{2}{c}{Spectroscopic}                            \\ 
$v \rm _{0}$(Si~II)	& 10,300$\pm$300 km s$^{-1}$\\
$\dot{v}$(Si~II)		&	69$\pm$4 km s$^{-1}$~d$^{-1}$\\
$R$(Si II)			  & 0.15$\pm$0.04\\
\hline
\end{tabular*}
\end{table}

\end{document}